\documentclass[longauth]{aa}

\usepackage{graphicx}
\usepackage{natbib}
\usepackage{scalerel}

\usepackage[table]{xcolor}

\usepackage{txfonts}
\usepackage[pdfencoding=auto,psdextra]{hyperref}
\hypersetup{
    colorlinks=true,
    linkcolor=blue,
    filecolor=magenta,      
    urlcolor=blue,
    citecolor=blue
}
\makeatletter
\renewcommand*\aa@pageof{, page \thepage{} of \pageref*{LastPage}}
\makeatother

\usepackage[utf8]{inputenc}

\usepackage[switch, modulo]{lineno}
\usepackage{euclid}
\usepackage{svg}
\usepackage{physics}

\begin{document}

\newcommand{\Omm}{\Omega_\mathrm{m}}
\newcommand{\Map}{M_\mathrm{ap}}
\newcommand{\MapEst}{\hat{M}_\mathrm{ap}}
\newcommand{\MapEstEps}{\hat{M}_{\mathrm{ap}, \epsilon}}

\newcommand{\Mperp}{M_{\perp}}
\newcommand{\MapMapMap}{\ensuremath{\expval{\Map^3}}}
\newcommand{\MapMap}{\expval{\Map^2}}

\newcommand{\MapMapMapEst}{\MapEst^3}
\newcommand{\MapMapMapEstEps}{\MapEstEps^3}

\newcommand{\MapMperpMperp}{\expval{\Map\Mperp^2}}
\newcommand{\astroang}[1]{\ang[angle-symbol-over-decimal]{#1}}
\newcommand{\ellvec}{\vec{\ell}}
\newcommand{\dirac}{\delta_\mathrm{D}}
\newcommand{\kronecker}[2]{\delta_{#1,#2}}
\newcommand{\Npix}{N_\mathrm{pix}}
\newcommand{\Nt}{N_\mathrm{t}}
\newcommand{\Apix}{A_\mathrm{pix}}
\newcommand{\varthetavec}{\vec{\vartheta}}
\newcommand{\varthetamax}{\vartheta_\mathrm{max}}
\newcommand{\I}{\mathrm{i}}
\newcommand{\E}{\mathrm{e}}
\newcommand{\etavec}{\vec{\eta}}
\newcommand{\alphavec}{\vec{\alpha}}
\newcommand{\qvec}{\vec{q}}
\newcommand{\realspace}{\mathbb{R}}
\newcommand{\Thetavec}{\vec{\Theta}}
\newcommand{\laila}[1]{\textcolor{magenta}{LL: #1}}
\newcommand{\kvec}{\vec{k}}
\newcommand{\svec}{\vec{s}}

\newcommand{\myexpval}[1]{\Big\langle #1 \Big\rangle}
\newcommand{\thetavec}{\vec{\theta}}

\newcommand{\MapMapEst}{\MapEst^2}

\newcommand{\TI}{T_{PPP, 1}}
\newcommand{\TII}{T_{PPP, 2}}
\newcommand{\TIV}{T_{BB}}
\newcommand{\TV}{T_{PT, 1}}
\newcommand{\TVI}{T_{PT, 2}}
\newcommand{\TVII}{T_{P_6}}
\newcommand{\rmg}{_\mathrm{g}}
%
%

\title{\Euclid and KiDS-1000: Quantifying the impact of source-lens clustering on cosmic shear analyses\thanks{This paper is published on behalf of the Euclid Consortium.}}    

   
\newcommand{\orcid}[1]{} 
\author{L.~Linke\orcid{0000-0002-2622-8113}\thanks{\email{Laila.Linke@uibk.ac.at}}\inst{\ref{aff1}}
\and S.~Unruh\orcid{0000-0002-3033-9967}\inst{\ref{aff2},\ref{aff3}}
\and A.~Wittje\orcid{0000-0002-8173-3438}\inst{\ref{aff3}}
\and T.~Schrabback\orcid{0000-0002-6987-7834}\inst{\ref{aff1}, \ref{aff2}}
\and S.~Grandis\orcid{0000-0002-4577-8217}\inst{\ref{aff1}}
\and M.~Asgari\orcid{0000-0002-3064-083X}\inst{\ref{aff4},\ref{aff5}}
\and A.~Dvornik\inst{\ref{aff6}}
\and H.~Hildebrandt\orcid{0000-0002-9814-3338}\inst{\ref{aff3}}
\and H.~Hoekstra\orcid{0000-0002-0641-3231}\inst{\ref{aff7}}
\and B.~Joachimi\orcid{0000-0001-7494-1303}\inst{\ref{aff8}}
\and R.~Reischke\inst{\ref{aff2}, \ref{aff3}}
\and J.~L.~van~den~Busch\orcid{0000-0001-9059-2553}\inst{\ref{aff3}}
\and A.~H.~Wright\orcid{0000-0001-7363-7932}\inst{\ref{aff3}}
\and P.~Schneider\orcid{0000-0001-8561-2679}\inst{\ref{aff2}}
\and N.~Aghanim\inst{\ref{aff9}}
\and B.~Altieri\orcid{0000-0003-3936-0284}\inst{\ref{aff10}}
\and A.~Amara\inst{\ref{aff11}}
\and S.~Andreon\orcid{0000-0002-2041-8784}\inst{\ref{aff12}}
\and N.~Auricchio\orcid{0000-0003-4444-8651}\inst{\ref{aff13}}
\and C.~Baccigalupi\orcid{0000-0002-8211-1630}\inst{\ref{aff14},\ref{aff15},\ref{aff16},\ref{aff17}}
\and M.~Baldi\orcid{0000-0003-4145-1943}\inst{\ref{aff18},\ref{aff13},\ref{aff19}}
\and S.~Bardelli\orcid{0000-0002-8900-0298}\inst{\ref{aff13}}
\and D.~Bonino\orcid{0000-0002-3336-9977}\inst{\ref{aff20}}
\and E.~Branchini\orcid{0000-0002-0808-6908}\inst{\ref{aff21},\ref{aff22},\ref{aff12}}
\and M.~Brescia\orcid{0000-0001-9506-5680}\inst{\ref{aff23},\ref{aff24},\ref{aff25}}
\and J.~Brinchmann\orcid{0000-0003-4359-8797}\inst{\ref{aff26},\ref{aff27}}
\and S.~Camera\orcid{0000-0003-3399-3574}\inst{\ref{aff28},\ref{aff29},\ref{aff20}}
\and V.~Capobianco\orcid{0000-0002-3309-7692}\inst{\ref{aff20}}
\and C.~Carbone\orcid{0000-0003-0125-3563}\inst{\ref{aff30}}
\and V.~F.~Cardone\inst{\ref{aff31},\ref{aff32}}
\and J.~Carretero\orcid{0000-0002-3130-0204}\inst{\ref{aff33},\ref{aff34}}
\and S.~Casas\orcid{0000-0002-4751-5138}\inst{\ref{aff35}}
\and F.~J.~Castander\orcid{0000-0001-7316-4573}\inst{\ref{aff36},\ref{aff37}}
\and M.~Castellano\orcid{0000-0001-9875-8263}\inst{\ref{aff31}}
\and S.~Cavuoti\orcid{0000-0002-3787-4196}\inst{\ref{aff24},\ref{aff25}}
\and A.~Cimatti\inst{\ref{aff38}}
\and G.~Congedo\orcid{0000-0003-2508-0046}\inst{\ref{aff39}}
\and C.~J.~Conselice\orcid{0000-0003-1949-7638}\inst{\ref{aff40}}
\and L.~Conversi\orcid{0000-0002-6710-8476}\inst{\ref{aff41},\ref{aff10}}
\and Y.~Copin\orcid{0000-0002-5317-7518}\inst{\ref{aff42}}
\and F.~Courbin\orcid{0000-0003-0758-6510}\inst{\ref{aff43}}
\and H.~M.~Courtois\orcid{0000-0003-0509-1776}\inst{\ref{aff44}}
\and A.~Da~Silva\orcid{0000-0002-6385-1609}\inst{\ref{aff45},\ref{aff46}}
\and H.~Degaudenzi\orcid{0000-0002-5887-6799}\inst{\ref{aff47}}
\and J.~Dinis\orcid{0000-0001-5075-1601}\inst{\ref{aff45},\ref{aff46}}
\and M.~Douspis\inst{\ref{aff9}}
\and F.~Dubath\orcid{0000-0002-6533-2810}\inst{\ref{aff47}}
\and X.~Dupac\inst{\ref{aff10}}
\and S.~Dusini\orcid{0000-0002-1128-0664}\inst{\ref{aff48}}
\and M.~Farina\orcid{0000-0002-3089-7846}\inst{\ref{aff49}}
\and S.~Farrens\orcid{0000-0002-9594-9387}\inst{\ref{aff50}}
\and S.~Ferriol\inst{\ref{aff42}}
\and P.~Fosalba\orcid{0000-0002-1510-5214}\inst{\ref{aff37},\ref{aff51}}
\and M.~Frailis\orcid{0000-0002-7400-2135}\inst{\ref{aff15}}
\and E.~Franceschi\orcid{0000-0002-0585-6591}\inst{\ref{aff13}}
\and M.~Fumana\orcid{0000-0001-6787-5950}\inst{\ref{aff30}}
\and S.~Galeotta\orcid{0000-0002-3748-5115}\inst{\ref{aff15}}
\and B.~Gillis\orcid{0000-0002-4478-1270}\inst{\ref{aff39}}
\and C.~Giocoli\orcid{0000-0002-9590-7961}\inst{\ref{aff13},\ref{aff52}}
\and A.~Grazian\orcid{0000-0002-5688-0663}\inst{\ref{aff53}}
\and F.~Grupp\inst{\ref{aff54},\ref{aff55}}
\and L.~Guzzo\orcid{0000-0001-8264-5192}\inst{\ref{aff56},\ref{aff12}}
\and S.~V.~H.~Haugan\orcid{0000-0001-9648-7260}\inst{\ref{aff57}}
\and W.~Holmes\inst{\ref{aff58}}
\and I.~Hook\orcid{0000-0002-2960-978X}\inst{\ref{aff59}}
\and F.~Hormuth\inst{\ref{aff60}}
\and A.~Hornstrup\orcid{0000-0002-3363-0936}\inst{\ref{aff61},\ref{aff62}}
\and P.~Hudelot\inst{\ref{aff63}}
\and K.~Jahnke\orcid{0000-0003-3804-2137}\inst{\ref{aff64}}
\and E.~Keih\"anen\orcid{0000-0003-1804-7715}\inst{\ref{aff65}}
\and S.~Kermiche\orcid{0000-0002-0302-5735}\inst{\ref{aff66}}
\and A.~Kiessling\orcid{0000-0002-2590-1273}\inst{\ref{aff58}}
\and M.~Kilbinger\orcid{0000-0001-9513-7138}\inst{\ref{aff50}}
\and T.~Kitching\orcid{0000-0002-4061-4598}\inst{\ref{aff67}}
\and B.~Kubik\orcid{0009-0006-5823-4880}\inst{\ref{aff42}}
\and K.~Kuijken\orcid{0000-0002-3827-0175}\inst{\ref{aff7}}
\and M.~K\"ummel\orcid{0000-0003-2791-2117}\inst{\ref{aff55}}
\and M.~Kunz\orcid{0000-0002-3052-7394}\inst{\ref{aff68}}
\and H.~Kurki-Suonio\orcid{0000-0002-4618-3063}\inst{\ref{aff69},\ref{aff70}}
\and S.~Ligori\orcid{0000-0003-4172-4606}\inst{\ref{aff20}}
\and P.~B.~Lilje\orcid{0000-0003-4324-7794}\inst{\ref{aff57}}
\and V.~Lindholm\orcid{0000-0003-2317-5471}\inst{\ref{aff69},\ref{aff70}}
\and I.~Lloro\inst{\ref{aff71}}
\and D.~Maino\inst{\ref{aff56},\ref{aff30},\ref{aff72}}
\and E.~Maiorano\orcid{0000-0003-2593-4355}\inst{\ref{aff13}}
\and O.~Mansutti\orcid{0000-0001-5758-4658}\inst{\ref{aff15}}
\and O.~Marggraf\orcid{0000-0001-7242-3852}\inst{\ref{aff2}}
\and K.~Markovic\orcid{0000-0001-6764-073X}\inst{\ref{aff58}}
\and N.~Martinet\orcid{0000-0003-2786-7790}\inst{\ref{aff73}}
\and F.~Marulli\orcid{0000-0002-8850-0303}\inst{\ref{aff74},\ref{aff13},\ref{aff19}}
\and R.~Massey\orcid{0000-0002-6085-3780}\inst{\ref{aff75}}
\and H.~J.~McCracken\orcid{0000-0002-9489-7765}\inst{\ref{aff63}}
\and E.~Medinaceli\orcid{0000-0002-4040-7783}\inst{\ref{aff13}}
\and S.~Mei\orcid{0000-0002-2849-559X}\inst{\ref{aff76}}
\and Y.~Mellier\inst{\ref{aff77},\ref{aff63}}
\and M.~Meneghetti\orcid{0000-0003-1225-7084}\inst{\ref{aff13},\ref{aff19}}
\and E.~Merlin\orcid{0000-0001-6870-8900}\inst{\ref{aff31}}
\and G.~Meylan\inst{\ref{aff43}}
\and M.~Moresco\orcid{0000-0002-7616-7136}\inst{\ref{aff74},\ref{aff13}}
\and L.~Moscardini\orcid{0000-0002-3473-6716}\inst{\ref{aff74},\ref{aff13},\ref{aff19}}
\and E.~Munari\orcid{0000-0002-1751-5946}\inst{\ref{aff15},\ref{aff14}}
\and R.~Nakajima\inst{\ref{aff2}}
\and R.~C.~Nichol\inst{\ref{aff11}}
\and S.-M.~Niemi\inst{\ref{aff78}}
\and J.~W.~Nightingale\orcid{0000-0002-8987-7401}\inst{\ref{aff4},\ref{aff79}}
\and C.~Padilla\orcid{0000-0001-7951-0166}\inst{\ref{aff80}}
\and S.~Paltani\orcid{0000-0002-8108-9179}\inst{\ref{aff47}}
\and F.~Pasian\orcid{0000-0002-4869-3227}\inst{\ref{aff15}}
\and K.~Pedersen\inst{\ref{aff81}}
\and V.~Pettorino\inst{\ref{aff78}}
\and S.~Pires\orcid{0000-0002-0249-2104}\inst{\ref{aff50}}
\and G.~Polenta\orcid{0000-0003-4067-9196}\inst{\ref{aff82}}
\and M.~Poncet\inst{\ref{aff83}}
\and L.~A.~Popa\inst{\ref{aff84}}
\and F.~Raison\orcid{0000-0002-7819-6918}\inst{\ref{aff54}}
\and R.~Rebolo\inst{\ref{aff85},\ref{aff86}}
\and A.~Renzi\orcid{0000-0001-9856-1970}\inst{\ref{aff87},\ref{aff48}}
\and J.~Rhodes\inst{\ref{aff58}}
\and G.~Riccio\inst{\ref{aff24}}
\and E.~Romelli\orcid{0000-0003-3069-9222}\inst{\ref{aff15}}
\and M.~Roncarelli\orcid{0000-0001-9587-7822}\inst{\ref{aff13}}
\and R.~Saglia\orcid{0000-0003-0378-7032}\inst{\ref{aff55},\ref{aff54}}
\and Z.~Sakr\orcid{0000-0002-4823-3757}\inst{\ref{aff88},\ref{aff89},\ref{aff90}}
\and D.~Sapone\orcid{0000-0001-7089-4503}\inst{\ref{aff91}}
\and B.~Sartoris\orcid{0000-0003-1337-5269}\inst{\ref{aff55},\ref{aff15}}
\and M.~Schirmer\orcid{0000-0003-2568-9994}\inst{\ref{aff64}}
\and A.~Secroun\orcid{0000-0003-0505-3710}\inst{\ref{aff66}}
\and G.~Seidel\orcid{0000-0003-2907-353X}\inst{\ref{aff64}}
\and S.~Serrano\orcid{0000-0002-0211-2861}\inst{\ref{aff37},\ref{aff92},\ref{aff36}}
\and C.~Sirignano\orcid{0000-0002-0995-7146}\inst{\ref{aff87},\ref{aff48}}
\and G.~Sirri\orcid{0000-0003-2626-2853}\inst{\ref{aff19}}
\and L.~Stanco\orcid{0000-0002-9706-5104}\inst{\ref{aff48}}
\and J.-L.~Starck\orcid{0000-0003-2177-7794}\inst{\ref{aff50}}
\and P.~Tallada-Cresp\'{i}\orcid{0000-0002-1336-8328}\inst{\ref{aff33},\ref{aff34}}
\and A.~N.~Taylor\inst{\ref{aff39}}
\and I.~Tereno\inst{\ref{aff45},\ref{aff93}}
\and R.~Toledo-Moreo\orcid{0000-0002-2997-4859}\inst{\ref{aff94}}
\and F.~Torradeflot\orcid{0000-0003-1160-1517}\inst{\ref{aff34},\ref{aff33}}
\and I.~Tutusaus\orcid{0000-0002-3199-0399}\inst{\ref{aff89}}
\and L.~Valenziano\orcid{0000-0002-1170-0104}\inst{\ref{aff13},\ref{aff95}}
\and T.~Vassallo\orcid{0000-0001-6512-6358}\inst{\ref{aff55},\ref{aff15}}
\and G.~Verdoes~Kleijn\orcid{0000-0001-5803-2580}\inst{\ref{aff96}}
\and A.~Veropalumbo\orcid{0000-0003-2387-1194}\inst{\ref{aff12},\ref{aff22},\ref{aff97}}
\and Y.~Wang\orcid{0000-0002-4749-2984}\inst{\ref{aff98}}
\and J.~Weller\orcid{0000-0002-8282-2010}\inst{\ref{aff55},\ref{aff54}}
\and G.~Zamorani\orcid{0000-0002-2318-301X}\inst{\ref{aff13}}
\and E.~Zucca\orcid{0000-0002-5845-8132}\inst{\ref{aff13}}
\and C.~Burigana\orcid{0000-0002-3005-5796}\inst{\ref{aff99},\ref{aff95}}
\and A.~Pezzotta\orcid{0000-0003-0726-2268}\inst{\ref{aff54}}
\and C.~Porciani\orcid{0000-0002-7797-2508}\inst{\ref{aff2}}
\and V.~Scottez\inst{\ref{aff77},\ref{aff100}}
\and M.~Viel\orcid{0000-0002-2642-5707}\inst{\ref{aff14},\ref{aff15},\ref{aff17},\ref{aff16},\ref{aff101}}
\and A.~M.~C.~Le~Brun\orcid{0000-0002-0936-4594}\inst{\ref{aff102}}}
										   
\institute{Universit\"at Innsbruck, Institut f\"ur Astro- und Teilchenphysik, Technikerstr. 25/8, 6020 Innsbruck, Austria\label{aff1}
\and
Universit\"at Bonn, Argelander-Institut f\"ur Astronomie, Auf dem H\"ugel 71, 53121 Bonn, Germany\label{aff2}
\and
Ruhr University Bochum, Faculty of Physics and Astronomy, Astronomical Institute (AIRUB), German Centre for Cosmological Lensing (GCCL), 44780 Bochum, Germany\label{aff3}
\and
School of Mathematics, Statistics and Physics, Newcastle University, Herschel Building, Newcastle-upon-Tyne, NE1 7RU, UK\label{aff4}
\and
E. A. Milne Centre, University of Hull, Cottingham Road, Hull, HU6 7RX, UK\label{aff5}
\and
Astronomisches Institut, Ruhr-Universit\"at Bochum, Universit\"atsstr. 150, 44801 Bochum, Germany\label{aff6}
\and
Leiden Observatory, Leiden University, Einsteinweg 55, 2333 CC Leiden, The Netherlands\label{aff7}
\and
Department of Physics and Astronomy, University College London, Gower Street, London WC1E 6BT, UK\label{aff8}
\and
Universit\'e Paris-Saclay, CNRS, Institut d'astrophysique spatiale, 91405, Orsay, France\label{aff9}
\and
ESAC/ESA, Camino Bajo del Castillo, s/n., Urb. Villafranca del Castillo, 28692 Villanueva de la Ca\~nada, Madrid, Spain\label{aff10}
\and
School of Mathematics and Physics, University of Surrey, Guildford, Surrey, GU2 7XH, UK\label{aff11}
\and
INAF-Osservatorio Astronomico di Brera, Via Brera 28, 20122 Milano, Italy\label{aff12}
\and
INAF-Osservatorio di Astrofisica e Scienza dello Spazio di Bologna, Via Piero Gobetti 93/3, 40129 Bologna, Italy\label{aff13}
\and
IFPU, Institute for Fundamental Physics of the Universe, via Beirut 2, 34151 Trieste, Italy\label{aff14}
\and
INAF-Osservatorio Astronomico di Trieste, Via G. B. Tiepolo 11, 34143 Trieste, Italy\label{aff15}
\and
INFN, Sezione di Trieste, Via Valerio 2, 34127 Trieste TS, Italy\label{aff16}
\and
SISSA, International School for Advanced Studies, Via Bonomea 265, 34136 Trieste TS, Italy\label{aff17}
\and
Dipartimento di Fisica e Astronomia, Universit\`a di Bologna, Via Gobetti 93/2, 40129 Bologna, Italy\label{aff18}
\and
INFN-Sezione di Bologna, Viale Berti Pichat 6/2, 40127 Bologna, Italy\label{aff19}
\and
INAF-Osservatorio Astrofisico di Torino, Via Osservatorio 20, 10025 Pino Torinese (TO), Italy\label{aff20}
\and
Dipartimento di Fisica, Universit\`a di Genova, Via Dodecaneso 33, 16146, Genova, Italy\label{aff21}
\and
INFN-Sezione di Genova, Via Dodecaneso 33, 16146, Genova, Italy\label{aff22}
\and
Department of Physics "E. Pancini", University Federico II, Via Cinthia 6, 80126, Napoli, Italy\label{aff23}
\and
INAF-Osservatorio Astronomico di Capodimonte, Via Moiariello 16, 80131 Napoli, Italy\label{aff24}
\and
INFN section of Naples, Via Cinthia 6, 80126, Napoli, Italy\label{aff25}
\and
Instituto de Astrof\'isica e Ci\^encias do Espa\c{c}o, Universidade do Porto, CAUP, Rua das Estrelas, PT4150-762 Porto, Portugal\label{aff26}
\and
Faculdade de Ci\^encias da Universidade do Porto, Rua do Campo de Alegre, 4150-007 Porto, Portugal\label{aff27}
\and
Dipartimento di Fisica, Universit\`a degli Studi di Torino, Via P. Giuria 1, 10125 Torino, Italy\label{aff28}
\and
INFN-Sezione di Torino, Via P. Giuria 1, 10125 Torino, Italy\label{aff29}
\and
INAF-IASF Milano, Via Alfonso Corti 12, 20133 Milano, Italy\label{aff30}
\and
INAF-Osservatorio Astronomico di Roma, Via Frascati 33, 00078 Monteporzio Catone, Italy\label{aff31}
\and
INFN-Sezione di Roma, Piazzale Aldo Moro, 2 - c/o Dipartimento di Fisica, Edificio G. Marconi, 00185 Roma, Italy\label{aff32}
\and
Centro de Investigaciones Energ\'eticas, Medioambientales y Tecnol\'ogicas (CIEMAT), Avenida Complutense 40, 28040 Madrid, Spain\label{aff33}
\and
Port d'Informaci\'{o} Cient\'{i}fica, Campus UAB, C. Albareda s/n, 08193 Bellaterra (Barcelona), Spain\label{aff34}
\and
Institute for Theoretical Particle Physics and Cosmology (TTK), RWTH Aachen University, 52056 Aachen, Germany\label{aff35}
\and
Institute of Space Sciences (ICE, CSIC), Campus UAB, Carrer de Can Magrans, s/n, 08193 Barcelona, Spain\label{aff36}
\and
Institut d'Estudis Espacials de Catalunya (IEEC),  Edifici RDIT, Campus UPC, 08860 Castelldefels, Barcelona, Spain\label{aff37}
\and
Dipartimento di Fisica e Astronomia "Augusto Righi" - Alma Mater Studiorum Universit\`a di Bologna, Viale Berti Pichat 6/2, 40127 Bologna, Italy\label{aff38}
\and
Institute for Astronomy, University of Edinburgh, Royal Observatory, Blackford Hill, Edinburgh EH9 3HJ, UK\label{aff39}
\and
Jodrell Bank Centre for Astrophysics, Department of Physics and Astronomy, University of Manchester, Oxford Road, Manchester M13 9PL, UK\label{aff40}
\and
European Space Agency/ESRIN, Largo Galileo Galilei 1, 00044 Frascati, Roma, Italy\label{aff41}
\and
Universit\'e Claude Bernard Lyon 1, CNRS/IN2P3, IP2I Lyon, UMR 5822, Villeurbanne, F-69100, France\label{aff42}
\and
Institute of Physics, Laboratory of Astrophysics, Ecole Polytechnique F\'ed\'erale de Lausanne (EPFL), Observatoire de Sauverny, 1290 Versoix, Switzerland\label{aff43}
\and
UCB Lyon 1, CNRS/IN2P3, IUF, IP2I Lyon, 4 rue Enrico Fermi, 69622 Villeurbanne, France\label{aff44}
\and
Departamento de F\'isica, Faculdade de Ci\^encias, Universidade de Lisboa, Edif\'icio C8, Campo Grande, PT1749-016 Lisboa, Portugal\label{aff45}
\and
Instituto de Astrof\'isica e Ci\^encias do Espa\c{c}o, Faculdade de Ci\^encias, Universidade de Lisboa, Campo Grande, 1749-016 Lisboa, Portugal\label{aff46}
\and
Department of Astronomy, University of Geneva, ch. d'Ecogia 16, 1290 Versoix, Switzerland\label{aff47}
\and
INFN-Padova, Via Marzolo 8, 35131 Padova, Italy\label{aff48}
\and
INAF-Istituto di Astrofisica e Planetologia Spaziali, via del Fosso del Cavaliere, 100, 00100 Roma, Italy\label{aff49}
\and
Universit\'e Paris-Saclay, Universit\'e Paris Cit\'e, CEA, CNRS, AIM, 91191, Gif-sur-Yvette, France\label{aff50}
\and
Institut de Ciencies de l'Espai (IEEC-CSIC), Campus UAB, Carrer de Can Magrans, s/n Cerdanyola del Vall\'es, 08193 Barcelona, Spain\label{aff51}
\and
Istituto Nazionale di Fisica Nucleare, Sezione di Bologna, Via Irnerio 46, 40126 Bologna, Italy\label{aff52}
\and
INAF-Osservatorio Astronomico di Padova, Via dell'Osservatorio 5, 35122 Padova, Italy\label{aff53}
\and
Max Planck Institute for Extraterrestrial Physics, Giessenbachstr. 1, 85748 Garching, Germany\label{aff54}
\and
Universit\"ats-Sternwarte M\"unchen, Fakult\"at f\"ur Physik, Ludwig-Maximilians-Universit\"at M\"unchen, Scheinerstrasse 1, 81679 M\"unchen, Germany\label{aff55}
\and
Dipartimento di Fisica "Aldo Pontremoli", Universit\`a degli Studi di Milano, Via Celoria 16, 20133 Milano, Italy\label{aff56}
\and
Institute of Theoretical Astrophysics, University of Oslo, P.O. Box 1029 Blindern, 0315 Oslo, Norway\label{aff57}
\and
Jet Propulsion Laboratory, California Institute of Technology, 4800 Oak Grove Drive, Pasadena, CA, 91109, USA\label{aff58}
\and
Department of Physics, Lancaster University, Lancaster, LA1 4YB, UK\label{aff59}
\and
Felix Hormuth Engineering, Goethestr. 17, 69181 Leimen, Germany\label{aff60}
\and
Technical University of Denmark, Elektrovej 327, 2800 Kgs. Lyngby, Denmark\label{aff61}
\and
Cosmic Dawn Center (DAWN), Denmark\label{aff62}
\and
Institut d'Astrophysique de Paris, UMR 7095, CNRS, and Sorbonne Universit\'e, 98 bis boulevard Arago, 75014 Paris, France\label{aff63}
\and
Max-Planck-Institut f\"ur Astronomie, K\"onigstuhl 17, 69117 Heidelberg, Germany\label{aff64}
\and
Department of Physics and Helsinki Institute of Physics, Gustaf H\"allstr\"omin katu 2, 00014 University of Helsinki, Finland\label{aff65}
\and
Aix-Marseille Universit\'e, CNRS/IN2P3, CPPM, Marseille, France\label{aff66}
\and
Mullard Space Science Laboratory, University College London, Holmbury St Mary, Dorking, Surrey RH5 6NT, UK\label{aff67}
\and
Universit\'e de Gen\`eve, D\'epartement de Physique Th\'eorique and Centre for Astroparticle Physics, 24 quai Ernest-Ansermet, CH-1211 Gen\`eve 4, Switzerland\label{aff68}
\and
Department of Physics, P.O. Box 64, 00014 University of Helsinki, Finland\label{aff69}
\and
Helsinki Institute of Physics, Gustaf H{\"a}llstr{\"o}min katu 2, University of Helsinki, Helsinki, Finland\label{aff70}
\and
NOVA optical infrared instrumentation group at ASTRON, Oude Hoogeveensedijk 4, 7991PD, Dwingeloo, The Netherlands\label{aff71}
\and
INFN-Sezione di Milano, Via Celoria 16, 20133 Milano, Italy\label{aff72}
\and
Aix-Marseille Universit\'e, CNRS, CNES, LAM, Marseille, France\label{aff73}
\and
Dipartimento di Fisica e Astronomia "Augusto Righi" - Alma Mater Studiorum Universit\`a di Bologna, via Piero Gobetti 93/2, 40129 Bologna, Italy\label{aff74}
\and
Department of Physics, Centre for Extragalactic Astronomy, Durham University, South Road, DH1 3LE, UK\label{aff75}
\and
Universit\'e Paris Cit\'e, CNRS, Astroparticule et Cosmologie, 75013 Paris, France\label{aff76}
\and
Institut d'Astrophysique de Paris, 98bis Boulevard Arago, 75014, Paris, France\label{aff77}
\and
European Space Agency/ESTEC, Keplerlaan 1, 2201 AZ Noordwijk, The Netherlands\label{aff78}
\and
Department of Physics, Institute for Computational Cosmology, Durham University, South Road, DH1 3LE, UK\label{aff79}
\and
Institut de F\'{i}sica d'Altes Energies (IFAE), The Barcelona Institute of Science and Technology, Campus UAB, 08193 Bellaterra (Barcelona), Spain\label{aff80}
\and
Department of Physics and Astronomy, University of Aarhus, Ny Munkegade 120, DK-8000 Aarhus C, Denmark\label{aff81}
\and
Space Science Data Center, Italian Space Agency, via del Politecnico snc, 00133 Roma, Italy\label{aff82}
\and
Centre National d'Etudes Spatiales -- Centre spatial de Toulouse, 18 avenue Edouard Belin, 31401 Toulouse Cedex 9, France\label{aff83}
\and
Institute of Space Science, Str. Atomistilor, nr. 409 M\u{a}gurele, Ilfov, 077125, Romania\label{aff84}
\and
Instituto de Astrof\'isica de Canarias, Calle V\'ia L\'actea s/n, 38204, San Crist\'obal de La Laguna, Tenerife, Spain\label{aff85}
\and
Departamento de Astrof\'isica, Universidad de La Laguna, 38206, La Laguna, Tenerife, Spain\label{aff86}
\and
Dipartimento di Fisica e Astronomia "G. Galilei", Universit\`a di Padova, Via Marzolo 8, 35131 Padova, Italy\label{aff87}
\and
Institut f\"ur Theoretische Physik, University of Heidelberg, Philosophenweg 16, 69120 Heidelberg, Germany\label{aff88}
\and
Institut de Recherche en Astrophysique et Plan\'etologie (IRAP), Universit\'e de Toulouse, CNRS, UPS, CNES, 14 Av. Edouard Belin, 31400 Toulouse, France\label{aff89}
\and
Universit\'e St Joseph; Faculty of Sciences, Beirut, Lebanon\label{aff90}
\and
Departamento de F\'isica, FCFM, Universidad de Chile, Blanco Encalada 2008, Santiago, Chile\label{aff91}
\and
Satlantis, University Science Park, Sede Bld 48940, Leioa-Bilbao, Spain\label{aff92}
\and
Instituto de Astrof\'isica e Ci\^encias do Espa\c{c}o, Faculdade de Ci\^encias, Universidade de Lisboa, Tapada da Ajuda, 1349-018 Lisboa, Portugal\label{aff93}
\and
Universidad Polit\'ecnica de Cartagena, Departamento de Electr\'onica y Tecnolog\'ia de Computadoras,  Plaza del Hospital 1, 30202 Cartagena, Spain\label{aff94}
\and
INFN-Bologna, Via Irnerio 46, 40126 Bologna, Italy\label{aff95}
\and
Kapteyn Astronomical Institute, University of Groningen, PO Box 800, 9700 AV Groningen, The Netherlands\label{aff96}
\and
Dipartimento di Fisica, Universit\`a degli studi di Genova, and INFN-Sezione di Genova, via Dodecaneso 33, 16146, Genova, Italy\label{aff97}
\and
Infrared Processing and Analysis Center, California Institute of Technology, Pasadena, CA 91125, USA\label{aff98}
\and
INAF, Istituto di Radioastronomia, Via Piero Gobetti 101, 40129 Bologna, Italy\label{aff99}
\and
Junia, EPA department, 41 Bd Vauban, 59800 Lille, France\label{aff100}
\and
ICSC - Centro Nazionale di Ricerca in High Performance Computing, Big Data e Quantum Computing, Via Magnanelli 2, Bologna, Italy\label{aff101}
\and
Laboratoire Univers et Th\'eorie, Observatoire de Paris, Universit\'e PSL, Universit\'e Paris Cit\'e, CNRS, 92190 Meudon, France\label{aff102}}    

%
%
%
%

%
%
\abstract{Cosmic shear is a powerful probe of cosmological models and the transition from current Stage-III surveys such as the Kilo-Degree Survey (KiDS) to the increased area and redshift range of Stage IV surveys such as \Euclid will significantly increase the precision of weak lensing analyses. However, with increasing precision, the accuracy of model assumptions needs to be evaluated. In this study, we quantify the impact of the correlated clustering of weak lensing source galaxies with the surrounding large-scale structure, known as source-lens clustering (SLC), which is commonly neglected. { We include the impact of realistic scatter in photometric redshift estimates, which impacts the assignment of galaxies to tomographic bins and increases the SLC}.
For this, we use simulated cosmological datasets with realistically distributed galaxies and measure shear correlation functions for both clustered and uniformly distributed source galaxies. Cosmological analyses are performed for both scenarios to quantify the impact of SLC on parameter inference for a KiDS-like and a \Euclid-like setting.
We find for Stage III surveys such as KiDS, SLC has a minor impact when accounting for nuisance parameters for intrinsic alignments and shifts of tomographic bins, as these nuisance parameters absorb the effect of SLC, thus changing their original meaning. { For KiDS (\Euclid), the inferred intrinsic alignment amplitude $A_\mathrm{IA}$ changes from $0.11_{-0.46}^{+0.44}$ ($-0.009_{-0.080}^{+0.079}$) for data without SLC to $0.28_{-0.44}^{+0.42}$ ($0.022_{-0.082}^{+0.081}$) with SLC.} However, fixed nuisance parameters lead to shifts in $S_8$ and $\Omega_\mathrm{m}$, emphasizing the need for including SLC in the modelling.  For \Euclid we find that $\sigma_8$, $\Omega_\mathrm{m}$, and $w_0$ are shifted by  0.19, 0.12, and 0.12 $\sigma$, respectively, {when including free nuisance parameters}, and by 0.20, 0.16, and 0.32 $\sigma$ when fixing the nuisance parameters. Consequently, SLC on its own has only a small impact on the inferred parameter inference when using uninformative priors for nuisance parameters. However, SLC might conspire with the breakdown of other modelling assumptions, such as magnification bias or source obscuration, which could collectively exert a more pronounced effect on inferred parameters.
}
    
%
%
%
%
   \titlerunning{Source-lens clustering in KiDS-1000 and \Euclid }
   \authorrunning{Linke et al. }
   
   \maketitle
%
%
%
%

   
\section{\label{sc:Intro}Introduction}

Weak gravitational lensing has emerged as a powerful tool in the field of cosmology, providing valuable insights into the nature of our Universe. In particular, the second-order correlation functions of weak lensing shear have become widely employed for precision cosmological measurements. These correlation functions rely on the matter power spectrum and are especially effective in determining the combined parameter $S_8=\sigma_8 (\Omega_\mathrm{m}/0.3)^{0.5}$, which represents a combination of the matter density parameter $\Omega_\mathrm{m}$ and the clustering parameter $\sigma_8$. Recent Stage III surveys, including the Hyper Suprime-Cam (HSC) survey \citep{Aihara2018}, the Kilo-Degree Survey (KiDS, \citealp{Kuijken2015}), and the Dark Energy Survey (DES, \citealp{Abbott2016, Becker2016}), have successfully measured $S_8$ with remarkable precision, thereby establishing cosmic shear as a reliable cosmological tool \citep{Asgari2021, Amon2022, Dalal2023, Li2023, DES_KiDS2023}.

However, intriguing trends have been observed in the results obtained from cosmic shear experiments. These surveys consistently yield lower values of $S_8$ compared to what would be expected based on measurements of the cosmic microwave background (CMB) using the Planck satellite under the framework of the cosmological standard model. Referred to as the $S_8$-tension, this discrepancy has a significance of $2-3 \sigma$, and raises the possibility of issues either in the standard model of cosmology or in the analyses of cosmic shear or the CMB \citep{diValentino2021, Abdalla2022}.

Several effects have been identified as potential sources for the tension observed in $S_8$ measurements. First, ill-understood astrophysical effects, such as the impact of baryonic matter or intrinsic alignments of galaxies, might bias the measurements from cosmic shear \citep{Semboloni2011, Troxel2015, Chisari2019}. Second, deviations from the cosmological standard model, such as modified gravity or exotic dark matter, could play a role \citep{Planck2016XIV, diValentino2016, Heimersheim2020}. { Third, observational systematics, such as the calibration of photometric redshift estimates and galaxy shape measurements can bias the measured cosmic shear signal (\citealp{Huterer2006},  \citealp{EP-Congedo}).} Finally, the apparent tension could be caused by simplifying assumptions made in the modelling of the cosmic shear signal. Examples of such assumptions include neglecting magnification bias \citep{Unruh2020, vonWietersheim-Kramsta2021, Duncan2022}, source obscuration \citep{Hartlap2011}, spatially varying survey depth \citep{Heydenreich2020}, or the intrinsically clustered positions of source galaxies \citep{Yu2015}. Understanding these factors is crucial for comprehending the nature of the $S_8$-tension, and for assessing the implications for Stage IV surveys such as \Euclid \citep{Laureijs11, EuclidSkyOverview} or the {Vera C. Rubin} Legacy Survey of Space and Time \citep[LSST][]{Ivezic2019}. In this study we focus specifically on the clustering of source galaxies with the lensing matter structures, which we refer to as source-lens clustering (SLC).

Conventionally, cosmic shear analyses assume that source galaxies are spatially uniformly distributed within the survey footprint, allowing the cosmic shear field to be sampled randomly across the sky. However, this assumption does not hold in reality. Source galaxies, like all galaxies, trace the underlying density field, meaning their positions correlate with the local matter structures. If the sources are divided into broad redshift bins, some sources in a bin will be at the same redshift as the matter causing a shear signal for other sources in the same bin.
Therefore, the lensing signal of the farther sources is correlated to the positions of the closer sources. This correlation contributes to the measured shear correlation functions and, if not taken into account, can, in principle, bias cosmological parameter inference \citep{Bernardeau1998}.

However, the magnitude of this effect is currently unclear. By considering a simple model that incorporates linear galaxy biases and an analytic form of the source redshift distribution, \citet{Deshpande-EP28} predicted that the effect of source-lens clustering would be significant for Stage IV surveys such as \Euclid and bias constraints on $\Omega_\mathrm{m}$ by more than $1\,\sigma$. However, based on analytical calculations, \citet{Krause2021} expect the effect to be dependent on fourth-order correlations and therefore negligible, at least for Stage III surveys. \citet{Yu2015} quantified the effect with $N$-body simulations and found a 1--10\% effect on the lensing power spectrum. However, they considered only two Gaussian tomographic bins, assumed a linear galaxy bias, and did not include realistic photometric redshift errors in their study. Thus, the magnitude of SLC for galaxies with realistic redshift distributions has not yet been determined. This paper aims to address this gap in knowledge.

A confounding factor in the discussion of SLC is that the non-uniform distribution of source galaxies causes several partially counteracting effects. First, the intrinsic clustering of sources in regions of higher density means that we sample the cosmic shear field predominantly in regions with higher shear signals. This effect, considered for example by \citet{Krause2021}, increases the measured cosmic shear signal compared to the theoretical expectation for unclustered sources. Second, the clustering causes a bias on the standard estimator for cosmic shear correlation functions. This estimator is only unbiased for uniformly distributed galaxies. A third effect occurs due to the noise in photometric redshift estimates. 
Due to the uncertainty of photometric redshifts, galaxies at low true redshifts might be assigned to higher tomographic bins. These galaxies carry less cosmic shear signal and therefore lower the overall signal of a tomographic bin. This decrease in signal is stronger in regions with high foreground matter densities since there are also more galaxies that can be assigned to higher tomographic bins. The correlation between the decrease in signal and the foreground matter distribution causes a decrease in the measured cosmic shear signal. Our goal here is to simultaneously constrain the impact of all these SLC effects on cosmological parameter estimation with cosmic shear.

To achieve this goal, we measured the shear correlation functions for clustered and uniformly distributed source galaxies in cosmological simulations with realistically distributed galaxies and Stage III- and Stage IV-like redshift distributions. We then performed cosmological analyses for the clustered and unclustered cases to assess the effect of source-lens clustering.

The structure of this paper is as follows. In Sect.~\ref{sc:Theo} we review the basics of weak gravitational lensing and cosmic shear and give a theoretical description of the source-lens clustering effect. Section \ref{sc:Data} describes the cosmological simulations used and our steps to obtain clustered and unclustered source galaxies. We describe our correlation function measurement and cosmological parameter inference in Sect. \ref{sc:Analysis}, and give the resulting parameter estimates in Sect. \ref{sc:Results}. We conclude with a discussion in Sect. \ref{sc:Discussion}.

\section{\label{sc:Theo} Theoretical background  }

\subsection{Second-order cosmic shear statistics}
Cosmological analyses of cosmic shear operate in the weak regime of gravitational lensing (for a review on weak lensing, see, for example, \citealp{Bartelmann2001}). They mostly analyse second-order statistics of the shear $\gamma$. The easiest of these statistics, conceptually speaking, are the shear correlation functions $\xi_+$ and $\xi_-$, defined by
\begin{equation}
\label{eq: definition xipm}
    \xi_{\pm}(\theta) = \expval{\gamma_\mathrm{t}\gamma_\mathrm{t}}(\theta) \pm \expval{\gamma_\times\gamma_\times}(\theta) \;,
\end{equation}
where $\gamma_\mathrm{t}$ and $\gamma_\times$ are the tangential and cross-component of the shear for a galaxy pair with angular separation $\theta$ {and the brackets denote an ensemble average}. { Assuming a flat sky, t}he shear $\gamma$ is related to the convergence $\kappa$ by the Kaiser-Squires relation \citep{Kaiser1993}
\begin{equation}
    \tilde{\gamma}(\ellvec) = \E^{2\I\,\phi_\ell} \tilde{\kappa}(\ellvec)\;,
\end{equation}
where the tilde denotes Fourier transforms and $\phi_\ell$ is the polar angle of {the wavevector} $\ellvec$. Therefore, using the lensing power spectrum $C(\ell)$, defined {with the Dirac delta `function' $\dirac$,} by
\begin{equation}
    (2\pi)^2 \dirac(\ellvec+\ellvec')\, C(\ell) = \expval{\hat{\kappa}(\ellvec')\, \hat{\kappa}(\ellvec)}\;,
\end{equation}
the shear correlation functions can be modelled as 
\begin{equation}
\label{eq: model xipm}
    \xi_{\pm}(\theta) = \int_0^\infty \frac{\dd{\ell}\,\ell}{2\pi} J_{0,4}(\ell\,\theta)\, C(\ell)\;,
\end{equation}
where $J_i$ is the $i$-th order Bessel function. Consequently, the impact of any modelling effect on the shear correlation function can be estimated equivalently on the lensing power spectrum.

The convergence $\kappa$ is a normalized surface mass density and thus related to the matter density contrast $\delta(\chi\thetavec, \chi)$ at angular position $\thetavec$ and comoving distance $\chi$ via
\begin{equation}
\label{eq: kappa}
    \kappa(\thetavec)=\int_0^\infty \dd{\chi}\,\delta(\chi\thetavec, \chi) \, \int_\chi^\infty \dd{\chi'} W(\chi, \chi')\,p(\chi'; \thetavec)\;,
\end{equation}
where $p(\chi; \thetavec)$ is the probability density of source galaxies at angular position $\thetavec$ and comoving distance $\chi$ and $W$ is the lensing efficiency kernel, {which is, assuming a flat Universe,}
\begin{equation}
    W(\chi, \chi') = \frac{3\Omega_\mathrm{m}H_0^2}{2c^2} \frac{\chi}{a(\chi)}\frac{\chi'-\chi}{\chi'}\;.
\end{equation}
The probability density $p$ is related to the number density $n$ of source galaxies by
\begin{equation}
    p(\chi; \thetavec) = \frac{n(\chi\,\thetavec, \chi)}{\int \dd\chi n(\chi\,\thetavec, \chi)}\;.
\end{equation}

Usually, one assumes that source galaxies are distributed uniformly across the survey area. Then, $p$ is only a function of the comoving distance $\chi$, and the convergence $\kappa_0(\thetavec)$ is given by
\begin{equation}
\label{eq: kappa0}
    \kappa_0(\thetavec)=\int_0^\infty \dd{\chi}  \delta(\chi\thetavec, \chi) \int_\chi^\infty \dd{\chi'} W(\chi, \chi')\, p(\chi') \;.
\end{equation}
Under the extended Limber approximation \citep{Limber1953, Kaiser1998, Loverde2008}, the lensing power spectrum then becomes
\begin{equation}
\label{eq: limber}
    C(\ell) = \int_0^\infty \dd{\chi} \frac{1}{\chi^2} \left[\int_\chi^\infty \dd{\chi'} p(\chi')\, W(\chi, \chi')\right]^2\, P\left(\frac{\ell+1/2}{\chi}, \chi\right)\;,
\end{equation}
where { $P(k, \chi)$} is the matter power spectrum. This form of the lensing power spectrum is usually assumed in cosmic shear analyses.

\subsection{The source-lens clustering effect}
\label{sc:Theo:SLC}
However, the assumption of uniformly distributed source galaxies is not correct. As is the case for any other galaxy population, the sources are tracers of the matter distribution and are therefore spatially clustered. This clustering invalidates Eq.~\eqref{eq: kappa0}, as the source density $n_\mathrm{s}$ now depends on angular position $\thetavec$. Instead, Eq.~\eqref{eq: kappa} needs to be used, which leads to different effects outlined in this section. 

\subsubsection{Estimator bias (EB)}
As noted by \citet{Yu2015}, SLC depends on the estimator used for the shear correlation functions. Different estimators incur different SLC-induced estimator biases. For example, when using an estimator based on pixelized shear maps, the noise in each pixel depends on the number of source galaxies in the pixel. The correlation between this number and the shear signal then leads to a correlation between the signal estimate and the noise \citep{Gatti2024}. Here, though, we consider a catalogue-based estimator for the shear correlation functions which does not require pixelization. This estimator $\hat{\xi}_\pm$ is 
\begin{equation}
    \hat{\xi}_\pm(\theta) = \frac{\sum_i \sum_j (\gamma_\mathrm{t}^i \gamma_\mathrm{t}^j\pm\gamma_\times^i \gamma_\times^j) \triangle(|\varthetavec_i -\varthetavec_j|; \theta)}{\sum_i \sum_j \triangle(|\varthetavec_i - \varthetavec_j|; \theta)}\;,
\end{equation}
where the sums go over all source galaxies, $\gamma_\mathrm{t}^i$ and $\gamma_\times^i$ are the tangential and cross shear of the $i$-th galaxy and $\triangle(\vartheta; \theta)$ is one if $\vartheta$ lies in the $\theta$-bin and zero otherwise. 
Using $\gamma_{\mathrm{t}/\times}(\varthetavec, \chi)$ for the tangential (cross) shear of a hypothetical source at angular position $\varthetavec$ and distance $\chi$, the expectation value of this estimator is
\begin{align}
    &\notag \langle \hat{\xi}_\pm \rangle (|\varthetavec-\varthetavec'|) \\
    &\label{eq:expvalue}= \Bigg\langle \Bigg\lbrace \int \dd{\chi_1}\int \dd{\chi_2}\int \dd[2]{\vartheta} p(\chi_1;\varthetavec)\, p(\chi_2;\varthetavec')\, \\
    &\notag\quad\quad\times \left[\gamma_\mathrm{t}(\varthetavec, \chi_1)\, \gamma_\mathrm{t}(\varthetavec', \chi_2) \pm \gamma_{\times}(\varthetavec, \chi_1)\, \gamma_{\times}(\varthetavec', \chi_2)  \right] \Bigg\rbrace \\
    &\notag \quad \times \left[\int \dd{\chi_1}\int \dd{\chi_2}\int \dd[2]{\vartheta} p(\chi_1;\varthetavec)\, p(\chi_2,\varthetavec')\right]^{-1} \Bigg\rangle\,.
\end{align}
If we neglect for the moment the correlation between the source density $n_\mathrm{s}$ and the shear (which we treat in the next two subsections), we can swap the averaging and the integrations such that the expectation value becomes
\begin{align}
    &\notag \langle \hat{\xi}_\pm \rangle (|\varthetavec-\varthetavec'|) \\
    &\label{eq:expvalueEBonly}= \Bigg[ \int \dd{\chi_1}\int \dd{\chi_2}\int \dd[2]{\vartheta} \langle p(\chi_1,\varthetavec)\, p(\chi_2,\varthetavec')\rangle \\
    &\notag\quad\quad\times \langle\gamma_\mathrm{t}(\varthetavec, \chi_1)\, \gamma_\mathrm{t}(\varthetavec', \chi_2) \pm \gamma_{\times}(\varthetavec, \chi_1)\, \gamma_{\times}(\varthetavec', \chi_2)  \rangle \Bigg] \\
    &\notag \quad \times \left[\int \dd{\chi_1}\int \dd{\chi_2}\int \dd[2]{\vartheta} \langle p(\chi_1,\varthetavec)\, p(\chi_2,\varthetavec')\rangle\right]^{-1}\,.
\end{align}
For unclustered sources, $p(\chi,\varthetavec)$ reduces to $p(\chi)$ and the expectation value in Eq.~\eqref{eq:expvalueEBonly} becomes exactly $\xi_\pm$. This occurs even for clustered sources in the limit of infinitely narrow redshift distributions, that is $p(\chi,\varthetavec) \propto \dirac(\chi-\chi')$, since then the $p$-correlations in numerator and denominator cancel. In reality, the shear correlation functions can only be measured for finitely broad $p(\chi)$, so in general, $\hat{\xi}_\pm$ is biased. { We refer to this bias as estimator bias (EB)}. 

{The estimator bias is also affected by shear weights. Usually, the measured galaxy ellipticities have an assigned weight that depends on the statistical error or bias of the shape measurement. This weight depends on properties of the considered galaxy, such as magnitude and size, but also on the environment of the galaxy, for example whether it is blended with another object. Consequently, the shear weight is correlated with the source number density, causing further bias to the estimator.}

\subsubsection{Intrinsic clustering (IC)}
For the EB effect we neglected that the shear signal itself is correlated with the source galaxy density. This correlation occurs, because due to source clustering the source redshift distribution is not uniform across the sky but depends on the local matter density contrast. Therefore,  the number density $n$ needs to be replaced by 
\begin{equation}
\label{eq: change n IC only}
    n(\chi) \rightarrow n(\chi)\left[1+\delta_\mathrm{g}(\chi\,\thetavec, \chi)\right]\;,
\end{equation}
where $\delta_\mathrm{g}$ is the (three-dimensional) galaxy number density contrast. This relation differs from the `projected' SLC ansatz \citep[e.g.][]{Schmidt2009, Deshpande-EP28}, where $\delta_\mathrm{g}(\chi\,\theta, \chi)$ is replaced by a projected density contrast $\delta_\mathrm{g}^\mathrm{p}(\theta)$ which no longer depends on $\chi$. However, the projected ansatz only agrees with Eq.~\eqref{eq: change n IC only} for very narrow bins, since then
 $\delta_\mathrm{g}^\mathrm{p}(\theta)\simeq \delta_\mathrm{g}(\bar{\chi}\,\theta, \bar{\chi})$, where $\bar{\chi}$ is the average distance for galaxies in the redshift bin \citep{Krause2021}.

Under the assumption of a linear galaxy bias $b$,
\begin{equation}
    n(\chi) \rightarrow n(\chi)\left[1+b\,\delta(\chi\,\thetavec, \chi)\right]\;.
\end{equation}
Applying this change to Eq.~\eqref{eq: kappa0} means that the true convergence $\kappa$ is given by
\begin{equation}
    \kappa(\thetavec) = \kappa_0(\thetavec) + \Delta \kappa(\thetavec)\;,
\end{equation}
with
\begin{equation}
    \Delta \kappa(\thetavec) = b\,\int_0^\infty \dd{\chi} \delta(\chi\thetavec, \chi) \int_\chi^\infty \dd{\chi'} W(\chi, \chi')\, p(\chi')\, \delta(\chi'\thetavec, \chi')\;.
\end{equation}
The additional term depends on both the density contrast $\delta(\chi\thetavec, \chi)$ in front of the source galaxies (i.e. the lens planes), and the density contrast $\delta(\chi'\thetavec, \chi')$ surrounding the source galaxies (i.e. in the `source planes'), which gives the effect its name: source-lens clustering.

As we show in Appendix \ref{app: modelling}, the $\Delta \kappa$ induces an additional term $\Delta C$ to the lensing power spectrum {\footnote{The contribution to the power spectrum we derive here follows the same form as the contribution to the shear power spectrum derived by \citet{Krause2021}. It deviates from the contribution by SLC found by \citet{Deshpande-EP28}, since they use the projected SLC ansatz.}} given by 
\begin{align}
\label{eq:Delta Cell final}
    \Delta C(\ellvec)
            &= b^2 \int_0^\infty \dd{\chi} \int_{\chi}^\infty \dd{\chi'}  \frac{ p^2(\chi')}{\chi'^2}\, \frac{ W^2(\chi, \chi')}{\chi^2}\,\\
            &\notag \quad \times   \int \frac{\dd[2]{L}}{(2\pi)^2}\, P\left(\frac{L}{\chi}, \chi\right)\,   P\left(\frac{|\ellvec-\vec{L}|}{\chi'}, \chi'\right) \, .
\end{align}
We note that this expression relies on the assumption of a linear galaxy bias, which might not be fulfilled at small scales. Therefore, instead of simply using Eq.~\eqref{eq:Delta Cell final} to predict the effect, we measure it in the following using cosmological simulations with realistic galaxy distributions.

\subsubsection{Tomographic bin contamination (TC)}
Another effect occurs when cosmic shear analyses are performed in tomographic bins divided by photometric redshift estimates. Photometric redshifts of individual galaxies are typically noisy. Therefore, a galaxy associated with a tomographic bin between $z_1$ and $z_2$ might have a true redshift well outside of this range. On average, this photometric redshift scatter is taken care of by using a $p(\chi)$ for each tomographic bin, which was calibrated with spectroscopic redshift estimates (see e.g. \citealp{Wright2020}). However, the deviation from the estimated $p(\chi)$ along each line-of-sight correlates with the galaxy number density. Along a line-of-sight with an overdensity in the front, there are more galaxies which can potentially be assigned to a higher tomographic bin, thus contaminating this bin. These contaminating galaxies carry a lower lensing signal. Consequently, in denser regions, the cosmic shear signal is suppressed. This effect also impacts weak lensing estimates of galaxy cluster masses \citep{Köhlinger2015}. Due to photometric redshift scatter, cluster member galaxies can be erroneously treated as background source galaxies. They carry no lensing signal due to the galaxy cluster, thus bias the mass estimate low. 

\begin{figure*}
    \includegraphics[width=\linewidth]{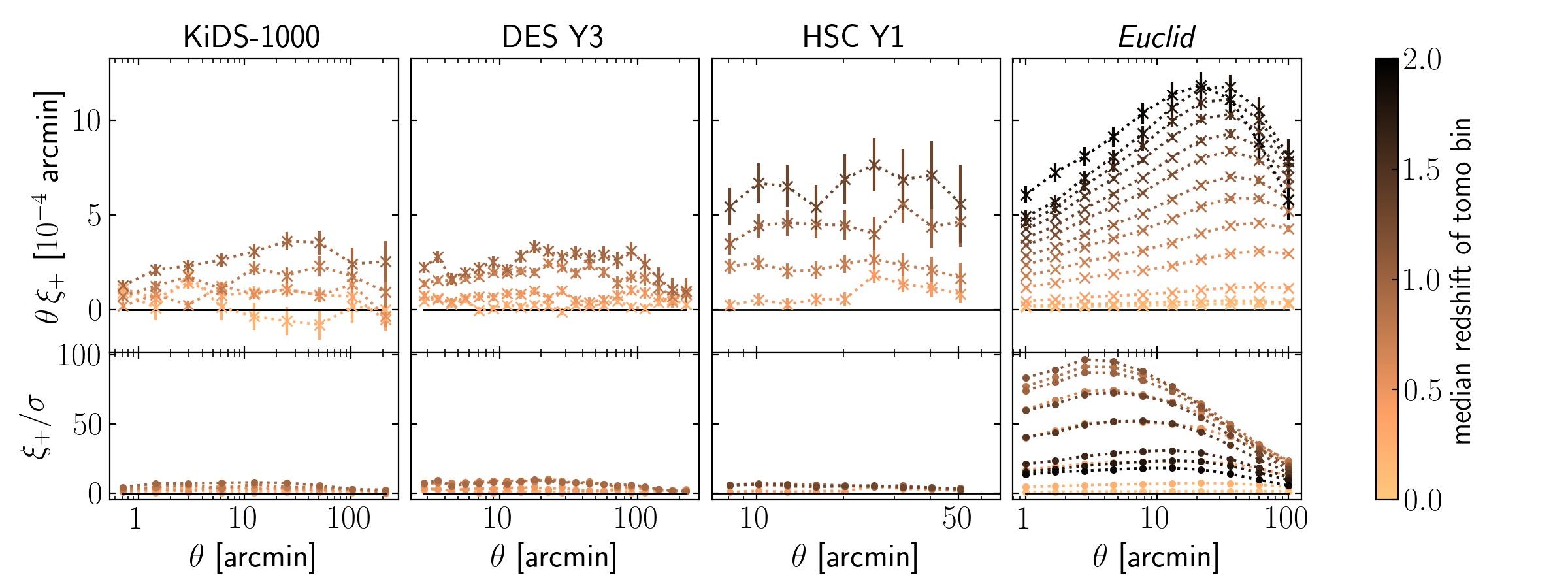}
    \caption{Measured and expected weak lensing signals for stage III surveys and \Euclid. Upper panel: Shear auto-correlations $\xi_+$ for the stage III surveys KiDS (from \citealp{Asgari2021}), DES (from \citealp{Amon2022}), and HSC (from \citealp{Hamana2020}) with their reported errorbars and the expectation for \Euclid. Error bars for \Euclid are the square root of the covariance matrix diagonal, calculated as described in Sect.~\ref{sc:Analysis:Model}. The colour indicates the median redshift of the tomographic bins, with darker colours corresponding to higher redshifts. Lower panel: S/N of $\xi_+$, given by dividing the measurement or prediction by the error bars.}
    \label{fig: xip_stageIII}
\end{figure*}

All SLC effects are expected to be much smaller than the cosmic shear signal itself. For example, since Eq.~\eqref{eq:Delta Cell final} depends on the fourth-order moment of the matter distribution, one could assume that this effect is negligible in a realistic weak lensing survey, where shape measurement uncertainties or the distribution of intrinsic galaxy shapes cause significant noise contributions. However, while this argument was made for stage III surveys \citep{Krause2021}, it does not necessarily hold for \Euclid. As illustration, we compare in Fig.~\ref{fig: xip_stageIII} the shear correlation function $\xi_+$ for auto-correlations of tomographic bins as measured by the stage III surveys KiDS (from \citealp{Asgari2021}), DES (from \citealp{Amon2022}), and HSC (from \citealp{Hamana2020}) with their reported error bars, to the expectation for \Euclid. Due to its greater depth, \Euclid will observe more tomographic bins at higher redshift, which carry larger cosmic shear signals. Due to its larger area and higher number density, the noise of the cosmic shear measurements will be significantly reduced, leading to an order of magnitude increase in S/N. This high precision requires us to have at least equally accurate models. The first step to these is understanding how model assumptions impact the cosmological inference. 

\section{\label{sc:Data} Simulated datasets}
We used two sets of simulated data to estimate the impact of source-lens clustering on cosmological analyses. One of these datasets, based on the Marenostrum Institut de Ciències de l'Espai (MICE) simulations, mimics the redshift distribution and tomographic set-up of the KiDS-1000, while the other, based on the \Euclid Flagship2 (FS2) simulation, mimics the expected redshift distribution and tomographic set-up of the \Euclid wide survey. We describe their creation and properties in the following.  Simulated galaxy catalogues with galaxy properties and weak lensing information from both MICE and the FS2 are accessible on CosmoHub\footnote{https://cosmohub.pic.es} \citep{Carretero2018, Tallada2020}.

\subsection{The Marenostrum Institut de Ciències de l'Espai (MICE) simulations}
\label{sc:Data:MICE}
We created a KiDS-like dataset based on the MICE simulation \citep{Fosalba2015a}, a dark-matter-only N-body simulation. The simulation used a flat $\Lambda$CDM cosmology with $\Omega_\mathrm{m}=0.25$, $\Omega_\Lambda=0.75$, $\Omega_\mathrm{b}=0.044$, $n_\mathrm{s}=0.95$, $\sigma_8=0.8$, $h=0.7$. It evolved $4096^3$ particles with a mass of $2.93  \times 10^{10} h^{-1} M_\odot$ inside a box with side length $3072\,h^{-1}\, \mathrm{Mpc}$ from an initial redshift of $z=100$ to today.  In the dark matter particle distribution, halos were identified using a friends-of-friends halo finder \citep{Crocce2015}. The halos were then populated up to redshift $z=1.4$ with galaxies, following a mixture of halo abundance matching and a halo occupation distribution model to match both the spatial and luminosity distribution of observed galaxies \citep{Carretero2015}.  We used the second version of this galaxy catalogue, called MICE2.

The MICE2 galaxies have assigned fluxes in many photometric bands. To generate KiDS-like photometry, we used the noiseless simulated photometry from MICE, similar to the approach presented in \citet{vandenBusch2020}. We chose the filters that are most similar to the KiDS-1000 filter set-up, which consists of the Very Large Telescope (VLT) Survey Telescope (VST) OmegaCAM $ugri$ bands and the Visible and Infrared Survey Telescope for Astronomy (VISTA) VIRCAM $ZYJHK_{\rm s}$ bands \citep{Kuijken2019}. For the $ugriZ$ bands, we found the Sloan Digital Sky Survey (SDSS) $ugriz$ filters provided in MICE to be most similar to the ones used in KiDS-1000. For the near-infrared imaging in the $Y$ band, the DES $y$ filter is used as a reasonable approximation. The filters of the VISTA $JHK_{\rm s}$ bands are readily available in MICE.
As presented in \citet{Fosalba2015b}, we applied a redshift-dependent evolution correction to the MICE magnitudes resulting in a set of magnitudes $m^{\rm evo}$. 
To estimate the observed magnitudes, we approximated the flux noise of KiDS-1000 \citep{Kuijken2019}. 

An effective projected galaxy radius size $r_{\rm eff}$ is estimated taking into account the half-light radius of a galaxy's bulge and disk and the bulge-to-total emitted flux fraction given in MICE.
The estimated projected galaxy size translates into a simulated aperture which mimics the Gaussian Aperture and Point Spread Function (GAaP) apertures used in KiDS \citep{Kuijken2008}. The aperture major and minor axis $A_{i,x}$ and $B_{i,x}$, respectively, are calculated per galaxy $i$ and per photometric filter $x$ through
\begin{align}
    A_{i,x} &= \min\left(\sqrt{r_{ \mathrm{eff},i}^2+\sigma_{\mathrm{PSF},x}^2+r_{\mathrm{min}}^2}, \ang{;;2.0}\right)\;, \\
    B_{i,x} &= \min\left(\sqrt{\left( \left[\frac{b}{ a}\right]_i r_{\mathrm{eff},i}  \right)^2+\sigma_{\mathrm{PSF},x}^2+r_{\mathrm{min}}^2}, \ang{;;2.0}\right)\;,
\end{align}
with the mean seeing $\sigma_{\mathrm{PSF},x}$ and the projected axis ratio of the bulge $\left[\frac{b}{a}\right]_i$ with the semi-minor and semi-major axis $b$ and $a$, respectively, which is given for all the MICE sources. The minimum radius $r_{\mathrm{min}}$ for the mocks was set to \ang{;;0.3}, which is the maximum difference in point spread function (PSF) size of the KiDS-1000 data between the different photometric bands. The maximum aperture axis is set to be \ang{;;2.0}, as for the KiDS-1000 data.

Considering the  limiting magnitudes $m_{\mathrm{lim},x}$ of the KiDS observations, we calculated the flux error
\begin{align}
    \Delta f_{i,x}= 10^{-0.4(m_{\mathrm{lim},x}-48.6)}  \sqrt{\frac{A_{i,x} B_{i,x}}{\sigma_{\mathrm{PSF},x}^2}}\;,
\end{align}
where we included the mock apertures $\pi A_{i,x} B_{i,x}$ to increase the noise for larger apertures compared to the PSF $ \pi \sigma_{\mathrm{PSF},x}^2$. 
Given the estimated noise and the evolution-corrected model fluxes $f^{\rm evo}$, a flux realization was computed from $f^{\rm obs}_{i,x} \sim \mathcal{N}(f^{\rm evo}_{i,x}, \Delta f_{i,x})$. Converting to magnitudes $m^{\rm obs}_{i,x}$ and their errors $\Delta m_{i,x}$, photometric redshift can be computed with the template fitting photo-$z$ algorithm \texttt{BPZ} \citep{Benitez2000}. 
We excluded all objects with a signal-to-noise ratio of less than 1 in the $r$ band since we label these galaxies as undetected, and therefore they would have no shape measurement.
Finally, we performed a $k$ Nearest Neighbours (kNN) matching between the mock catalogue and the KiDS data based on the 9 magnitudes $ugrizYJHK_{\rm s}$ to assign a mock shape weight \verb|recal_weight| from the real galaxies to the mock ones. This is then used to select the objects with non-zero shape weights, which were used as the source sample in KiDS. 

The MICE simulations were designed for application to gravitational lensing surveys, and thus the lensing observables shear and convergence were computed for the simulation as described in \citet{Fosalba2015b} using the approach by \citet{Fosalba2008} under the Born approximation. From this, one obtains full sky maps of the lensing convergence for sources at 265 different redshifts between 0 and 1.4. These $\kappa$-maps were converted to maps of the weak lensing shear $\gamma$, using first their decomposition in harmonic space,
\begin{align}
    \kappa(\varthetavec) &= \sum_{\ell=0}^\infty \sum_{m=-\ell}^\ell \hat{\kappa}_{\ell m}\, Y_{\ell m}(\varthetavec)\;,\\
    \gamma(\varthetavec) &= \sum_{\ell=2}^\infty \sum_{m=-\ell}^\ell \hat{\gamma}_{\ell m}\, {}_2Y_{\ell m}(\varthetavec)\;,
\end{align}
{where we neglect $B$-modes and use the Laplace spherical harmonics $Y_{\ell m}$ and the spin-$s$-weighted spherical harmonics ${}_sY_{\ell m}$. Then, we apply}
the inverse Kaiser--Squires relation \citep{Kaiser1993},
\begin{equation}
    \hat{\gamma}_{\ell m} = -\sqrt{\frac{(\ell+2)(\ell-1)}{\ell(\ell+1)}}\hat{\kappa}_{\ell m}\;.
\end{equation}
Each galaxy at angular position $\varthetavec$ and redshift $z$ in the MICE2 catalogue was then assigned the shear at position $\varthetavec$ of the weak lensing map for redshift $z$. {The galaxies were also assigned deflected positions, arising from magnification of the sources by the density field. As we are not interested in magnification effects on the cosmic shear signal, we use throughout the original, unmagnified positions of the galaxies.}

In the following, we use both the MICE2 catalogue and the shear maps to create clustered and unclustered source galaxy samples, as detailed in Sect.~\ref{sc:Data:samples}. For this, it is sufficient for us to use a rather small area, namely the rectangular region with right ascension $\alpha \in (\ang{40}, \ang{50})$ and declination $\delta \in (\ang{20}, \ang{50})$.

\begin{figure}
    \centering
    \includegraphics[width=0.9\linewidth]{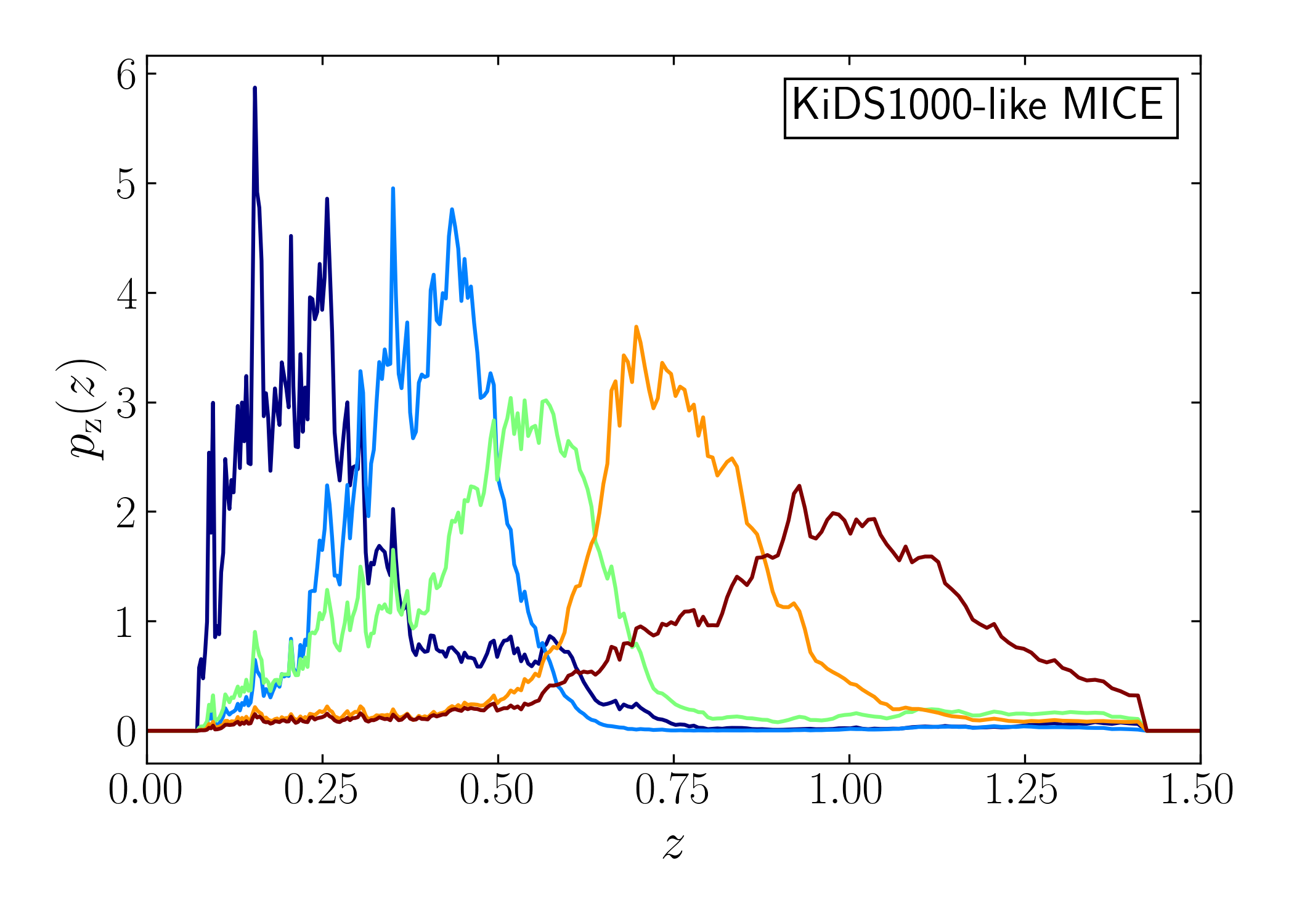}
    \includegraphics[width=0.9\linewidth]{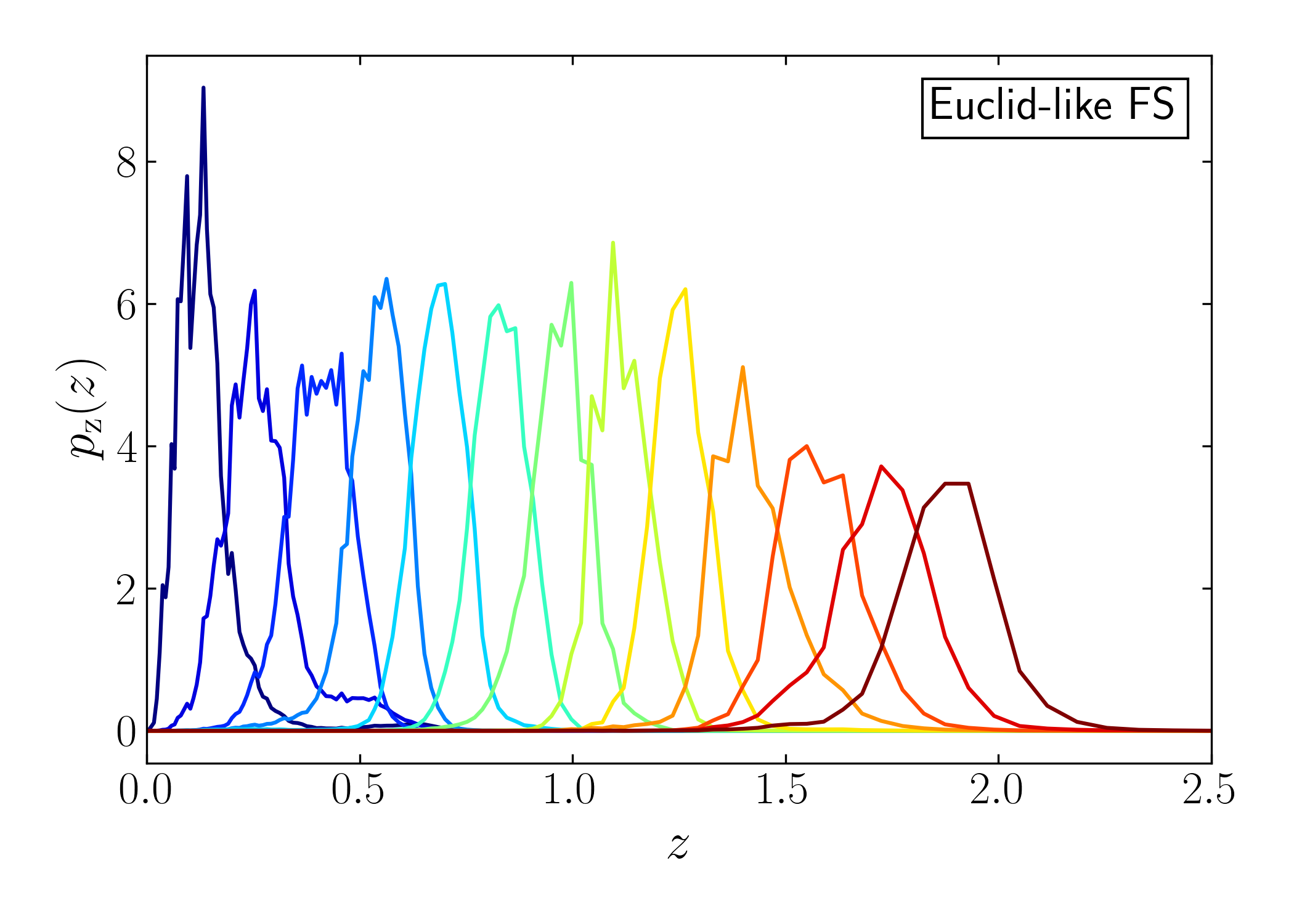}
    
    \caption{Redshift distribution of KiDS-1000-like galaxies in the MICE  (top) and \Euclid-like galaxies in the FS2 (bottom).}
    \label{fig:nz}
\end{figure}

\subsection{The \Euclid flagship simulation 2}
\label{sc:Data:FS}

Since the MICE galaxies are limited to $z<1.4$, it is not possible to construct a \Euclid -like sample from them, as \Euclid will observe a significant number of galaxies at higher redshifts. Therefore, we used the \Euclid flagship simulations 2 (FS2, \citealp{EuclidSkyFlagship}) to create a \Euclid-like sample. 

The FS2 used  $16\,000^3$ particles with a mass of $10^9\,h^{-1}\,M_\odot$ and evolved them in a simulation box with side length $3600\,h^{-1}$ Mpc with the PKDGRAV3 code \citep{Potter2016}. This larger box size allows for ray tracing and galaxy distribution up to $z=3.0$. The simulation assumed a flat $\Lambda$CDM cosmology with $\Omega_\mathrm{m}=0.319$, $\Omega_\mathrm{b}=0.049$, $\Omega_{\Lambda} = 0.681 - \Omega_\gamma - \Omega_{\nu}$, with a photon density parameter $\Omega_\gamma = 0.00005509$, and a neutrino density parameter $\Omega_{\nu} = 0.00140343$. The reduced Hubble constant is $h = 0.67$, the scalar spectral index of the initial fluctuations $n_\mathrm{s} = 0.96$, and its amplitude $A_\mathrm{s} = 2.1 \times 10^{-9}$, which corresponds to $\sigma_8 \simeq 0.813$.

Dark matter halos were found using the friends-of-friends halo finder ROCKSTAR \citep{Behroozi2013}. These halos were populated with central and satellite galaxies according to a halo occupation distribution. Central galaxies were placed at the halo centres, while satellite galaxies are distributed inside the halo following a triaxial Navarro-Frenk-White profile. The galaxies were assigned luminosities with an abundance-matching approach. 

The galaxies were assigned photometric redshifts using the nearest-neighbour photometric redshift (NNPZ) pipeline, which estimates redshift probability distributions for each galaxy \citep[see][for a comparison of this and other redshift estimation algorithms]{Desprez-EP10}. We used the mode of these probability distributions as photometric redshift estimate. The pipeline used a reference sample of two million objects simulated at the depth of the \Euclid calibration fields. Since the full depth of these calibration fields will only be available towards the end of \Euclid's observing run, the accuracy of the derived photometric redshifts is optimistic for the first \Euclid data release. However, it is realistic for the final \Euclid data releases, which have the highest statistical power and are therefore the most strongly affected by subtle effects such as SLC. 

Similarly to the MICE, full-sky shear and convergence maps were computed for the FS2 with the ray-tracing approach by \citet{Fosalba2008}. This approach produced shear and convergence maps at 115 source redshifts between $z=2.3$ and today, given as healpix maps with $N_\mathrm{side}=8192$, which corresponds to an angular resolution of $\simeq \ang{;0.43;}$. While the shear maps were computed for the full-sky, galaxy catalogues are (at time of writing) only available for an octant, stretching for right ascensions between \ang{145} and \ang{235} and declinations between \ang{0} and \ang{90}. Therefore, our analysis is limited to this area.

\subsection{Creating clustered and unclustered galaxy samples}
\label{sc:Data:samples}

Both the MICE and the FS2 contain realistically clustered source galaxies. We divided these galaxies into different tomographic bins based on their photometric redshifts. For the KiDS-like set-up, we used the five bins defined by \citet{Hildebrandt2021} and also used by \citet{Asgari2021}, while for the \Euclid-like set-up, we used the 13 tomographic bins defined by \citet{Pocino-EP12}, shown in Table \ref{tab:tomo bins}. We show the distributions of true redshifts for these bins in Fig.~\ref{fig:nz}. Both samples exhibit galaxy biases between 0.5 and 1.5 (increasing with redshift), which is comparable to what is expected in a real lensing survey (see App. \ref{app: bias})

We created two new galaxy catalogues for each tomographic bin based on the shear maps and the original catalogue. The first of these, which we refer to as the `clustered' catalogue, was used purely for validation purposes. To create it, we took the angular positions $\thetavec$ and redshift $z$ of each galaxy in the original catalogue and assigned it the shear $\gamma$ at position $\thetavec$ on the shear map closest in redshift to $z$. 

The second catalogue we created, which we refer to as `unclustered', is created similarly to the clustered catalogue but crucially without using the galaxies' angular positions. Instead, for each galaxy in the catalogue we drew a random angular position from a uniform distribution and assigned the shear at this new position and the galaxies' redshift to the galaxy. In this way, the galaxies in the unclustered catalogue have the same average number density and redshift distribution as those in the clustered and original catalogue but uniformly distributed positions. The SLC effect is thus present only in the clustered catalogue.

\begin{table}
  \caption{Photometric redshift ranges of tomographic bins for the KiDS-like and the \Euclid-like sample}
    \centering
    \begin{tabular}{c|cc}
    \hline
         Bin & KiDS-like & \Euclid-like \\
         \hline
        1 & (0.1,0.3] & (0,0.15] \\
        2 & (0.3, 0.5] & (0.15, 0.31]\\
        3& (0.5,0.7] & (0.31, 0.46]\\
        4&(0.7,0.9] & (0.46, 0.63]\\
        5 & (0.9,1.2] & (0.63, 0.77]\\
        6 & & (0.77, 0.92]\\
        7 & & (0.92, 1.08]\\
        8 & & (1.08, 1.23]\\
        9 & & (1.23, 1.38]\\
        10 & & (1.38, 1.54]\\
        11 & & (1.54, 1.69]\\
        12 & & (1.69, 1.85]\\
        13 & & (1.85, 2.0]\\
         \hline
    \end{tabular}
  
    \label{tab:tomo bins}
\end{table}

To test the impact of the individual SLC effects (see Sect.~\ref{sc:Theo:SLC}), we created two more catalogues from MICE. The first of these consists of taking the true galaxy positions on the sky and assigning to them the shear at positions rotated $90^\circ$ away with respect to the survey patch center. In this way, the clustering of the source galaxies is no longer correlated to the shear signal. This eliminates the IC and TC effects but retains the EB since the sources still have a non-zero angular correlation function $\omega$.

The last catalogue we created is designed to include the EB and the IC effects but not the TC. For each tomographic bin, we randomly subselected galaxies from the full MICE catalogue such that their true redshifts follow the redshift distribution of the bin but without using the photometric redshifts. In that way, the selection of galaxies in a tomographic bin is no longer dependent on the photometric redshift scatter, eliminating the TC effect.

\section{\label{sc:Analysis} Cosmological analysis}
\subsection{COSEBI estimation, modelling and covariance}
\label{sc:Analysis:Model}
To match the fiducial cosmological analysis of \citet{Asgari2021}, we performed our cosmological inference using the Complete Orthogonal Sets of E/B-Integrals (COSEBIs, \citealp{Schneider2010}). These consist of the $E$-modes $E_n$ and the $B$-modes $B_n$, where, to first order, only $E$-modes can be generated by weak lensing. They can be obtained as weighted integrals over the shear correlation functions $\xi_+$ and $\xi_-$ with
\begin{align}
\label{eq:E_n_defintion}
E_n &= \frac{1}{2} \int_{\theta_\mathrm{min}}^{\theta_\mathrm{max}} \dd \theta \,\theta \left[ T_{+n}(\theta) \xi_+(\theta) + T_{-n}(\theta) \xi_-(\theta) \right]\, , \\
\label{eq:B_n_defintion}
B_n &= \frac{1}{2} \int_{\theta_\mathrm{min}}^{\theta_\mathrm{max}} \dd \theta \,\theta \left[ T_{+n}(\theta) \xi_+(\theta) - T_{-n}(\theta) \xi_-(\theta) \right]\, ,
\end{align}
where $T_{\pm n}(\theta)$ are filter functions defined such that $\theta$ is bounded by $\theta_\mathrm{min}$ and $\theta_\mathrm{max}$ \citep{Schneider2010}. 

To obtain the COSEBIs, we first measured the shear correlation functions $\xi_+$ and $\xi_-$ for each combination of the tomographic bins. We used $4000$ radial bins spaced between \ang{;0.5;} and \ang{;300;} logarithmically for both the KiDS and \Euclid-like case. The measurements are conducted with \verb|treecorr| \citep{Jarvis2004}. The correlation functions are then converted to the first 5 COSEBIs. We chose this number $n_\mathrm{max}$ of COSEBIs for two reasons. First, the information content as a function of $n_\mathrm{max}$ saturates between $n_\mathrm{max}=5$ and $n_\mathrm{max}=10$ \citep{Asgari2012}. Second, the signal-to-noise ratio diminishes with higher $n$ and even for the \Euclid-like case, the signal-to-noise ratio for $E_6$ is lower than 3.8 for all redshift bins. Thus, we do not expect a significant improvement when including more COSEBIs. Nevertheless, determining the optimal choice of $n_\mathrm{max}$, along with the considered angular scales is an important goal for future work.  We modelled $E_n$ using
\begin{align}
E_n &= \int_{0}^{\infty} \frac{\dd \ell \, \ell}{2\pi} C (\ell) \, W_n(\ell) \, , 
\end{align}
where the $W_n$ are Hankel transforms of the $T_{\pm n}$ and given in \citet{Asgari2012}. The lensing power spectrum $C(\ell)$ was modelled as in  \citet{Joachimi2021} from Eq.~\eqref{eq: limber}.  We used the redshift distributions $p_z(z)$ displayed in Fig.~\ref{fig:nz}, but included free parameters $\delta z$ for shifts in the $p_z(z)$. This $\delta z$ are the difference
between the mean of the estimated $p_z(z)$ of each tomographic bin and the true redshift distribution. We used the linear power spectrum model from CAMB \citep{Lewis2000,  Howlett2012}. 

\begin{figure}
    \centering
    \includegraphics[width=0.9\linewidth]{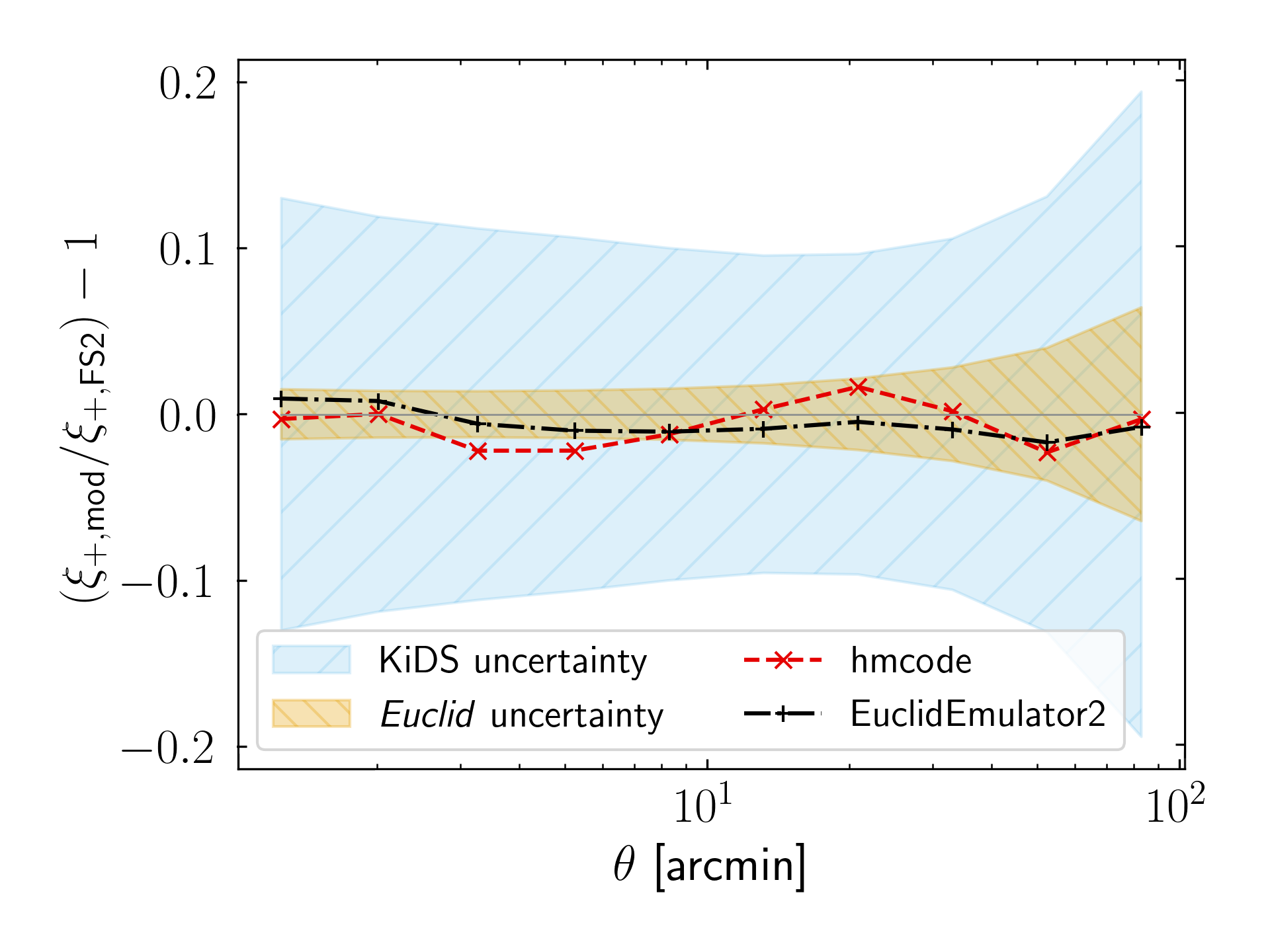}
    \caption{Fractional difference between $\xi_+$ modelled with HMCode-2020 (red, dashed) and the EuclidEmulator (black, dash-dotted) to the measurements in the FS2 for sources at $0.9<z<1.2$. The blue area corresponds to the KiDS-1000 uncertainty and the yellow area to the \Euclid uncertainty. The correlation function is binned here in 10 logarithmic bins between \ang{;1.25;} and \ang{;83;}.}
    \label{fig: EE2 HMCode}
\end{figure}

We modelled the non-linear matter power spectrum  $P(k)$ using HMCode-2020 \citep{Mead2021} for the KiDS-like analysis. We fixed the baryon feedback parameters $\eta$ and $A_\mathrm{b}$ to 0 and 3.13, respectively, corresponding to the dark-matter-only case. However, we could no use the same prescription for the \Euclid-like case. As shown in Fig.~\ref{fig: EE2 HMCode}, the measured shear correlation functions for the KiDS-like case show a deviation of up to 2\% to the HMCode-based prediction for the simulation cosmology for sources at redshifts between 0.9 and 1.2, which is well within the measurement uncertainty of KiDS-1000. For \Euclid, though, this deviation is of the order of the measurement uncertainty, indicating that the model is not accurate enough. Instead, for the \Euclid-set-up, we used the non-linear matter power spectrum predicted by the EuclidEmulator2 \citep{Knabenhans-EP9}, which agrees with the measurements within 1\%, which is within the uncertainty.

Though the EuclidEmulator has better accuracy than the HMCode model, it is calibrated on a smaller range of cosmological parameters than the HMCode. Therefore, we must assume tight priors for the inference in the \Euclid-like set-up. Furthermore, we could not use the EuclidEmulator for the KiDS-like set-up since the required tight prior range would bias the cosmological inference.

We further included the impact of intrinsic alignments on the cosmic shear power spectrum via the non-linear alignment (NLA) model (\citealp{Bridle2007, Kirk2012}; see Equations 13--16 in \citealt{Joachimi2021}), which gave us the free parameter $A_\mathrm{IA}$ characterizing the strength of intrinsic alignments. 

For the cosmological inference, we also required a covariance estimate. We modelled the covariance analytically, using the code described in \citet{Reischke2024}, which follows the approach in  \citet{Joachimi2021}. To summarize, we first calculated the three parts of the covariance of the lensing power spectrum, which are the Gaussian part, depending only on the matter power spectrum, the dispersion of intrinsic galaxy shapes and source galaxy number density; the intra-survey non-Gaussian part, depending on the matter trispectrum at modes smaller than the survey; and the super-sample covariance, depending on modes larger than the survey. The sum of these terms was then converted to a covariance estimate for the COSEBIs by convolution with a suitable kernel function following the approach by \citet{Asgari2021}. The analytical covariance does not include the impact of SLC. However, we show in App.~\ref{app: covariance} that this impact is negligible.

For the covariance estimate, we used the component-wise shape noise $\sigma_\epsilon=0.28, 0.27, 0.28, 0.27$, and $0.28$ for each bin of the KiDS-like set-up. For the \Euclid-like set-up, we assumed $\sigma_\epsilon=0.21$ for all bins, which is the same as was assumed in \citet{Blanchard-EP7} and \citet{Deshpande-EP28}.

\subsection{Cosmological parameter estimation}

Our cosmological parameter estimation was conducted using \verb|cosmosis| \citep{Zuntz2015} and used the priors in Table \ref{tab:priors}. For the KiDS-1000-like set-up, we sampled the posterior for five cosmological parameters ($\Omega_\mathrm{c}\,h^2$, $\Omega_\mathrm{b}\,h^2$, $S_8$, $h$, $n_\mathrm{s}$) and six nuisance parameters ($A_\mathrm{IA}$ and the redshift distribution shifts for each tomographic bin). We used the same priors as \citet[see Table~\ref{tab:priors}]{Asgari2021}, except for the parameter $\Omega_\mathrm{b}\, h^2$, for which the true value in the MICE, $0.255$, is very close to the prior boundary in \citet{Asgari2021}, $0.26$. We therefore extended the prior on the right-hand side.
 
For the \Euclid-like case, we sampled over the power spectrum normalization $A_\mathrm{s}$ instead of $S_8$ to use the fiducial parameter of the EuclidEmulator. We also included two parameters for the equation of state $w$ of dark energy, namely $w_0$ and $w_a$ defined by
\begin{equation}
    w(z) = w_0+w_a \frac{z}{1+z}\,,
\end{equation}
and the redshift distribution shifts of eight more tomographic bins. Due to the limited range of the EuclidEmulator, we used the tighter priors in Table~\ref{tab:priors} for the \Euclid-like set-up. We also changed the prior for $\delta z$. While for the KiDS-like analysis, we assumed a correlated prior on $\delta z$, using the correlation matrix by \citet{Hildebrandt2021}, for \Euclid, we assumed an uncorrelated Gaussian prior. For the $\Euclid$ priors, we chose a width of $0.001 (1+\Bar{z})$.

For both analyses, we assumed a Gaussian likelihood and used the nested sampler \verb|nautilus| \citep{Lange2023}. We also performed cosmological parameter estimation where we fixed all nuisance parameters to zero. This tests whether the parameter constraints are biased due to SLC if we strictly assume no intrinsic alignments to occur. From the nested sampler, we obtained estimates for the parameter values that maximize the marginal posterior distributions and their $1\sigma$ uncertainties. However, these generally differ from the parameter values that maximize the full (high-dimensional) posterior, known as Maximum A-Posteriori (MAP). We therefore ran an optimization procedure using the Nelder--Mead algorithm \citep{Nelder1965} to find the MAP for each analysis, following the same approach as in \citet{Asgari2020}.

\begin{table}
    \centering
    \caption{Priors for sampling parameters. }
    \begin{tabular}{c|cc} 
    \hline
    Parameter & KiDS-like set-up  & \Euclid-like set-up \\
    \hline
     $S_8$    & $ \mathcal{U}(0.1, 1.3)$ & -- \\
     $\ln(A_\mathrm{s})$ & -- & $\mathcal{U}(-20.1, -19.8)$\\
     $\Omega_\mathrm{c}\, h^2$   & $\mathcal{U}(0.051, 0.255)$ & $\mathcal{U}(0.107, 0.126)$ \\
     $\Omega_\mathrm{b}\, h^2$ & $\mathcal{U}(0.019, 0.026)$ & $\mathcal{U}(0.0214, 0.0223)$ \\
     $h$ &  $\mathcal{U}(0.64, 0.82)$ & $\mathcal{U}(0.61, 0.73)$\\
     $n_\mathrm{s}$ & $\mathcal{U}(0.84, 1.1)$ & $\mathcal{U}(0.92, 1.0)$\\
     $w_0$ & -- &  $\mathcal{U}(-1.3, -0.7)$\\
     $w_a$ & -- & $\mathcal{U}(-0.7, 0.7)$\\
     $A_\mathrm{IA}$ & $\mathcal{U}(-6, 6)$ & $\mathcal{U}(-6, 6)$\\
$\delta z$ & $\mathcal{N}(\mu, \mathsf{C})$ &$\mathcal{N}[0, 0.001 (1+\Bar{z})]$ \\
     \hline
    \end{tabular}
    \tablefoot{We sample in $S_8$ for the KiDS-like and in $\ln(A_\mathrm{s})$ for the $\Euclid$-like set-up. For the KiDS-like set-up the priors are the same as in \citet{Asgari2021} except for $\Omega_\mathrm{b}\, h^2$, which we extend to higher values. For the \Euclid-like set-up, the priors are given by the parameter range of the EuclidEmulator. The $\Bar{z}$ denotes the average redshift of the tomographic bin}
    \label{tab:priors}
\end{table}

\section{\label{sc:Results} Results}
\subsection{KiDS-1000-like set-up}

\begin{figure*}
\centering
    \includegraphics[width=0.9\linewidth, trim={1.5cm 2.5cm 0 2cm}, clip]{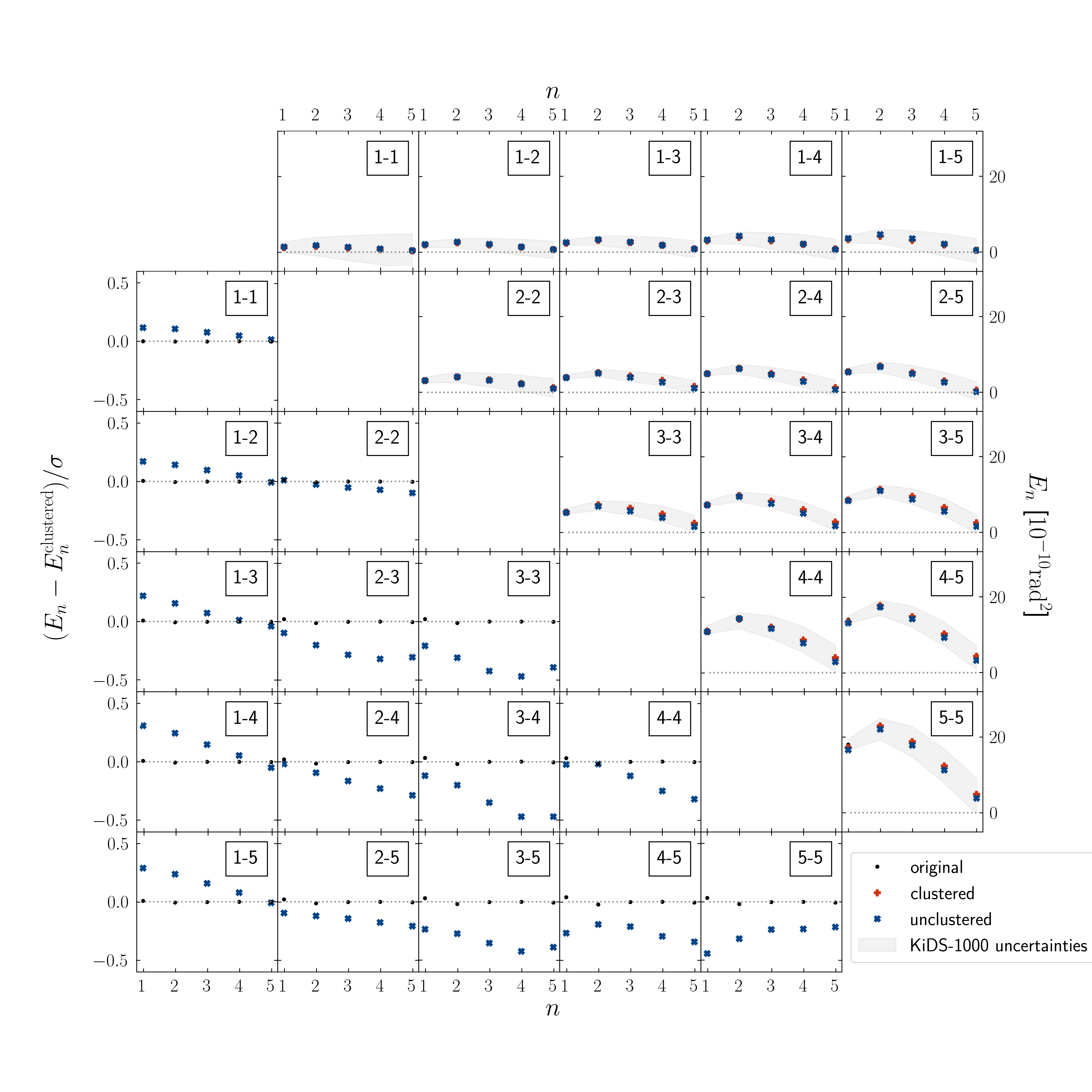}
    \caption{Measurement for the KiDS-like set-up. The $x$-axis denotes the COSEBI order $n$, each panel shows the measurement for one combination of tomographic bins, whose numbers are given in the upper right corner of each panel. Upper right: COSEBI $E$-modes $E_n$ measured for the original MICE galaxies (black points), the clustered catalogue (red plusses) and the unclustered catalogue (blue crosses) with the KiDS-1000 uncertainty (grey band). Lower left: Difference between $E_n$ for the original and clustered sources (black points) and between the clustered and unclustered sources (blue crosses), divided by the KiDS-1000 uncertainty. {The original sources have shear and position taken directly from the MICE galaxy catalogue, while the clustered sources have their position taken from the catalogue and the shear assigned from the shear maps. The unclustered sources have uniformly distributed positions and shear assigned from the shear maps.}}
    \label{fig: COSEBIs MICE}
\end{figure*}

 We first consider the KiDS-1000-like set-up. In the upper right corner of Fig.~\ref{fig: COSEBIs MICE}, we show the $E_n$ measured for the MICE simulations using the original, clustered, and unclustered source galaxy catalogues, together with the KiDS-1000 uncertainties. We note that these uncertainties are not error estimates for the measurements in the mock catalogues, since they were obtained from a different survey area and did not include shape noise. We also note that the $B$-modes $B_n$ were consistent with zero for all three cases. We consider in the lower left half of Fig.~\ref{fig: COSEBIs MICE} the difference between the clustered and the unclustered $E_n$, and between the original and unclustered $E_n$ normalized by the KiDS-1000 uncertainty.

The COSEBIs for the original and clustered source catalogue are identical. This suggests that our method of assigning shears from the MICE shear map to galaxy positions agrees with the original creation of the MICE catalogue.

The impact of SLC is given by the difference between the clustered and the unclustered $E_n$ normalized by the KiDS-1000 uncertainty. The effect shows a dependence on tomographic bins. While SLC decreases the $E_n$ for the lowest tomographic bin, it increases them at higher bins. This increase is stronger for larger $n$. The deviation between the $E_n$ for the clustered and unclustered sources is strongest for the lowest tomographic bin (up to $20\%$). However, since the uncertainty is also the largest there, the ratio of deviation to uncertainty peaks for the combination of the third and fourth bin. For all tomographic bins, the SLC has an effect of less than $0.5\,\sigma$. 

\begin{figure*}
\sidecaption
    \includegraphics[width=12cm]{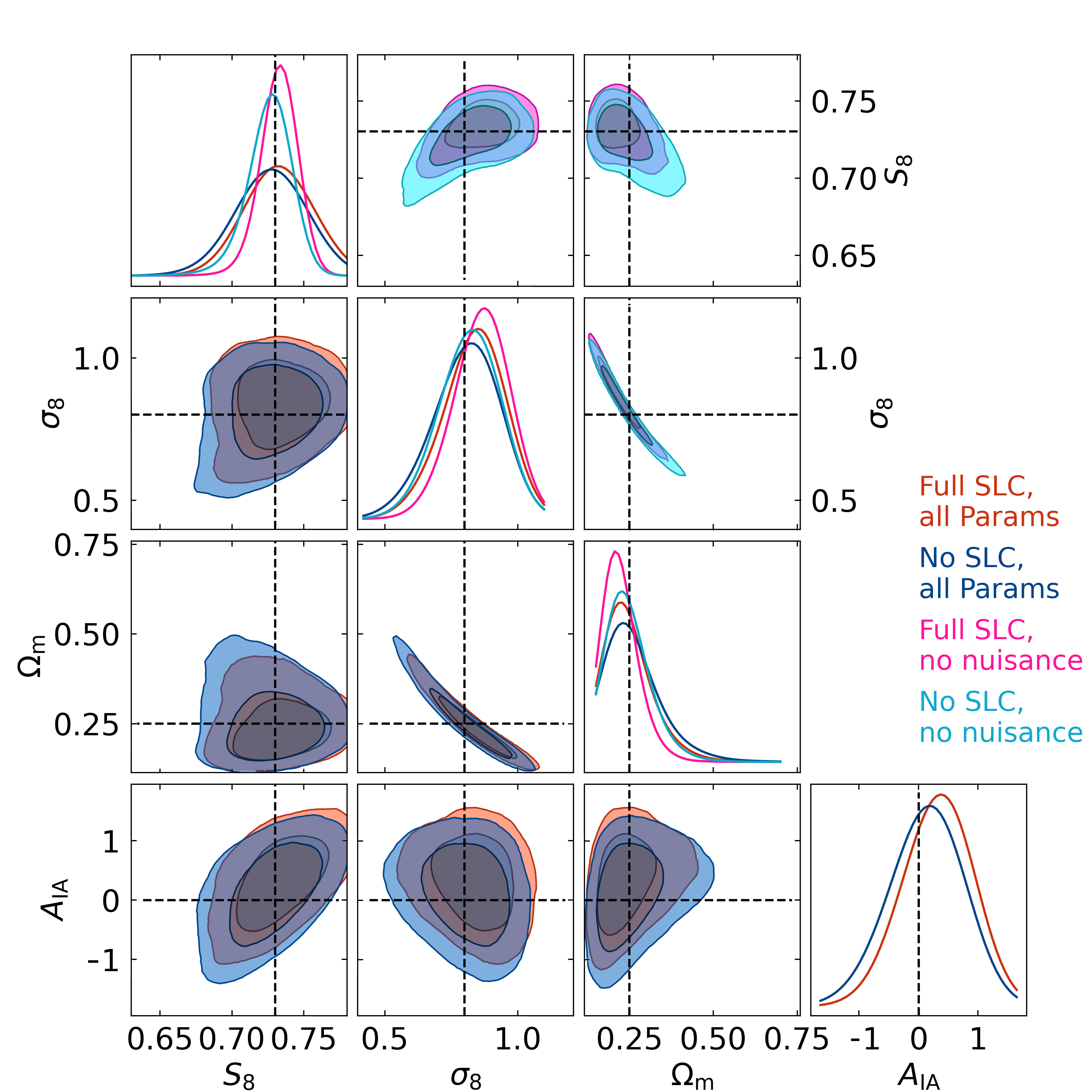}
    \caption{Parameter constraints for the KiDS-like set-up using all parameters (lower corner) and with fixed nuisance parameters (upper corner), either for a data vector including SLC (red and pink) or without SLC (dark blue and cyan). Constraints for all sampled parameters can be found in Tables \ref{tab: MAP MICE full} and \ref{tab: MAP MICE no nuis}.}
    \label{fig: MCMC MICE}
\end{figure*}

Using the COSEBIs, we first perform a cosmological inference with all nuisance parameters. In Table \ref{tab: MAP MICE full}  we report the maxima of the marginalized posteriors together with their $1\sigma$ uncertainty and the MAP values. We also give the shifts in the maxima, normalized by the $1\sigma$ uncertainties and the shifts in the MAP due to the inclusion of SLC in the data vector. Figure~\ref{fig: MCMC MICE} shows the corresponding posteriors for $S_8$, $\sigma_8$, $\Omega_\mathrm{m}$ and $A_\mathrm{IA}$ in dark blue and red. The posteriors for all varied parameters are given in Appendix~\ref{app: contours}.

\begin{table*}
	\caption{Maxima of marginal posteriors with $1\sigma$ uncertainties, MAP, and shifts due to SLC for the KiDS-like set-up with varying nuisance parameters}
	\label{tab: MAP MICE full}
	\centering
	
	\begin{tabular}{c|c|ccc|ccc}
		\hline
		Parameter                                & Truth   & \multicolumn{3}{c|}{max. + marginal $\sigma$}                           & \multicolumn{3}{c}{MAP}                         \\
		&         & clustered                 & unclustered               & $\Delta/\sigma$ & clustered & unclustered & $\Delta_\mathrm{MAP}$ \\ \hline
		$\Omega_\mathrm{c}\,h^2$ {[}$10^{-3}${]} & $101$   & $73_{-19}^{+37}$          & $77_{-25}^{+41}$          & $-0.14$         & $76$      & $79$        & $-3$                  \\
		$\Omega_\mathrm{b}\,h^2$ {[}$10^{-4}${]} & 216     & $387_{-211}^{+75}$        & $391_{-131}^{+59}$        & $-0.041$        & $231$     & $246$       & $-15$                 \\
		$h$                                      & $0.700$ & $0.739_{-0.071}^{+0.061}$ & $0.734_{-0.077}^{+0.054}$ & 0.076           & $0.704$   & $0.701$     & $0.003$               \\
		$n_\mathrm{s}$                           & $0.95$  & $0.896_{-0.047}^{0.085}$  & $0.897_{-0.050}^{0.083}$  & $-0.015$        & $0.874$   & $0.884$     & $-0.010$              \\
		$S_8$ {[}$10^{-3}${]}                    & $730$   & $731\pm17$                & $728\pm18$                & $0.18$          & $726$     & $725$       & $1$                  \\ \hline
		$\sigma_8$ {[}$10^{-3}${]}               & $800$   & $850_{-120}^{+110}$       & $830_{-130}^{+110}$       & $0.17$          & $884$     & $879$       & $5$                   \\
		$A_\mathrm{s}$ {[}$10^{-10}${]}          & $21$    & $36_{-25}^{+44}$          & $30_{-24}^{+41}$          & $0.17$          & $50$      & $47$        & $3$                   \\
		$\Omega_\mathrm{m}$ {[}$10^{-3}${]}      & $250$   & $220_{-58}^{+74}$         & $228_{-64}^{+87}$         & $-0.12$         & $205$     & $209$       & $4$                   \\ \hline
		$A_\mathrm{IA}$                          & 0       & $0.28_{-0.44}^{+0.42}$    & $0.11_{-0.46}^{+0.44}$    & $\mathbf{0.39}$ & $0.21$    & $-0.03$     & $0.24$                \\
		$\delta z_1$ {[}$10^{-3}${]}             & 0       & $-1_{-12}^{+11}$          & $0_{-12}^{+11}$           & $-0.087$        & $-1$      & $0$         & $-1$                  \\
		$\delta z_2$ {[}$10^{-3}${]}             & 0       & $5\pm12$                  & $4_{-11}^{+12}$           & $0.083$         & $6$       & $6$         & $0$                   \\
		$\delta z_3$ {[}$10^{-3}${]}             & 0       & $-2\pm12$                 & $-3\pm12$                 & $-0.083$        & $2$       & $1$         & $1$                   \\
		$\delta z_4$ {[}$10^{-3}${]}             & 0       & $3\pm9.1$                 & $0_{-10}^{+9.1}$          & $\mathbf{0.33}$ & $-3$      & $-7$        & $4$                   \\
		$\delta z_5$ {[}$10^{-3}${]}             & 0       & $-1_{-11}^{+10}$          & $0_{-11}^{+10}$           & $0.083$         & $-1$      & $0$         & $-1$                  \\ \hline
	\end{tabular}
	
	\tablefoot{$\Delta/\sigma$ is the difference between the maxima of the marginal posterior for clustered and unclustered sources, divided by the mean of the upper and lower 1$\sigma$ uncertainty of clustered sources.  Bold values indicate differences we consider significant (i.e. $|\Delta/\sigma|>0.25$). $\Delta_\mathrm{MAP}$ is the difference between the MAP for clustered and unclustered sources.}

\end{table*}

Similarly to \citet{Deshpande-EP28}, we consider shifts of more than $0.25\sigma$ significant, since then, if the marginalized posteriors were Gaussian, the contours with and without shift overlap less than 90\%. We find that, when including the nuisance parameters, none of the cosmological parameters show a significant shift due to SLC. The largest shift occurs for $\sigma_8$, which is shifted by $0.19\sigma$ towards higher values. This likely occurs to compensate for the signal increase in higher tomographic bins due to SLC. 

The MAP values are shifted by similar absolute amounts as the maxima of the marginalized posteriors.  Furthermore, all shifts have the same sign. This indicates that SLC does not have a strong influence on the degeneracy direction between parameters and simply shifts the multidimensional posterior to new values, since otherwise the marginalized posteriors would be affected differently from the MAP.

While the cosmological parameters are relatively unaffected by SLC, the nuisance parameters $A_\mathrm{IA}$ and $\delta z_4$ are shifted significantly ($0.39 \sigma$ and $0.33 \sigma$, respectively) to higher values. These shifts compensate for two different redshift regimes. A larger $A_\mathrm{IA}$ decreases the cosmic shear signal, in particular at low redshifts, which is similar to the decrease in $E_n$ for the lowest tomographic bin due to SLC. A larger $\delta z_4$ increases the signal for $E_n$ including the fourth tomographic bin, similar to the increase in signal due to SLC for the higher tomographic bins. Therefore, these two nuisance parameters neutralize the effect of SLC on the cosmological parameters, even though their physical meaning is completely unrelated to SLC.

This raises the question of whether the cosmological analysis would still be insensitive to SLC if we excluded the nuisance parameters. This scenario is particularly interesting because future cosmic shear analyses are expected to use informative priors for effects such as intrinsic alignment, constrained from independent observations \citep[e.g.][]{Johnston2019, Fortuna2021}. These priors are significantly tighter than the broad flat prior used here. To study whether SLC is still unimportant for the analyses when the nuisance parameters are not varied, we perform the second cosmological inference with $A_\mathrm{IA}$ and $\delta z_i$ fixed to zero. The resulting maxima of the marginalized posteriors and the MAP values are given in Table~\ref{tab: MAP MICE no nuis}.

\begin{table*}
    \caption{Maxima of marginal posteriors with $1\sigma$ uncertainties, MAP, and shifts due to SLC for the KiDS-like set-up with fixed nuisance parameters}
    \label{tab: MAP MICE no nuis}
	\centering

	\begin{tabular}{c|c|ccc|ccc}
		\hline
		\multicolumn{1}{c|}{Parameter}           & \multicolumn{1}{c|}{Truth} & \multicolumn{3}{c|}{max. + marginal $\sigma$}                            & \multicolumn{3}{c}{MAP}                         \\
		&                            & clustered                 & unclustered               & $\Delta/\sigma$  & clustered & unclustered & $\Delta_\mathrm{MAP}$ \\ \hline
		$\Omega_\mathrm{c}\,h^2$ {[}$10^{-3}${]} & $101$                      & $72_{-18}^{+29}$          & $80_{-25}^{+36}$          & $\mathbf{-0.34}$ & $98.0$    & $107.3$     & $-9.3$                \\
		$\Omega_\mathrm{b}\,h^2$ {[}$10^{-4}${]} & 216                        & $391_{-119}^{+72}$        & $408_{-130}^{+61}$        & $-0.18$          & $252$     & $252$       & $0$                   \\
		$h$                                      & $0.700$                    & $0.749_{-0.064}^{+0.047}$ & $0.734\pm0.055$           & $\mathbf{0.27}$  & $0.660$   & $0.654$     & $0.06$                \\
		$n_\mathrm{s}$                           & $0.95$                     & $0.893_{-0.047}^{+0.084}$ & $0.892_{-0.046}^{+0.084}$ & $0.015$          & $0.906$   & $0.905$     & $0.01$                \\
		$S_8$ {[}$10^{-3}${]}                    & $730$                      & $734_{-12}^{11}$          & $729_{-14}^{+13}$         & $\mathbf{0.42}$  & $735$     & $732$       & $3$                   \\ \hline
		$\sigma_8$ {[}$10^{-3}${]}               & $800$                      & $877_{-102}^{+94}$        & $840_{-120}^{+100}$       & $\mathbf{0.38}$  & $921$     & $862$       & $59$                  \\
		$A_\mathrm{s}$ {[}$10^{-10}${]}          & $21$                       & $43_{-27}^{+41}$          & $30_{-23}^{+36}$          & $\mathbf{0.38}$  & $81$      & $57$        & $24$                  \\
		$\Omega_\mathrm{m}$ {[}$10^{-3}${]}      & $250$                      & $206_{-44}^{+53}$         & $221_{-50}^{+69}$         & $\mathbf{-0.31}$ & $191$     & $216$       & $-25$                 \\ \hline
	\end{tabular}

		\tablefoot{Columns defined as in Table \ref{tab: MAP MICE full}.}
\end{table*}

\begin{figure*}
    \centering
    \includegraphics[width=\linewidth, trim={1.5cm 2.5cm 0 2cm}, clip]{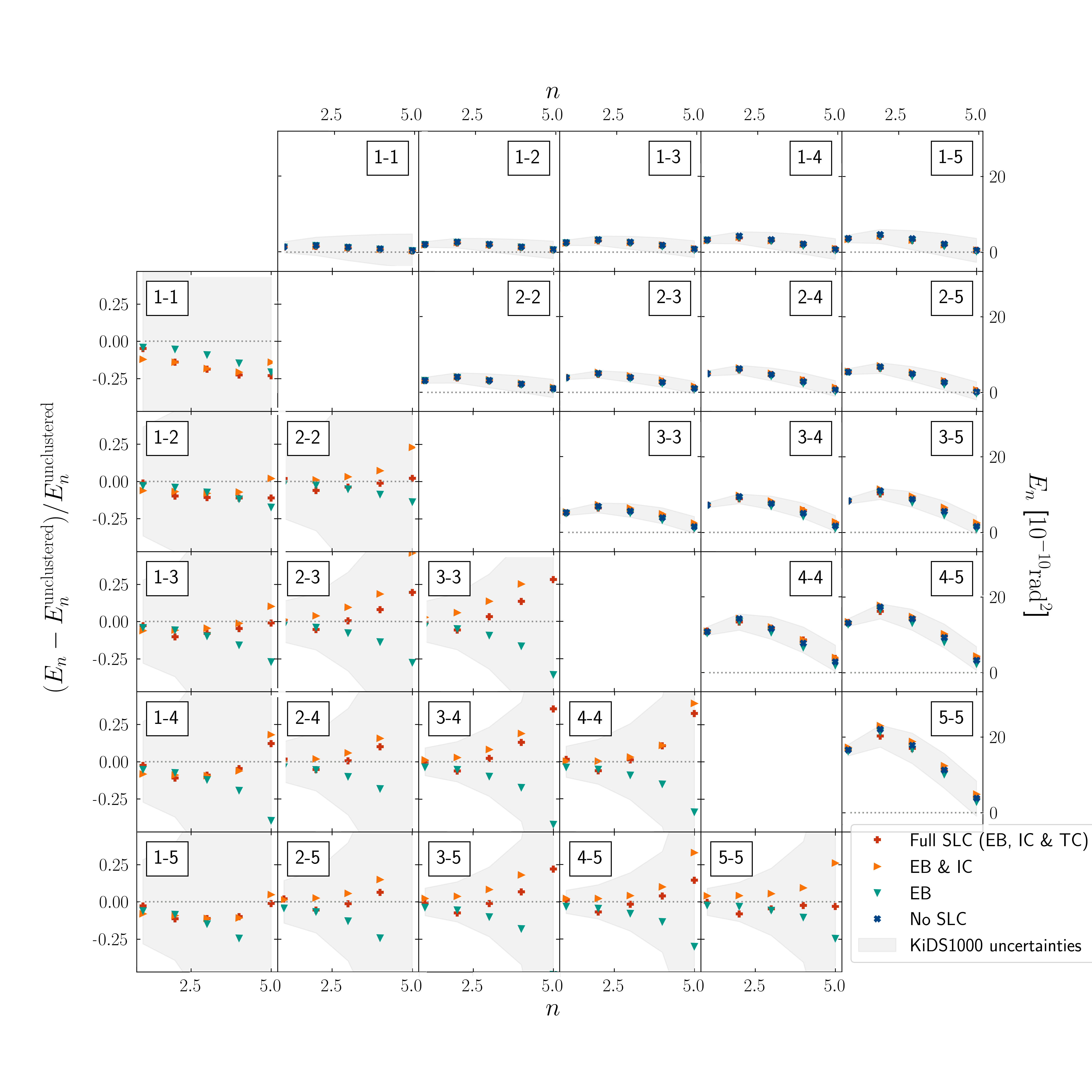}
    \caption{COSEBIs for the KiDS-like set-up when different SLC effects are included, either the full SLC (red crosses), the estimator bias (EB) and intrinsic clustering (IC; orange triangles), only the estimator bias (EB; green triangles), and no SLC (blue dots). \textit{Upper right}: COSEBIs. \textit{Lower left}: Fractional difference to COSEBIs for unclustered sources (i.e. without SLC).}
    \label{fig: MICE cosebis individual effects}
\end{figure*}

\begin{figure*}
    \centering
    \includegraphics[width=0.9\linewidth]{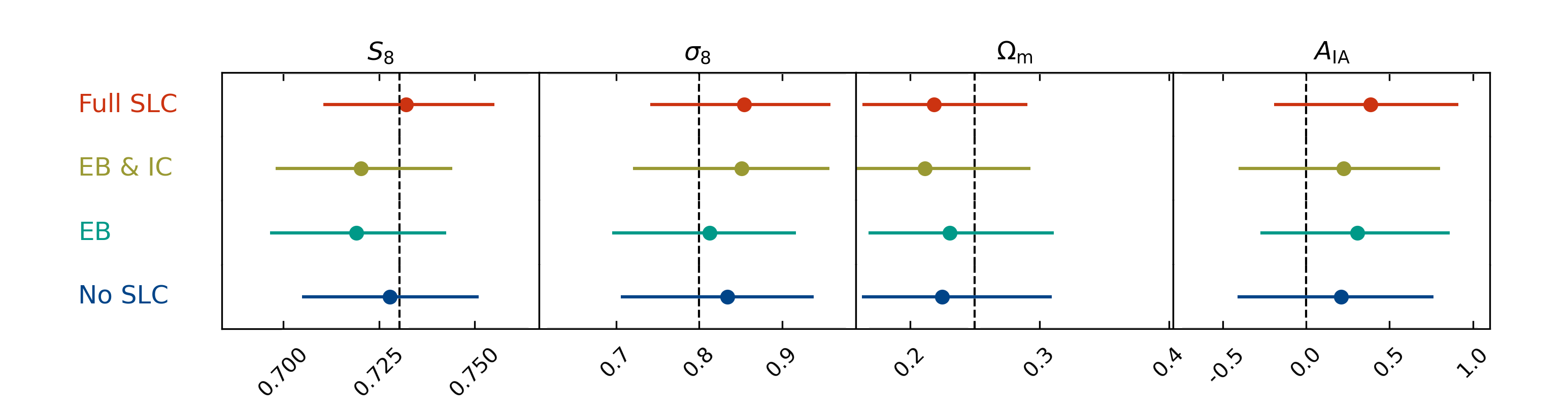}
    \caption{Parameter constraints for the KiDS-like set-up, for a data vector including the full SLC, the estimator bias (EB) and intrinsic clustering (IC), only the estimator bias (EB), and without any SLC.}
    \label{fig: MCMC MICE individual effects}
\end{figure*}

The posteriors for  $S_8$, $\sigma_8$, and $\Omega_\mathrm{m}$ are shown in Fig.~\ref{fig: MCMC MICE} as pink and cyan contours and the full posteriors are in Appendix~\ref{app: contours}. Neglecting SLC would now lead to shifts of $0.42 \sigma$ in $S_8$ and $0.31 \sigma$ in $\Omega_\mathrm{m}$, with $S_8$ shifted to higher and $\Omega_\mathrm{m}$ shifted to lower values. These increases in the effect of SLC are partly due to the decreased parameter uncertainties compared to the analysis with nuisance parameters, which increase the significance of the parameter shifts. However, the absolute values of the shifts also increase, for example $S_8$ shifts by $0.7\%$ for fixed nuisance parameters and $0.4\%$ for variable nuisance parameters. This confirms that SLC would bias the cosmological inference stronger if nuisance parameters were fixed.

Both with and without nuisance parameters, the MICE simulation's true cosmological parameters are within the cosmological inference's uncertainties. However, in both cases, the values for the unclustered data vector is closer to the true values. This is unsurprising, as the unclustered data vector corresponds to the model assumption of unclustered source galaxies. The remaining difference between the inferred parameters for the unclustered data vector and the true cosmology is likely due to the fact that the simulation is a single realization of the cosmology and the limited accuracy of the model.

We now investigate the impact of the different SLC effects discussed in Sect.~\ref{sc:Theo:SLC}. For this, we show in Fig.~\ref{fig: MICE cosebis individual effects} the COSEBIs when different parts of the SLC are included, as well as the fractional difference to the signal for unclustered sources. The largest impact on the signal is due to the estimator bias (EB), which suppresses the signal for all tomographic bins. The IC and TC effects partially counteract each other. While the IC increases the signal, as expected from Eq.~\eqref{eq:Delta Cell final}, the contamination by lower redshift galaxies decreases it. The scale dependence of the SLC effects is investigated using shear correlation functions in Appendix \ref{app: ind Effects Xip}.

To see how the individual SLC effects affect the cosmological parameter constraints, Fig.~\ref{fig: MCMC MICE individual effects} shows the corresponding parameter constraints. For $S_8$ the strongest impact occurs due to EB, which lowers the estimated value. This is in line with the effect on the COSEBIs -- lowering the cosmic shear signal is equivalent to lowering $S_8$. The IC slightly increases $S_8$ again, as expected since it increases the cosmic shear signal. Interestingly, $S_8$ increases even more, once the TC effect is included. This seems counterintuitive since TC generally lowers the shear signal. However, simultaneously to increasing $S_8$, the effect also increases the intrinsic alignment parameter $A_\mathrm{IA}$, which lowers the signal. Since $A_\mathrm{IA}$ mainly causes a decrease in the signal at low tomographic bins, where the suppression due to TC is strongest, while $S_8$ mainly increases the signal at higher redshifts, these two parameters together can match the total effect by the contamination.

\subsection{\Euclid-like set-up}

\begin{figure*}
\centering
    \includegraphics[width=0.9\linewidth, trim={1.5cm 2cm 1.5cm 2cm}, clip]{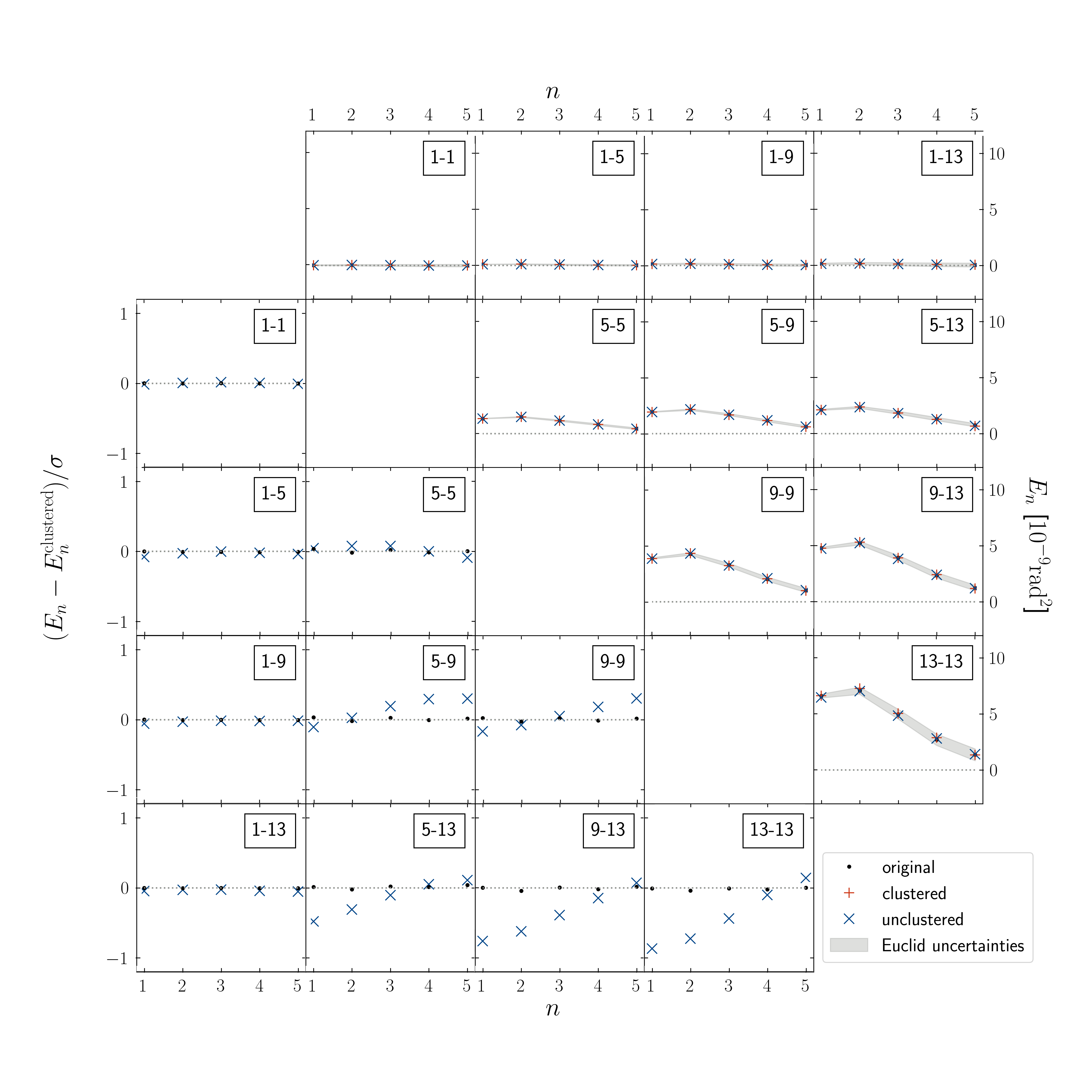}
        \caption{Measurement for the \Euclid-like set-up. \textit{Upper right}: COSEBI $E$-modes $E_n$ measured for the original FS2 galaxies (black points), the clustered catalogue (red plusses) and the unclustered catalogue (blue crosses) with the \Euclid uncertainty (grey band) for a subset of the 13 tomographic bins. \textit{Lower left}: Difference between $E_n$ for the original and clustered sources (black points) and between the clustered and unclustered sources (blue crosses), divided by \Euclid uncertainty. The full measurement for all tomographic bins is in Fig.~\ref{fig: cosebisEuclid_all}.}
    \label{fig: COSEBIs Euclid}
        
\end{figure*}

Finally, we consider the \Euclid-like set-up. Figure~\ref{fig: COSEBIs Euclid} shows the measured COSEBIs for the clustered and unclustered set-up, as well as the differences for a subset of the tomographic bins. The full measurements for all tomographic bins are provided in Fig.~\ref{fig: cosebisEuclid_all}.  The measurements for the clustered and original catalogue agree perfectly, which confirms that the creation of the two mock catalogues is consistent. The difference between the clustered and unclustered case is smaller for the \Euclid-like set-up than for the KiDS-1000-like set-up. This is likely due to the narrower redshift distributions $p_z(z)$, with shorter tails and fewer outliers. Consequently, there is less overlap between source galaxies in a tomographic bin and the matter structures lensing other galaxies in the same tomographic bin, which reduces the impact of SLC. Nevertheless, since the statistical uncertainty for \Euclid is much smaller than for Stage III surveys such as KiDS, the difference between the two cases approaches $1\sigma$ for the highest tomographic bins, while it remains below $0.5\sigma$ for KiDS-1000.

\begin{figure*}
    \includegraphics[width=0.9\linewidth]{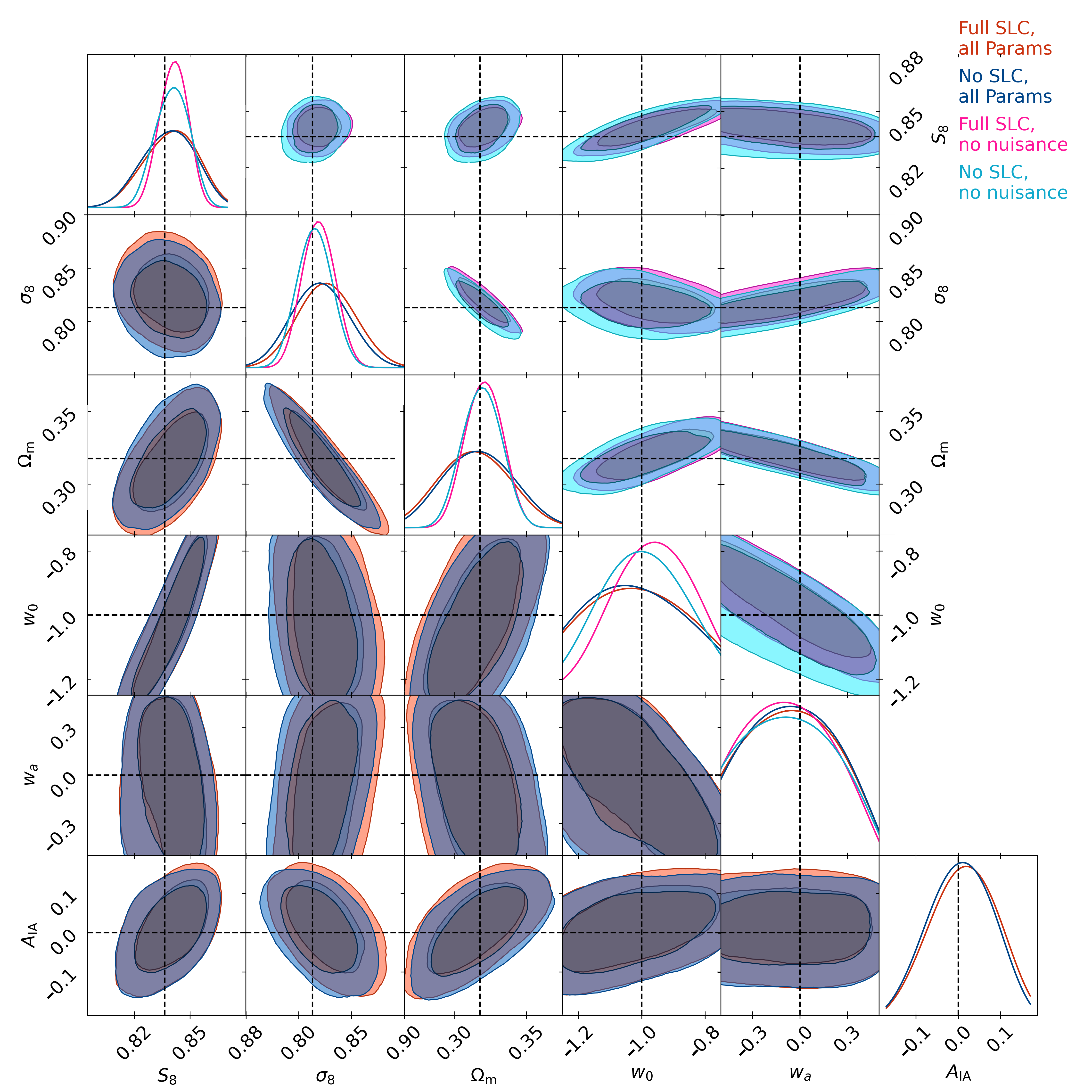}
    \caption{Parameter constraints for the \Euclid-like set-up using all parameters (lower corner) and with fixed nuisance parameters (upper corner), either for a data vector including SLC (dark blue and cyan) or without SLC (red and pink). Constraints for all sampled parameters can be found in Tables \ref{tab: MAP Euclid full} and \ref{tab: MAP Euclid no nuis}.}
    \label{fig: MCMC Euclid}
\end{figure*}

Figure \ref{fig: MCMC Euclid} shows the parameter constraints for the most important cosmological parameters (red and blue constraints), while the maxima of the marginalized posteriors and the MAP values for all parameters are given in Table \ref{tab: MAP Euclid full}. We find lower shifts in parameter estimates due to SLC than for the KiDS-like set-up, which is likely due to the narrower redshift bins. This provides an additional motivation for analysing \Euclid with more than ten tomographic bins, as already suggested by \citet{Pocino-EP12}.

\begin{table*}
	    \caption{Maxima of marginal posteriors with $1\sigma$ credible intervals, MAP, and shifts due to SLC for the \Euclid-like set-up with varying nuisance parameters}
	\label{tab: MAP Euclid full}
	\centering
	\begin{tabular}{c|c|ccc|ccc}
		\hline
		Parameter                                & Truth  & \multicolumn{3}{c|}{max. + marginal $\sigma$}                            & \multicolumn{3}{c}{MAP}                         \\
		&        & clustered                 & unclustered                & $\Delta/\sigma$ & clustered & unclustered & $\Delta_\mathrm{MAP}$ \\ \hline
		$\Omega_\mathrm{c}\,h^2$ {[}$10^{-3}${]} & 120.5  & $120.8_{-6.2}^{+4.1}$     & $121.8_{-6.0}^{+4.1}$      & $-0.10$         & 120.3     & 120.6       & $-0.3$                \\
		$\Omega_\mathrm{b}\,h^2$ {[}$10^{-4}${]} & 220.0  & $217.7_{-4.0}^{+1.8}$     & $217.4_{-2.1}^{+3.7}$      & $0.10$          & $219.8$   & $219.7$     & $-0.1$                \\
		$h$                                      & 0.670  & $0.674_{-0.028}^{+0.029}$ & $0.670_{-0.027}^{+0.030}$  & $0.14$          & $0.670$   & $0.669$     & $0.01$                \\
		$n_\mathrm{s}$                           & 0.960  & $0.980_{-0.025}^{+0.017}$ & $0.982_{-0.024}^{+0.016}$  & $-0.095$        & $0.968$   & $0.965$     & $0.003$               \\
		$A_\mathrm{s}$ {[}$10^{-10}${]}          & 21.0   & $21.7_{-2.3}^{2.2}$       & $21.4_{-2.4}^{+2.3}$       & $0.13$          & $21.5$    & $21.1$      & $0.4$                 \\
		$w_0$                                    & $-1.000$ & $-1.04_{-0.16}^{+0.18}$   & $-1.06_{-0.151}^{+0.19}$   & $0.12$          & $-1.00$   & $-1.00$     & $0$                   \\
		$w_a$                                    & 0.000  & $-0.04_{-0.39}^{+0.34}$   & $-0.06\pm0.36$             & $0.05$          & $-0.05$   & $-0.08$     & $0.03$                \\ \hline
		$S_8$ {[}$10^{-3}${]}                    & 836.5  & $841_{-16}^{+12}$         & $843_{-15}^{+13}$          & $-0.14$         & $841$     & $844$       & $-3$                  \\
		$\sigma_8$ {[}$10^{-3}${]}               & 813.0  & $825_{-25}^{+27}$         & $820_{-27}^{+26}$          & $0.19$          & $822$     & $818$       & $4$                   \\
		$\Omega_\mathrm{m}$ {[}$10^{-3}${]}      & 317.6  & $313_{-24}^{+25}$         & $316\pm 24$                & $-0.12$         & $318$     & $320$       & $-2$                  \\ \hline
		$A_\mathrm{IA}$                          & 0      & $0.022_{-0.082}^{0.081}$  & $-0.009_{-0.080}^{+0.079}$ & $\mathbf{0.39}$ & 0.027     & $-0.002$    & $0.029$               \\
		$\delta z_1$ {[}$10^{-4}${]}             & 0      & $2.6_{-9.9}^{+5.4}$       & $0.5_{-8.5}^{+6.8}$        & $\mathbf{0.27}$ & 7.4       & 2.8         & 2.8                   \\
		$\delta z_2$ {[}$10^{-4}${]}             & 0      & $-0.5_{-7.0}^{+8.0}$      & $-1.2_{-7.0}^{+8.2}$       & 0.093           & 5.6       & 5.0         & 0.6                   \\
		$\delta z_3$ {[}$10^{-4}${]}             & 0      & $-1.2_{-6.5}^{+8.1}$      & $-2.0_{-5.8}^{+8.9}$       & $0.11$          & $-4.0$    & $-4.6$      & $0.6$                 \\
		$\delta z_4$ {[}$10^{-4}${]}             & 0      & $3.1_{-8.5}^{+5.9}$       & $1.0_{-8.0}^{+6.5}$        & $\mathbf{0.29}$ & 5.3       & 3.0         & 2.3                   \\
		$\delta z_5$ {[}$10^{-4}${]}             & 0      & $-0.2_{-7.3}^{+7.4}$      & $1.9_{-8.6}^{+6.1}$        & $-0.23$         & $-2.9$    & $-0.9$      & $-2.0$                \\
		$\delta z_6$ {[}$10^{-4}${]}             & 0      & $0.8_{-8.0}^{+6.8}$       & $2.4_{-8.7}^{+6.0}$        & $-0.22$         & $-2.5$    & $-1.3$      & $-1.2$                \\
		$\delta z_7$ {[}$10^{-4}${]}             & 0      & $-1.6_{-6.7}^{+8.3}$      & $-0.8_{-6.8}^{+8.3}$       & $-0.11$         & $-1.9$    & 0.1         & $-2.0$                \\
		$\delta z_8$ {[}$10^{-4}${]}             & 0      & $0.0_{-7.4}^{+7.7}$       & $-1.8_{-6.3}^{+8.8}$       & 0.24            & 0.5       & $-2.2$      & $2.6$                 \\
		$\delta z_9$ {[}$10^{-4}${]}             & 0      & $-2.2_{-6.0}^{+9.4}$      & $-1.1_{-6.6}^{+8.8}$       & $-0.14$         & $-1.0$    & $-3.9$      & 2.9                   \\
		$\delta z_{10}$ {[}$10^{-4}${]}          & 0      & $0.6_{-8.2}^{+7.0}$       & $-1.5_{-6.5}^{+8.8}$       & $\mathbf{0.27}$ & $4.0$     & $2.3$       & 1.7                   \\
		$\delta z_{11}$ {[}$10^{-4}${]}          & 0      & $0.7_{-8.4}^{+7.0}$       & $2.5_{-9.5}^{+5.9}$        & $-0.23$         & $-1.6$    & $-0.3$      & $-1.3$                \\
		$\delta z_{12}$ {[}$10^{-4}${]}          & 0      & $1.4_{-9.1}^{+6.4}$       & $1.8_{-9.3}^{+6.0}$        & $-0.052$        & $4.6$     & $5.0$       & $-0.4$                \\
		$\delta z_{13}$ {[}$10^{-4}${]}          & 0      & $-1.0_{-10.4}^{+5.1}$     & $-3.0_{-5.1}^{+10.3}$      & $\mathbf{0.26}$ & $-1.9$    & $-5.1$      & 3.2                   \\ \hline
	\end{tabular}

\tablefoot{Columns defined as in Table \ref{tab: MAP MICE full}.}
\end{table*}

Similarly to the KiDS set-up, the nuisance parameters are more strongly affected than the cosmological ones. Several redshift bin shifts $\delta z$ vary by more than $0.25\sigma$ between the cases with and without SLC. Consequently, these parameters again account for the SLC, while the more physical parameters remain stable. 

The MAP values show similar absolute shifts due to SLC as the maxima of the marginalized posteriors, which also show the same sign. The shift in $S_8$ is slightly larger for the MAP than for the marginalized posterior ($-2.5\times 10^{-3}$ compared to $-2.0\times 10^{-3}$) while the shift in $\Omega_\mathrm{m}$ is slightly smaller  ($-1.7\times 10^{-3}$ compared to $-3.0\times 10^{-3}$). Overall, the similarity of the absolute shifts suggests that, similar to the KiDS-set-up, SLC mainly moves the full posterior to a different parameter set and does not strongly influence the degeneracy directions between parameters.

We investigate the effect of neglecting the nuisance parameters in Fig.~\ref{fig: MCMC Euclid} as pink and cyan contours. Here, we fix the $\delta z$ and $A_\mathrm{IA}$ parameters to zero. In this set-up the shifts due to the unaccounted SLC increase, with the shift for $\sigma_8$ and $\Omega_\mathrm{c}\,h^2$ grazing the significance threshold of $0.25\sigma$ and the shift in $w_0$ becoming larger than this threshold. This increase in significance is partly due to the strongly decreased parameter uncertainties. However, for example for $w_0$, neglecting the nuisance parameters also increases the absolute value of the shift, from $0.02$ to $0.04$. Consequently, for \Euclid, it is vital to either use nuisance parameters to account for unmodelled biases or to correctly include SLC in the model.

\begin{table*}
	    \caption{Maxima of marginal posteriors with $1\sigma$ credible intervals, MAP, and shifts due to SLC for the \Euclid-like set-up for fixed nuisance parameters}
\label{tab: MAP Euclid no nuis}
\centering
		\begin{tabular}{c|c|ccc|ccc}
			\hline
			Parameter                                & Truth  & \multicolumn{3}{c|}{max. + marginal $\sigma$}                           & \multicolumn{3}{c}{MAP}                         \\
			&        & clustered                 & unclustered               & $\Delta/\sigma$ & clustered & unclustered & $\Delta_\mathrm{MAP}$ \\ \hline
			$\Omega_\mathrm{c}\,h^2$ {[}$10^{-3}${]} & 120.5  & $120.5_{-6.0}^{+4.2}$     & $121.3_{-6.4}^{+4.5}$     & $-0.24$         & 121.1     & 121.4       & 0.3                   \\
			$\Omega_\mathrm{b}\,h^2$ {[}$10^{-4}${]} & 220.0  & $217.4_{-2.0}^{+3.8}$     & $217.8_{-2.4}^{+3.4}$     & $-0.14$         & 219.1     & 218.4       & 0.7                   \\
			$h$                                      & 0.670  & $0.667_{-0.018}^{+0.019}$ & $0.664\pm0.019$           & 0.16            & 0.672     & 0.665       & 0.007                 \\
			$n_\mathrm{s}$                           & 0.960  & $0.975_{-0.020}^{+0.018}$ & $0.977_{-0.020}^{+0.017}$ & $-0.11$         & 0.962     & 0.965       & $-0.003$              \\
			$A_\mathrm{s}$ {[}$10^{-10}${]}          & 21.0   & $21.3\pm1.8$              & $21.2_{-1.6}^{+1.8}$      & 0.056           & $21.4$    & $21.2$      & 0.2                   \\
			$w_0$                                    & $-1.000$ & $-0.96_{-0.12}^{+0.13}$   & $-1.00_{-0.13}^{+0.14}$   & $\mathbf{0.32}$ & $-0.96$   & $-0.98$     & 0.02                  \\
			$w_a$                                    & 0.000  & $-0.10_{-0.36}^{+0.35}$   & $-0.08_{-0.38}^{+0.36}$   & $-0.056$        & $-0.04$   & $-0.02$     & $-0.02$               \\ \hline
			$S_8$ {[}$10^{-3}${]}                    & 836.5  & $841.9_{-7.7}^{+7.0}$     & $841.3_{-9.3}^{+8.5}$     & 0.082           & 843.4     & 849.3       & $-0.6$                \\
			$\sigma_8$ {[}$10^{-3}${]}               & 813.0  & $819\pm15$                & $815\pm16$                & 0.2             & 819       & 816         & 3                     \\
			$\Omega_\mathrm{m}$ {[}$10^{-3}${]}      & 317.6  & $321_{-13}^{+12}$         & $319\pm13$                & 0.16            & 325       & 318         & 7                     \\ \hline
		\end{tabular}
\tablefoot{Columns defined as in Table \ref{tab: MAP MICE full}.}
\end{table*}

\begin{figure*}
    \includegraphics[width=\linewidth]{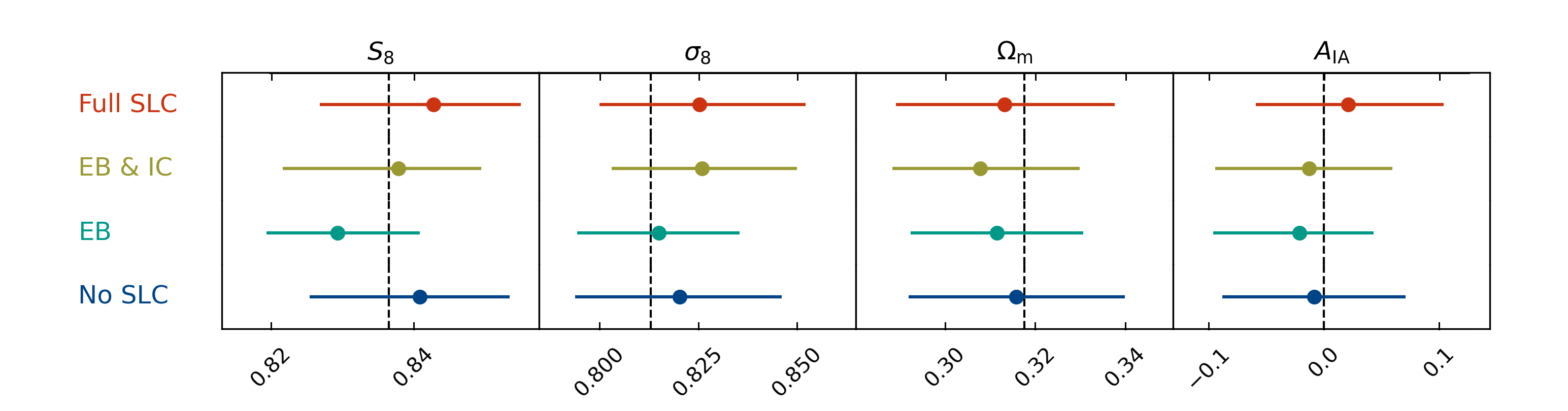}
        \caption{Parameter constraints for \Euclid-like set-up, for a data vector including the full SLC, the estimator bias (EB) and intrinsic clustering (IC), only the estimator bias (EB), and without any SLC.}
    \label{fig: MCMC FS individual effects}
\end{figure*}

Finally, we also test the impact of the different SLC effects for the \Euclid-like set-up. Figure~\ref{fig: MCMC FS individual effects} shows the obtained parameter constraints for $S_8$, $\sigma_8$, $\Omega_\mathrm{m}$ and $A_\mathrm{IA}$ when excluding parts of the SLC effects. The effects qualitatively induce the same changes in the parameter constraints as for the KiDS-like set-up. The EB effect decreases $S_8$, while the IC and TC effects increase it. The intrinsic alignment parameter remains stable under the EB and IC effects but shifts to higher values for the TC effect. Again, the EB has the strongest impact on the $S_8$ parameter.

\section{\label{sc:Discussion} Discussion}
In this work we investigated the impact of the clustering of source galaxies with lensing matter structures, known as source-lens clustering (SLC), on cosmological parameter inference with second-order statistics of cosmic shear. We considered realistically clustered mock galaxies in the MICE and FS2 simulations, as well as uniformly distributed sources, and measured the COSEBIs for a KiDS-1000-like and a \Euclid-like set-up. We then performed a cosmological inference for both clustered and unclustered sources using nested sampling of the posterior and including nuisance parameters for intrinsic alignments and shifts of the source redshift distributions. In summary, we found that if nuisance parameters are taken into account, SLC has only a minor impact on cosmological inference for Stage III, making it an unlikely candidate to resolve the `$S_8$-tension'.

For KiDS-1000, we found little impact of SLC on cosmological parameters if nuisance parameters are taken into account. The SLC causes a small decrease (within $0.5\sigma$) in the cosmic shear signal for the lowest tomographic bin and a similarly small increase for the higher tomographic bins. These shifts are `absorbed' by changes in the intrinsic alignment parameter $A_\mathrm{IA}$ and the redshift distribution biases $\delta z$, leading to cosmological parameters, such as $S_8$ and $\Omega_\mathrm{m}$, to be shifted by less than $0.2 \sigma$. Thus, it is valid for Stage III surveys to neglect SLC in the modelling as long as nuisance parameters for intrinsic alignments and shifts of the tomographic bins are included. We thus confirm the expectation outlined in \citet{Krause2021} that SLC has only a minor impact on cosmological inference with Stage III surveys and can be neglected in their modelling.

This picture changes if nuisance parameters are fixed and not varied with the analysis. In this case, we see that $S_8$ is shifted by $0.42\sigma$ and $\Omega_\mathrm{m}$ by $0.31 \sigma$ for the KiDS-1000-like set-up. Consequently, when neglecting nuisance parameters, SLC should be included in the modelling of the cosmic shear signal. Furthermore, we advise against interpreting parameters such as $A_\mathrm{IA}$ as `quantifying intrinsic alignment'. As shown here, effects completely unrelated to intrinsic alignments can affect this parameter substantially. A further example of this was found by \citet{Li2021}, who found that the intrinsic alignment nuisance parameter is strongly correlated with the redshift uncertainty for low-redshift tomographic bins and fixing the assumed redshift uncertainty to zero could lead to an unphysically high inferred $A_\mathrm{IA}$ for blue galaxies. Furthermore, \citet{Leonard2024} found that the redshift scaling of the intrinsic alignment amplitude is strongly degenerate with the mean and variance of the redshift distribution. The interplay between redshift distribution uncertainties and intrinsic alignment constraints is further dependent on the considered intrinsic alignment model and can be affected by the considered priors \citep{Fischbacher2022}.

Different intrinsic alignment models might further complicate the interplay between intrinsic alignment parameters and SLC. Here, we only investigated the NLA model, but more complicated models, such as the Tidal Alignment - Tidal Torquing (TATT) model \citep{Blazek2019}, include more free parameters that can further compensate for the impact of SLC. 

The fact that unmodelled systematic effects such as SLC change the meaning of the nuisance parameters can be an issue when deciding on priors for these parameters. Ideally, effects such as intrinsic alignment could be constrained from observations using physical models \citep[e.g.][]{Fortuna2021}. These constraints would provide tight priors on the nuisance parameters. However, this is no longer straightforward if the parameters are meant to also capture other effects such as SLC. Consequently, if observational motivated priors for nuisance parameters are used, effects such as SLC need to be carefully included in the modelling. 

The parameter uncertainties we find for the KiDS-1000-like set-up are of the same order as those found in \citet{Asgari2021} except for $\Omega_\mathrm{b}h^2$, whose uncertainty is more than twice the one from \citet{Asgari2021}. This is due to the larger prior for this parameter we assume here, as $\Omega_\mathrm{b}$ cannot be constrained from a KiDS-1000-like survey. Since cosmic shear is almost insensitive to the baryonic matter density $\Omega_\mathrm{b}$ the constraints are completely prior-dominated. Our larger prior thus also leads to a larger posterior. In general, our confidence intervals are slightly larger than those by \citet{Asgari2021}, which is likely because we are using the sampler \verb|nautilus| to sample the posterior, while they use \verb|multinest| \citep{Feroz2019}, which can underestimate posteriors \citep{Lemos2023}. Furthermore, we also do not vary the parameter $A_\mathrm{b}$ for baryonic feedback, which in the simulation is exactly zero.

For KiDS, we investigated the impact of different parts of the SLC effect. We found that the largest effect is due to the estimator bias, which occurs due to the neglect of the source galaxies angular correlation function and decreases the measured signal. The intrinsic clustering of the sources and the contamination of tomographic bins due to photometric redshift errors partly counteract each other, leaving the estimator bias as the most important contribution. 

For the \Euclid-like set-up, we see a similar effect as for the KiDS-1000-like set-up. However, the absolute shift in the signal due to SLC is smaller. This is likely due to the narrower $p_z(z)$ with shorter tails. These lead to less scattering between tomographic redshift bins, thus reducing the tomographic bin contamination (TC) effect. The $p_z(z)$ are still broader than the typical galaxy correlation length, which is of the order of 1 Mpc, so the intrinsic clustering remains. 

We also find that the impact on the cosmological parameter analysis for \Euclid, aside from the nuisance parameters, is small. $S_8$ is shifted only by $0.14 \sigma$ and $w_0$ by $0.12 \sigma$ if nuisance parameters are taken into account. This indicates that SLC might even be neglected for a Stage IV survey. Again, the picture would change if nuisance parameters were fixed in the analysis. Then, the shift in $w_0$ becomes significant ($0.32 \sigma$) and shifts in $\sigma_8$ and $\Omega_\mathrm{m}$ also increase. 

This small effect of SLC on cosmological parameters is in apparent conflict with \citet{Deshpande-EP28}, who found that neglecting SLC can cause biases of more than $1\sigma$ on $\Omega_\mathrm{m}$ and $\sigma_8$. However, their analysis relied on modelling SLC analytically, while we directly measured the effect using realistically clustered sources. The analytic modelling naturally included simplifying assumptions, such as a modelled redshift distribution, linear galaxy bias and idealized photometric redshift uncertainties. Our analysis did not require these and could be seen as a more realistic extension of their analysis.

We note that our uncertainties on the cosmological parameters are larger than those quoted in \citet{Deshpande-EP28}. This is likely due to three reasons.
First, we perform a full sampling of the posterior instead of a Fisher analysis as in \citet{Deshpande-EP28}. Fisher analyses can underestimate a posterior's width, leading to tighter parameter constraints \citep{Wolz2012}. 
Second, we include more parameters in our analysis: the intrinsic alignment parameter $A_\mathrm{IA}$ and the $p_z(z)$-shifts for the tomographic bins. Including nuisance parameters reduces the constraining power on the cosmological parameters, leading to larger uncertainties.
Third, we use a different observable, namely COSEBIs, instead of the lensing power spectrum. Consequently, our analysis uses other (and differently weighted) length scales, and the constraints are not directly comparable. As shown by \citet{Asgari2021}, the COSEBIs used here are most sensitive to $\ell$ between 10 and 1500, while \citet{Deshpande-EP28} considered $\ell$ up to 5000. COSEBIs also provide a separation into $E$- and $B$-modes, while an analysis using the $C(\ell)$ directly, implicitly assumes vanishing $B$-modes, which leads to tighter parameter constraints.

We note that the photometric redshift estimates for our simulated \Euclid-like sample were derived assuming the full depth of the \Euclid calibration fields. These will not be fully available for the first \Euclid data releases, so the photometric redshifts will show a larger scatter, increasing the TC effect. However, the statistical uncertainty for the first data release will also be significantly larger due to the smaller survey area, so a subtle effect such as SLC decreases in importance.

{ Another confounding factor to the SLC that has not been considered here is the impact of blending. Since galaxies have a finite extent on the sky, galaxies that are close by might not be detected as individual objects, which biases shear estimates (Euclid Collaboration: Congedo et al., in prep). This bias is correlated with the distribution of sources, as stronger source clustering leads to more blended objects, leading to a further correlation between the estimated shear and the source distribution. Studying the impact of blending requires not only simulated sources with a realistic spatial distribution, provided here by the FS2 and MICE simulations but also a realistic (observed) size distribution obtained from image simulations, as created for example by \citet{Jansen2024}. We leave the investigation of the impact of blending on SLC to future work.}

Accounting for SLC in the theoretical modelling remains a challenge. The intrinsic clustering (IC) effect can be estimated from first principles, as discussed in \citet{Krause2021} and Appendix \ref{app: modelling}. However, this is actually the smallest contributor to the total SLC. Modelling the estimator bias (EB) and TC effects is more involved, as they require realistic models of the (three-dimensional) source galaxy distribution $n(\chi\varthetavec, \chi)$. In principle, this could be measured alongside the sources' shears, but for photometric samples, the distance estimate will not be accurate enough. Instead, a better understanding of how the scatter in photometric redshift estimates impacts the number density of selected sources, conditioned on the overall galaxy number density, could help in modelling these effects more thoroughly.

Finally, we note that our analysis was concerned with second-order cosmic shear statistics. \Euclid will achieve its tightest constraints on cosmological parameters from second-order statistics by combining the cosmic shear analysis with measurements of galaxy clustering and the cross-correlation of shear and galaxy positions. These constraints will be tighter than for cosmic shear alone (e.g., about a factor 2 in $\Omega_\mathrm{m}$ for a flat $\Lambda$CDM setting, \citealp{Blanchard-EP7}). Consequently, the impact of SLC becomes more significant.

Furthermore, higher-order statistics, which could further improve the cosmological constraints, might be even stronger impacted by SLC. Recently, \citet{Gatti2024} investigated the impact of SLC on higher-order shear statistics and found that it can significantly bias measurements and cosmological inferences. However, this bias is caused partly by the choice of estimator for the higher-order statistics, which relies on shear or convergence maps. Since the SLC impacts the number of sources in each pixel of the map, it also affects the noise of the estimate. The correlation between the source number and the signal then introduces a correlation of the signal variance with the signal itself and, thus, an additional third-order correlation. The effect needs to be modelled for every statistic involving shear or convergence maps but does not occur for statistics measured directly from galaxy catalogues. Thus, we expect SLC to have a negligible effect on higher-order statistics obtained from correlation functions, such as third-order cosmic shear \citep{Schneider2005, Heydenreich2023}. We leave the proper investigation of SLC for third-order shear statistics measured from correlation functions to future work.

\begin{acknowledgements}
We thank the anonymous referee for their helpful suggestions.
\AckEC  

LL is supported by the Austrian Science Fund (FWF) [ESP 357-N]. The Innsbruck authors acknowledge support from the Austrian Research Promotion Agency (FFG) and the Federal Ministry of the Republic of Austria for Climate Action, Environment, Mobility, Innovation and Technology (BMK) via grants 899537 and 900565. SU acknowledges support from the Max Planck Society and the Alexander von Humboldt Foundation in the framework of the Max Planck-Humboldt Research Award endowed by the Federal Ministry of Education and Research. H. Hildebrandt is supported by a DFG Heisenberg grant (Hi 1495/5-1), the DFG Collaborative Research Center SFB1491, as well as an ERC Consolidator Grant (No. 770935). MA is supported by the UK Science and Technology Facilities Council (STFC) under grant numbers ST/Y002652/1 and the Royal Society under grant numbers RGSR2222268 and ICAR1231094.
This work has made use of CosmoHub.

CosmoHub has been developed by the Port d'Informació Científica (PIC), maintained through a collaboration of the Institut de Física d'Altes Energies (IFAE) and the Centro de Investigaciones Energéticas, Medioambientales y Tecnológicas (CIEMAT) and the Institute of Space Sciences (CSIC \& IEEC), and was partially funded by the "Plan Estatal de Investigación Científica y Técnica y de Innovación" program of the Spanish government.

Based on data products from observations made with ESO Telescopes at the La Silla Paranal Observatory under programme IDs 177.A-3016, 177.A-3017, 177.A-3018, 179.A-2004, 298.A-5015. 
\end{acknowledgements}

%
%

\bibliography{my, Euclid}

%

\begin{appendix}
  \onecolumn 
  
\section{Additional term to the convergence power spectrum due to IC source-lens clustering}
\label{app: modelling}
This appendix calculates the expected additional term $\Delta C$ to the lensing power spectrum $C(\ell)$ due to the IC part of source-lens clustering. As discussed in Sect.~\ref{sc:Theo:SLC}, the effect leads to an additional term $\Delta \kappa$ to the lensing convergence. Consequently, the convergence correlation function $\xi_\mathrm{\kappa}(|\thetavec_1-\thetavec_2|)=\expval{\kappa(\thetavec_1)\, \kappa(\thetavec_2)}$ contains additional terms and is
\begin{equation}
    \expval{\kappa(\thetavec_1)\, \kappa(\thetavec_2)} = \expval{\kappa_0(\thetavec_1)\, \kappa_0(\thetavec_2)} + \underbrace{\expval{\kappa_0(\thetavec_1)\, \Delta \kappa(\thetavec_2)}}_{\mathcal{A}(\thetavec_1, \thetavec_2)} + \expval{\Delta \kappa(\thetavec_1)\, \kappa_0(\thetavec_2)} + \underbrace{\expval{\Delta \kappa(\thetavec_1)\, \Delta \kappa(\thetavec_2)}}_{\mathcal{B}(\thetavec_1, \thetavec_2)}\;.
\end{equation}

We are first considering $\mathcal{A}$, which is
\begin{align}
    \mathcal{A}(\thetavec_1, \thetavec_2) = b \int_0^\infty \dd{\chi_1} \int_0^\infty \dd{\chi_2}  \int_{\chi_1}^\infty \dd{\chi'_1} \int_{\chi_2}^\infty \dd{\chi'_2} n(\chi'_1)\, n(\chi'_2)\, W(\chi_1, \chi_1')\, W(\chi_2, \chi_2')\,
    \expval{\delta(\chi_1\thetavec_1, \chi_1)\,\delta(\chi_2\thetavec_2, \chi_2)\,\delta(\chi'_2\thetavec_2, \chi'_2)}\;.
\end{align}
This term vanishes under the Limber approximation. This can be seen qualitatively by considering the three-point correlation function. The correlation function will only give a significant contribution if the densities are evaluated at similar distances. Therefore, $\chi_1\simeq \chi_2 \simeq \chi'_2$. However, for $\chi_2=\chi'_2$, the lensing efficiency $W(\chi_2, \chi'_2)$ vanishes. Consequently, $\mathcal{A}$ becomes zero. 

To see this more clearly, we are Fourier transforming $\mathcal{A}$ to $\Tilde{\mathcal{A}}$, which is
\begin{align}
    \Tilde{\mathcal{A}}(\ellvec_1, \ellvec_2) &= \int \dd[2]{\theta_1}\int \dd[2]{\theta_2} \E^{-\I \thetavec_1\cdot \ellvec_1-\I \thetavec_2\cdot \ellvec_2} \mathcal{A}(\thetavec_1, \thetavec_2)\\
    \notag &= b \,\int_0^\infty \dd{\chi_1}\int_0^\infty \dd{\chi_2} \int_{\chi_1}^\infty\dd{\chi'_1}\, \int_{\chi_2}^\infty\dd{\chi'_2} p(\chi'_1)\, p(\chi'_2)\, W(\chi_1, \chi'_1)\,W(\chi_2, \chi'_2)\\
    \notag &\quad \times \int \frac{\dd{k_{1z}}}{2\pi} \int \frac{\dd{k_{2z}}}{2\pi}\, \E^{\I\, k_{1z}(\chi_1-\chi'_2)}\, \E^{\I\, k_{2z}(\chi_2-\chi'_2)}\, (2\pi)^2 \dirac\left(\frac{\ellvec_1}{\chi_1}+\frac{\ellvec_2}{\chi_2}+\frac{\ellvec_2}{\chi'_2}\right)\\
    &\notag\quad \times B\left[\sqrt{\frac{|\ellvec_1|^2}{\chi^2_1} +k^2_{1z}}, \sqrt{\frac{|\ellvec_2|^2}{\chi^2_2} +k^2_{2z}}, \sqrt{\frac{|\ellvec_2|^2}{{\chi'_2}^2} + (k_{1z}+k_{2z})^2}, \chi_1, \chi_2, \chi'_2\right]\;,
\end{align}
where $B$ is the (three-dimensional) matter bispectrum. Under the Limber approximation, the bispectrum simplifies as
\begin{equation}
 B\left[\sqrt{\frac{|\ellvec_1|^2}{\chi^2_1} +k^2_{1z}}, \sqrt{\frac{|\ellvec_2|^2}{\chi^2_2} +k^2_{2z}}, \sqrt{\frac{|\ellvec_2|^2}{{\chi'_2}^2} + (k_{1z}+k_{2z})^2}, \chi_1, \chi_2, \chi'_2\right]\rightarrow  B\left(\frac{\ell_1}{\chi_1}, \frac{\ell_2}{\chi_2}, \frac{\ell_2}{\chi'_2}, \chi_1, \chi_2, \chi'_2\right)\;.
\end{equation}
Then, $\Tilde{\mathcal{A}}$ becomes
\begin{align}
    &\Tilde{\mathcal{A}}(\ellvec_1, \ellvec_2) \\
    \notag &=(2\pi)^2 b \int_0^\infty \dd{\chi_1}\int_0^\infty \dd{\chi_2}\,  \int_{\chi_1}^\infty\dd{\chi'_1}\, \int_{\chi_2}^\infty\dd{\chi'_2} p(\chi'_1)\, p(\chi'_2)\, W(\chi_1, \chi'_1)\,W(\chi_2, \chi'_2)\,  \dirac(\chi_1-\chi'_2)\, \dirac(\chi_2-\chi'_2)\, \dirac\left(\frac{\ellvec_1}{\chi_1}+\frac{\ellvec_2}{\chi_2}+\frac{\ellvec_2}{\chi'_2}\right)\\
    \notag &\quad \times B\left(\frac{\ell_1}{\chi_1}, \frac{\ell_2}{\chi_2}, \frac{\ell_2}{\chi'_2}, \chi_1, \chi_2, \chi'_2\right)\\
    \notag &= (2\pi)^2b \,\int_0^\infty \dd{\chi_2}\,\int_{\chi_2}^\infty\dd{\chi'_1}\,  p(\chi'_1)\,p(\chi_2)\,  W(\chi_2, \chi'_1)\,W(\chi_2, \chi_2)\,  \dirac\left(\frac{\ellvec_1+2\ellvec_2}{\chi_2}\right)\, B\left(\frac{\ell_1}{\chi_2}, \frac{\ell_2}{\chi_2}, \frac{\ell_2}{\chi_2}, \chi_2, \chi_2, \chi_2\right)\\
    &= 0\;,
\end{align}
where we used in the last step that $W(\chi_2, \chi_2)=0$.

Consequently, the only term that can impact the convergence power spectrum is $\mathcal{B}$. This term is given by
\begin{align}
    \mathcal{B}(\thetavec_1, \thetavec_2) = b^2 \!\!\! \int_0^\infty \!\!\! \dd{\chi_1} \int_0^\infty \!\!\! \dd{\chi_2}  \int_{\chi_1}^\infty \!\!\! \dd{\chi'_1} \int_{\chi_2}^\infty \!\!\! \dd{\chi'_2} n(\chi'_1)\, n(\chi'_2)\, W(\chi_1, \chi_1')\, W(\chi_2, \chi_2')\,
    \expval{\delta(\chi_1\thetavec_1, \chi_1)\,\delta(\chi_2\thetavec_2, \chi_2)\,\delta(\chi'_1\thetavec_1, \chi'_1)\,\delta(\chi'_2\thetavec_2, \chi'_2)}\;.
\end{align}
The four-point correlation function can be decomposed into its connected and unconnected part,
\begin{align}
     \expval{\delta(\chi_1\thetavec_1, \chi_1)\,\delta(\chi_2\thetavec_2, \chi_2)\,\delta(\chi'_1\thetavec_1, \chi'_1\,\delta(\chi'_2\thetavec_2, \chi'_2)} &=
         \expval{\delta(\chi_1\thetavec_1, \chi_1)\,\delta(\chi_2\thetavec_2, \chi_2)\,\delta(\chi'_1\thetavec_1, \chi'_1)\,\delta(\chi'_2\thetavec_2, \chi'_2)}_\mathrm{c}\\
        &\notag+ \expval{\delta(\chi_1\thetavec_1, \chi_1)\,\delta(\chi'_2\thetavec_2, \chi'_2)}\expval{\delta(\chi'_1\thetavec_1, \chi'_1)\,\delta(\chi_2\thetavec_2, \chi_2)}\\
         &\notag+ \expval{\delta(\chi_1\thetavec_1, \chi_1)\,\delta(\chi'_1\thetavec_1, \chi'_1)}\expval{\delta(\chi_2\thetavec_2, \chi_2)\,\delta(\chi'_2\thetavec_2, \chi'_2)}\\
         &\notag+ \expval{\delta(\chi_1\thetavec_1, \chi_1)\,\delta(\chi_2\thetavec_2, \chi_2)}\expval{\delta(\chi'_1\thetavec_1, \chi'_1)\,\delta(\chi'_2\thetavec_2, \chi'_2)}\;.
\end{align}
Using the same arguments as above, only the last summand can contribute to $\mathcal{B}$, since the lensing efficiency suppresses all correlations with $\chi_a\simeq\chi'_a$. Consequently,
\begin{align}
    \mathcal{B}(\thetavec_1, \thetavec_2) = b^2\!\!\! \int_0^\infty\!\!\! \dd{\chi_1} \int_0^\infty\!\!\! \dd{\chi_2} \int_{\chi_1}^\infty\!\!\! \dd{\chi'_1} \int_{\chi_2}^\infty\!\!\! \dd{\chi'_2} n(\chi'_1)\, n(\chi'_2)\, W(\chi_1, \chi_1')\, W(\chi_2, \chi_2')
    \langle{\delta(\chi_1\thetavec_1, \chi_1)\,\delta(\chi_2\thetavec_2, \chi_2)}\rangle\langle{\delta(\chi'_1\thetavec_1, \chi'_1)\,\delta(\chi'_2\thetavec_2, \chi'_2)}\rangle\,.
\end{align}
This is the full impact of source-lens clustering on the convergence correlation function under the assumptions of the Limber approximation.

To find the impact $\Delta C(\ellvec)$ on the convergence power spectrum $C(\ell)$, we Fourier-transform $\mathcal{B}$, so
\begin{align}
    \Delta C(\ellvec_1) (2\pi)^2 \dirac(\ellvec_1+\ellvec_2) &=  \int \dd[2]{\theta_1}\int \dd[2]{\theta_2} \E^{-\I \thetavec_1\cdot \ellvec_1-\I \thetavec_2\cdot \ellvec_2} \mathcal{B}(\thetavec_1, \thetavec_2)\\
    &= b^2 \int_0^\infty \dd{\chi_1} \int_0^\infty \dd{\chi_2}  \int_{\chi_1}^\infty \dd{\chi'_1} \int_{\chi_2}^\infty \dd{\chi'_2} n(\chi'_1)\, n(\chi'_2)\, W(\chi_1, \chi_1')\, W(\chi_2, \chi_2')\\
    &\notag \quad \times \int \frac{\dd[2]{k_{\perp}}}{(2\pi)^2}\, \int \frac{\dd[2]{k'_{\perp}}}{(2\pi)^2} \int \frac{\dd{k_{z}}}{2\pi} \int \frac{\dd{k'_{z}}}{2\pi} P\left[|(\kvec_\perp, k_z)|\right]\, P\left[|(\kvec'_\perp, k'_z)|\right] \, \E^{\I k_{z}\,(\chi_1-\chi_2) + \I k'_{z}\, (\chi'_1-\chi'_2)}\\
    &\notag \quad \times (2\pi)^4\, \dirac(\kvec_\perp\,\chi_1+\kvec'_\perp\,\chi'_1-\ellvec_1)\, \dirac(-\kvec_\perp\,\chi_2-\kvec'_\perp\,\chi'_2-\ellvec_2)\;.
\end{align}
Under the Limber approximation, we replace $P\left[|(\kvec_\perp, k_z)|\right]$ by $P(k_\perp)$, so we can execute the $k_z$ integrals, leading to
\begin{align}
    \Delta C(\ellvec_1) (2\pi)^2 \dirac(\ellvec_1+\ellvec_2) 
    &= b^2 \int_0^\infty \dd{\chi_1} \int_0^\infty \dd{\chi_2} \int_{\chi_1}^\infty \dd{\chi'_1} \int_{\chi_2}^\infty \dd{\chi'_2}  n(\chi'_1)\, n(\chi'_2)\, W(\chi_1, \chi_1')\, W(\chi_2, \chi_2')\, \dirac(\chi_1-\chi_2)\, \dirac(\chi'_1-\chi'_2)\\
    &\notag \quad \times \int \frac{\dd[2]{k_{\perp}}}{(2\pi)^2}\, \int \frac{\dd[2]{k'_{\perp}}}{(2\pi)^2} P(k_\perp)\, P(k'_\perp) \, (2\pi)^4\, \dirac(\kvec_\perp\,\chi_1+\kvec'_\perp\,\chi'_1-\ellvec_1)\, \dirac(-\kvec_\perp\,\chi_2-\kvec'_\perp\,\chi'_2-\ellvec_2)\\
        &= b^2 \int_0^\infty \dd{\chi}   \int_{\chi}^\infty \dd{\chi'} n^2_\mathrm{s}(\chi')\,  W^2(\chi, \chi')\, \frac{1}{\chi'^2} \int \frac{\dd[2]{k_{\perp}}}{(2\pi)^2}\, P(k_\perp)\, P\left(\frac{|\ellvec_1-\kvec_\perp\,\chi|}{\chi'}\right) \, (2\pi)^2\,  \dirac(-\ellvec_1-\ellvec_2)\\
            &= (2\pi)^2\,  \dirac(\ellvec_1+\ellvec_2)\, b^2  \int_0^\infty \dd{\chi} \int_{\chi}^\infty \dd{\chi'} \int \frac{\dd[2]{L}}{(2\pi)^2}\, \frac{n^2_\mathrm{s}(\chi')}{\chi'^2}  \frac{W^2(\chi, \chi')}{\chi^2}\,  P\left(\frac{L}{\chi}\right)\, P\left(\frac{|\ellvec_1-\vec{L}|}{\chi}\right) \;, 
\end{align}
where $\vec{L}=\kvec_\perp\,\chi$. Comparing this expression to Equations (47) and (48) of \citet{Krause2021} shows that 
\begin{equation}
    \Delta C(\ellvec) = \Delta C_{EE}(\ellvec) + \Delta C_{BB}(\ellvec)\;,
\end{equation}
where $\Delta C_{EE}$ and $\Delta C_{BB}$ are their additional terms to the cosmic shear $E$- and $B$-mode power spectra due to source-lens-clustering without magnification bias. Consequently, we confirm the expression for the source-lens clustering effect in \citet{Krause2021}. This is different from the expression given in \citet{Deshpande-EP28}. There, they modelled the angular position-dependent $n(\chi\,\thetavec, \chi)$ by
\begin{equation}
    n(\chi\,\thetavec, \chi) = n(\chi) \left[1+b\,\kappa(\thetavec)\right]\;,
\end{equation}
which is different from our Eq.~\eqref{eq: change n IC only}.

\section{Galaxy bias of considered simulated galaxies}
\label{app: bias}
The clustering of galaxies is strongly dependent on the galaxy bias, which is the ratio between the galaxy and matter densities. To enable comparisons of our results to other galaxy samples, we report here the biases of the considered source galaxies.

The bias $b$ is defined by the square of the ratio between the galaxy power spectrum $P_\mathrm{g}(k)$ and the matter power spectrum $P(k)$,
\begin{equation}
    b(k) = \sqrt{\frac{P_\mathrm{g}(k)}{P(k)}}\;.
\end{equation}

To measure the bias, we follow \citet{Jullo2012} and \citet{Simon2018} and consider the aperture number count $\mathcal{N}$, defined as
\begin{equation}
    \mathcal{N}(\theta;\vartheta) = \int \dd[2]{\vartheta'} \kappa_\mathrm{g}(|\varthetavec-\varthetavec'|)\, U_\theta(\vartheta)\;,
\end{equation}
where $\kappa_\mathrm{g}$ is the two-dimensional galaxy overdensity and $U_\theta$ is a compensated filter of scale radius $\theta$, that is $\int \dd[2]{\vartheta} U_\theta(\vartheta) = 0$.
The variance of the aperture number count,
\begin{equation}
    \langle \mathcal{N}^2 \rangle (\theta) = \langle \mathcal{N}^2(\theta;\vartheta) \rangle\;,
\end{equation}
can be related to the two-dimensional galaxy power spectrum $C_\mathrm{g}(\ell)$ and the three-dimensional $P_\mathrm{g}(k)$ by \citep{Schneider2002}
\begin{align}
    \langle \mathcal{N}^2 \rangle (\theta) 
    &= 2\pi \int_0^\infty \dd{\ell} \ell\, \hat{U}_\theta(\ell)\, C_\mathrm{g}(\ell)\\
        &= 2\pi \int_0^\infty \dd{\ell} \ell\, \hat{U}_\theta(\ell)\, \int_0^\infty \dd{\chi}\, \frac{p^2(\chi)}{\chi^2}\, P_\mathrm{g}\left(\frac{\ell+1/2}{\chi}, \chi\right)\\
               &= 2\pi \int_0^\infty \dd{\ell} \ell\, \hat{U}_\theta(\ell)\, \int_0^\infty \dd{\chi}\, \frac{p^2(\chi)}{\chi^2}\, b^2\left(\frac{\ell+1/2}{\chi}\right)\,P\left(\frac{\ell+1/2}{\chi}, \chi\right)\;,
\end{align}
where hats denote Fourier transform and we applied the Limber approximation.

A common assumption is a linear, scale-independent bias, that is, $b(k) = b$ for all $k$. Then,
\begin{align}
    \langle \mathcal{N}^2 \rangle (\theta) 
               &= b^2\, 2\pi \int_0^\infty \dd{\ell} \ell\, \hat{U}_\theta(\ell)\, \int_0^\infty \dd{\chi}\, \frac{p^2(\chi)}{\chi^2}\,P\left(\frac{\ell+1/2}{\chi}, \chi\right)\\
               &= b^2\, \langle \mathcal{N}^2 \rangle_{b=1}(\theta)\;.
\end{align}
Then, to estimate $b^2$ one needs to measure $\langle \mathcal{N}^2 \rangle (\theta) $ and divide it by a modelled $\langle \mathcal{N}^2 \rangle_{b=1}(\theta)$.

To model $\langle \mathcal{N}^2 \rangle(\theta)_{b=1}$, we use the same non-linear matter power spectrum as described in Sect.~\ref{sc:Analysis:Model} (i.e. HMCode-2020 for the KiDS-like and EuclidEmulator for the \Euclid-like set-up). 
We measure $\langle \mathcal{N}^2 \rangle(\theta)$ by first measuring the angular clustering function $w(\vartheta)$ for each tomographic bin with a Landy--Szalay estimator. Then, we use the relation of $\langle \mathcal{N}^2 \rangle(\theta)$ to $\omega$, 
\begin{equation}
    \langle \mathcal{N}^2 \rangle(\theta) = \int \dd{\vartheta} \frac{\vartheta}{2\theta^2}\, T_\theta(\vartheta)\, \omega(\theta)\;,
\end{equation}
where $T$ can be inferred from the filter function $U$. 
We use the filter from \citet{Crittenden2002},
\begin{equation}
    U_\theta(\vartheta) = \frac{1}{2\pi}\, \left[1-\frac{\vartheta^2}{2\theta^2}\right]\, \E^{-\vartheta^2/(2\theta^2)}\;,
\end{equation}
\begin{equation}
    T_\theta(\vartheta) = \frac{1}{128}\left[\left(\frac{\vartheta}{\theta}\right)^4 - 16 \left(\frac{\vartheta}{\theta}\right)^2 + 32 \right]\,\E^{-\vartheta^2/(2\theta^2)}\;.
\end{equation}

\begin{figure}
    \centering
    \includegraphics[width=0.49\linewidth]{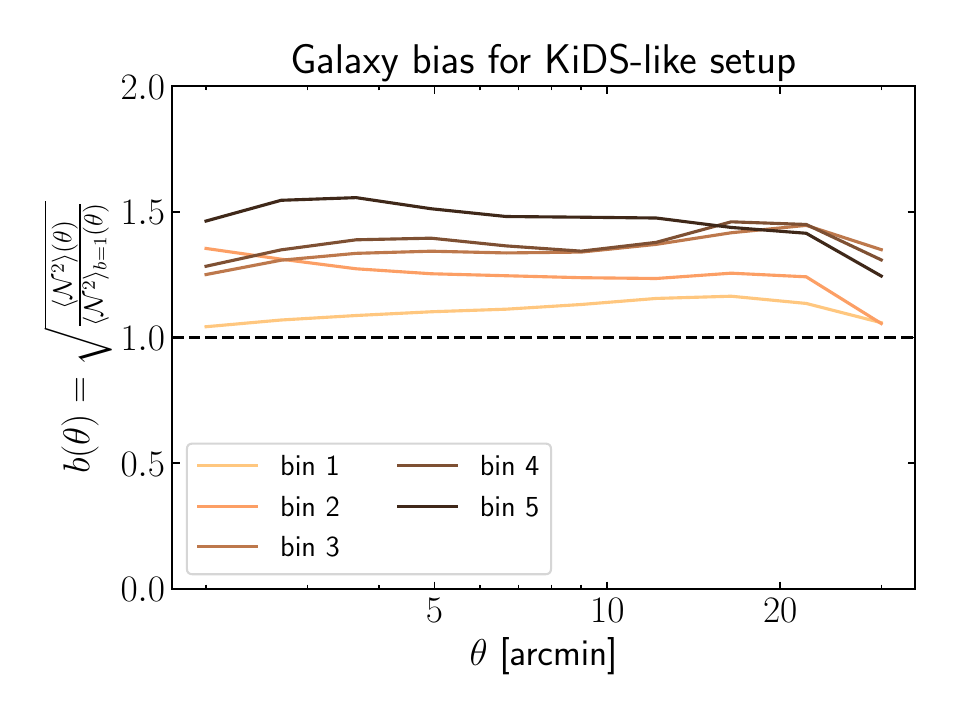}
    \includegraphics[width=0.49\linewidth]{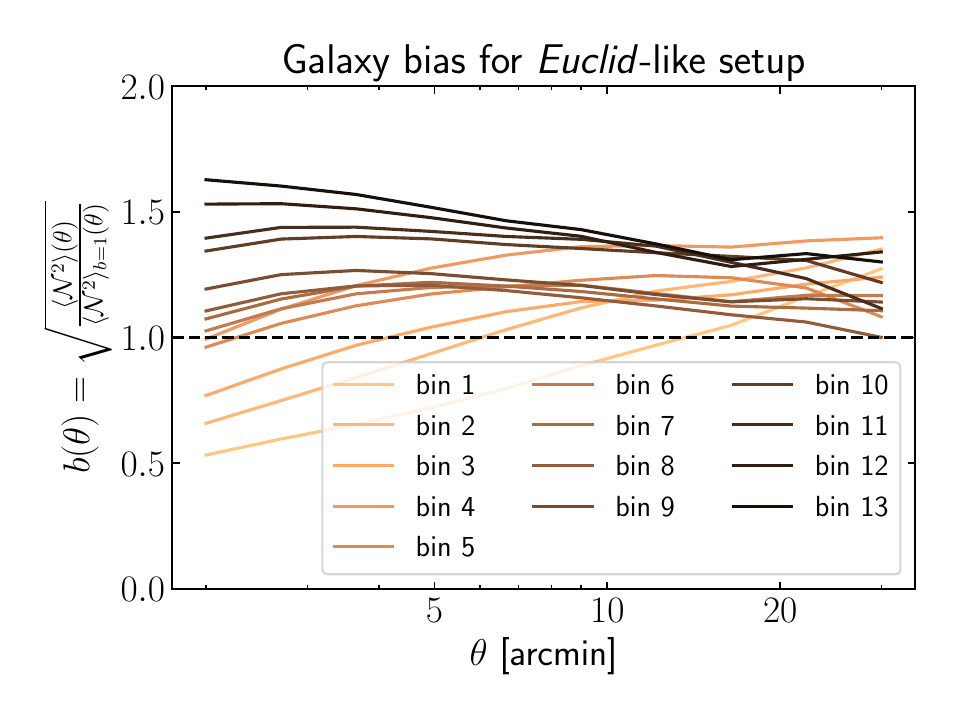}
    \caption{Galaxy biases inferred from the aperture number count for the KiDS-like (left) and \Euclid-like (right) set-up. Each line corresponds to one tomographic bin with darker colours referring to higher mean redshifts.}
    \label{fig: bias}
\end{figure}

The resulting values for $b$ for each tomographic bin are shown in Fig.\ref{fig: bias}. The galaxy bias stays in all cases between 0.5 and 1.5, which is similar to what is expected for cosmic shear source galaxy samples. For both the KiDS and the \Euclid-like set-up, the bias increases with tomographic bin. This is expected since we are using a magnitude-selected sample. At higher redshifts, we include more galaxies with higher luminosity and stellar mass. These galaxies tend to have higher biases, as they predominantly reside in denser matter environments.

\section{Impact of SLC on the cosmic shear covariance}
\label{app: covariance}

We investigate the impact of SLC on the covariance by estimating covariances directly from the galaxy catalogues. For this, we use the jackknife resampling method, as implemented in \verb|treecorr| \citep{Jarvis2004}. We choose 1200 jackknife patches of the same area and measure the shear correlation functions $\xi_+$ and $\xi_-$ in each patch for all tomographic redshift combinations, both for the clustered and the unclustered sources. The jackknife resampling algorithm then takes 1200 averages over all but one of the individual patch estimates, leaving out a different patch every time. The sample covariance of these averages is our covariance estimate (aside from a constant prefactor that depends on the number of patches).

\begin{figure}
    \centering
    \includegraphics[width=\linewidth]{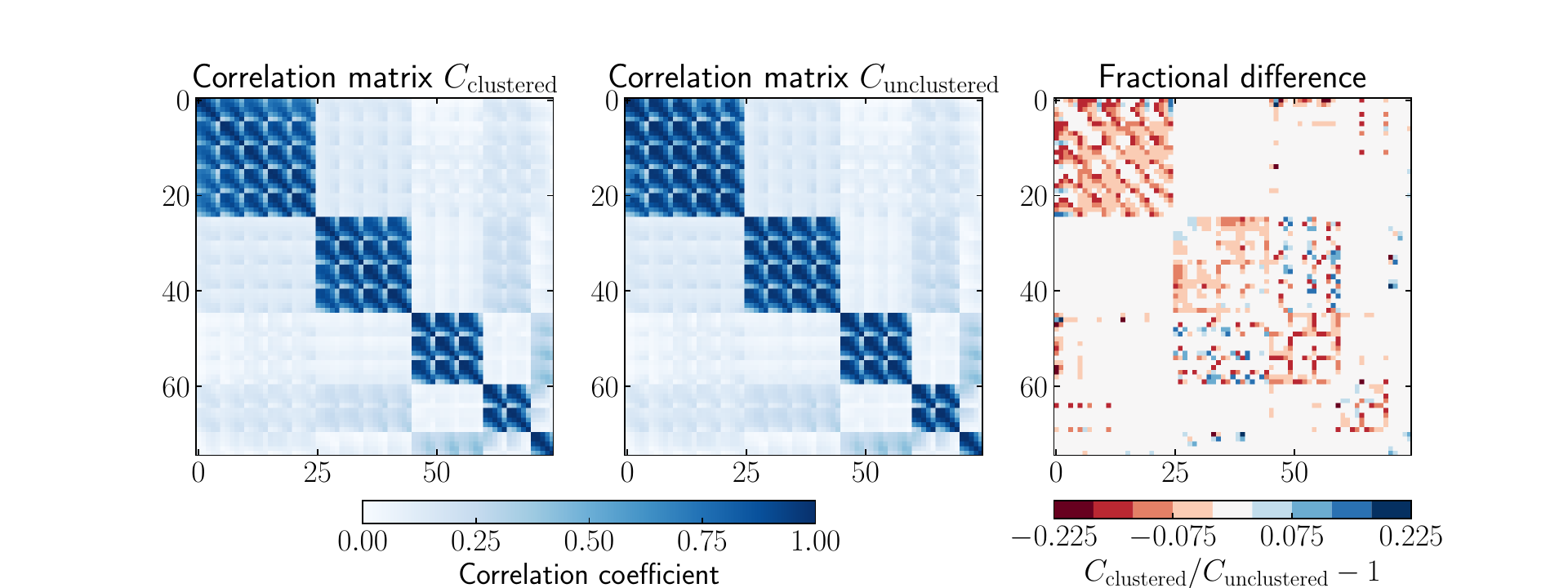}
    \caption{Correlation matrices for $\xi_+$ for the KiDS-like set-up, using either the clustered or the unclustered source galaxies, and the fractional difference between the two cases.}
    \label{fig: corrMats}
\end{figure}

Figure~\ref{fig: corrMats} shows the correlation matrices for the clustered and unclustered sources, as well as their fractional difference for the KiDS-like set-up. For clarity, we only show the matrices for $\xi_+$, our conclusions also hold for $\xi_-$.

The two matrices show the same dependence on angular scale and redshift bin. All individual elements agree better than 15\%, with 95\% of elements agreeing better than 5\%. This indicates that SLC has only a minor impact on the covariance and our simplification in the main analysis to omit the SLC in the covariance is justified.

We note that covariances estimated from jackknife resampling can be biased on large scales, since they assume that the individual patches are uncorrelated. However, since SLC is a small-scale effect, we expect the bias due to the jackknife resampling to be the same for the covariance with clustered and with unclustered source galaxies.

\section{Impact of individual SLC effects on shear correlation functions}
\label{app: ind Effects Xip}

\begin{figure*}
    \centering
    \includegraphics[width=\linewidth]{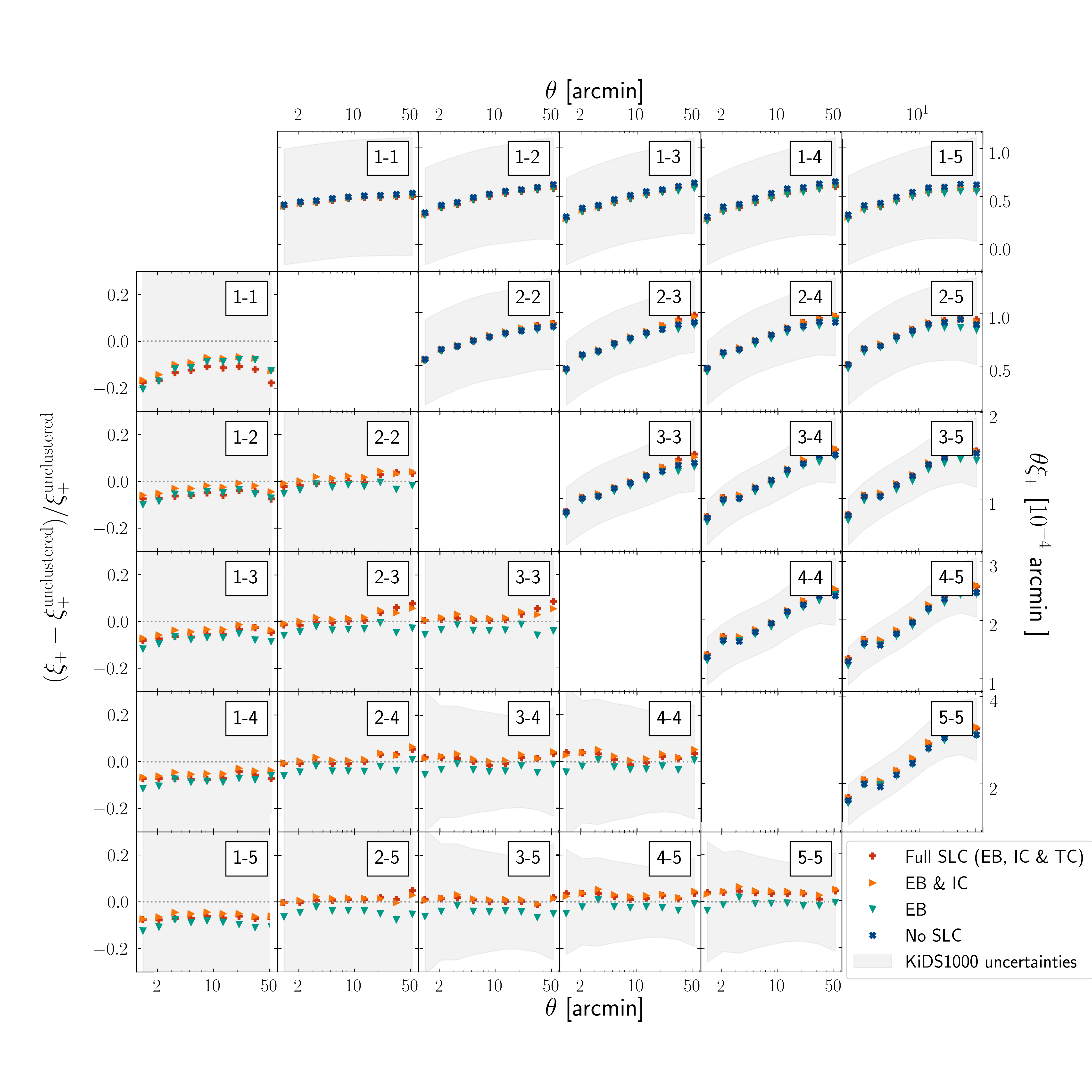}
    \caption{Shear correlation function $\xi_+$ for the KiDS-like set-up when different SLC effects are included, either the full SLC (red crosses), the estimator bias (EB) and intrinsic clustering (IC; orange triangles), only the estimator bias (EB; green triangles), and no SLC (blue dots). \textit{Upper right}: Shear correlation functions. \textit{Lower left}: Fractional difference to $\xi_+$ for unclustered sources (i.e. without SLC).}
    \label{fig: MICE xip individual effects}
\end{figure*}

The scale dependence of the SLC effects is easier considered when using shear correlation functions than COSEBIs, since they directly give the dependence on angular scales. We show in Fig.~\ref{fig: MICE xip individual effects} the shear correlation functions $\xi_+$ when different parts of the SLC are included, as well as the fractional difference to the $\xi_+$ for unclustered sources. Similarly to the COSEBIs, the estimator bias (EB) has the largest impact on the signal, which suppresses the signal across all scales. It has the strongest effect at small scales for the lowest tomographic bin, where source galaxy positions are show the strongest correlation. The intrinsic clustering (IC) effect, which partially cancels the EB effect, is also strongest at small scales, so the combination of EB and IC is roughly scale-independent. The TC effect has a similar strength across all scales.

\section{Additional figures}
\label{app: contours}
For completeness, we report in this section the posteriors for all cosmological and nuisance parameters, which were varied in the cosmological inference, as well as the measured COSEBIs for all tomographic bins in the \Euclid-like set-up.
\begin{figure}
    \centering
    \includegraphics[width=\linewidth]{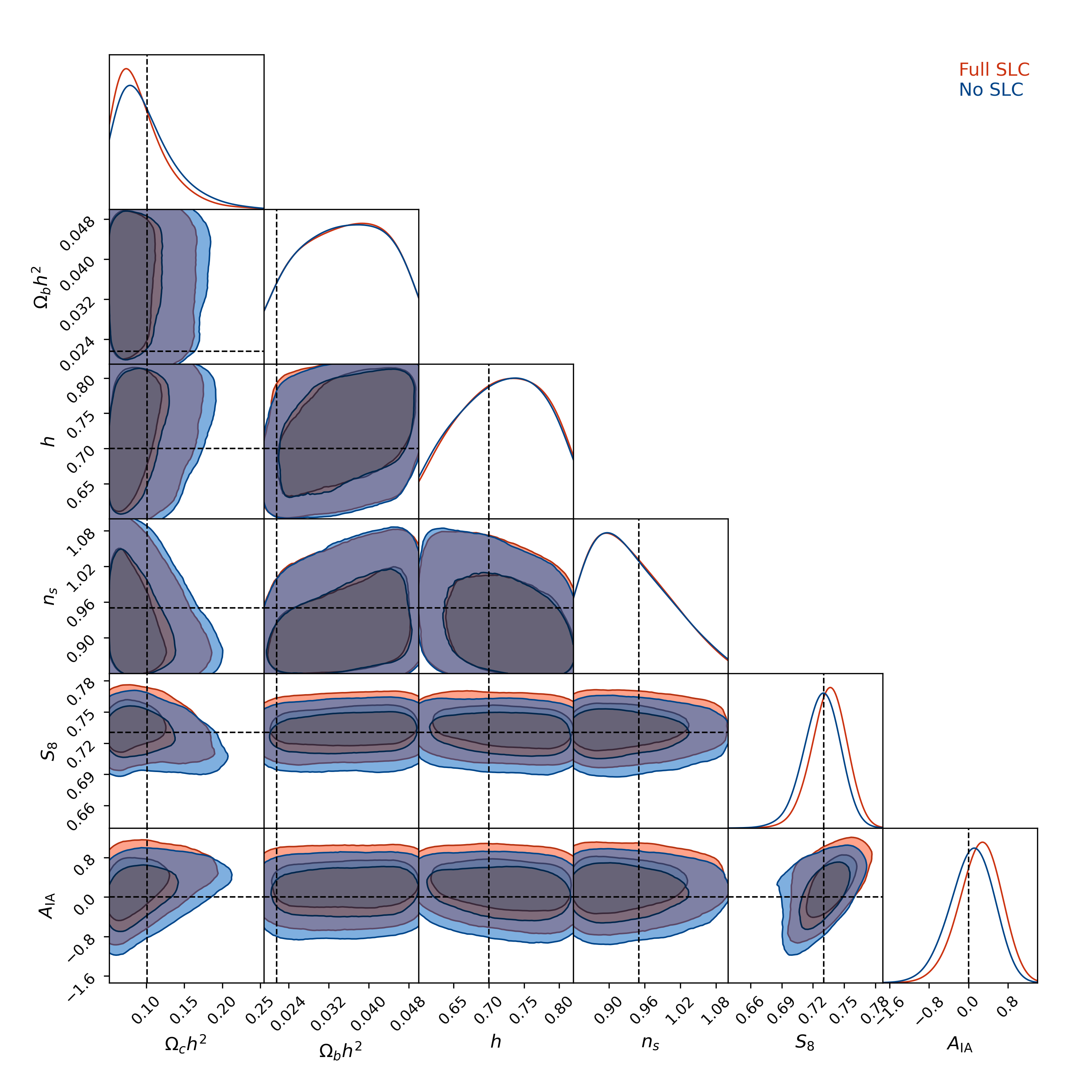}
    \caption{Cosmological parameter constraints for the KiDS-like set-up with nuisance parameters.}
\end{figure}

\begin{figure}
    \centering
    \includegraphics[width=\linewidth]{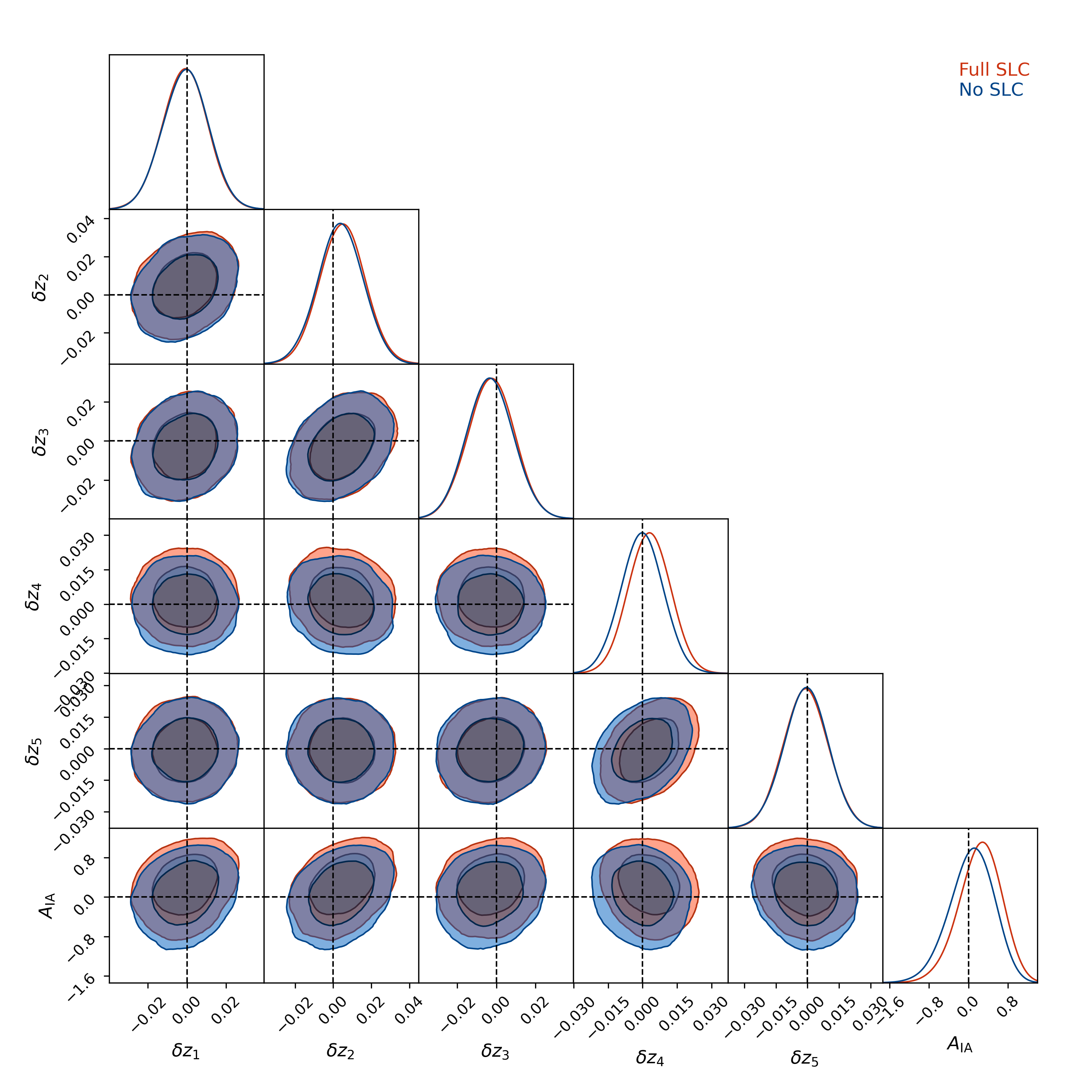}
    \caption{Nuisance parameter constraints for the KiDS-like set-up.}
\end{figure}

\begin{figure}
    \centering
    \includegraphics[width=\linewidth]{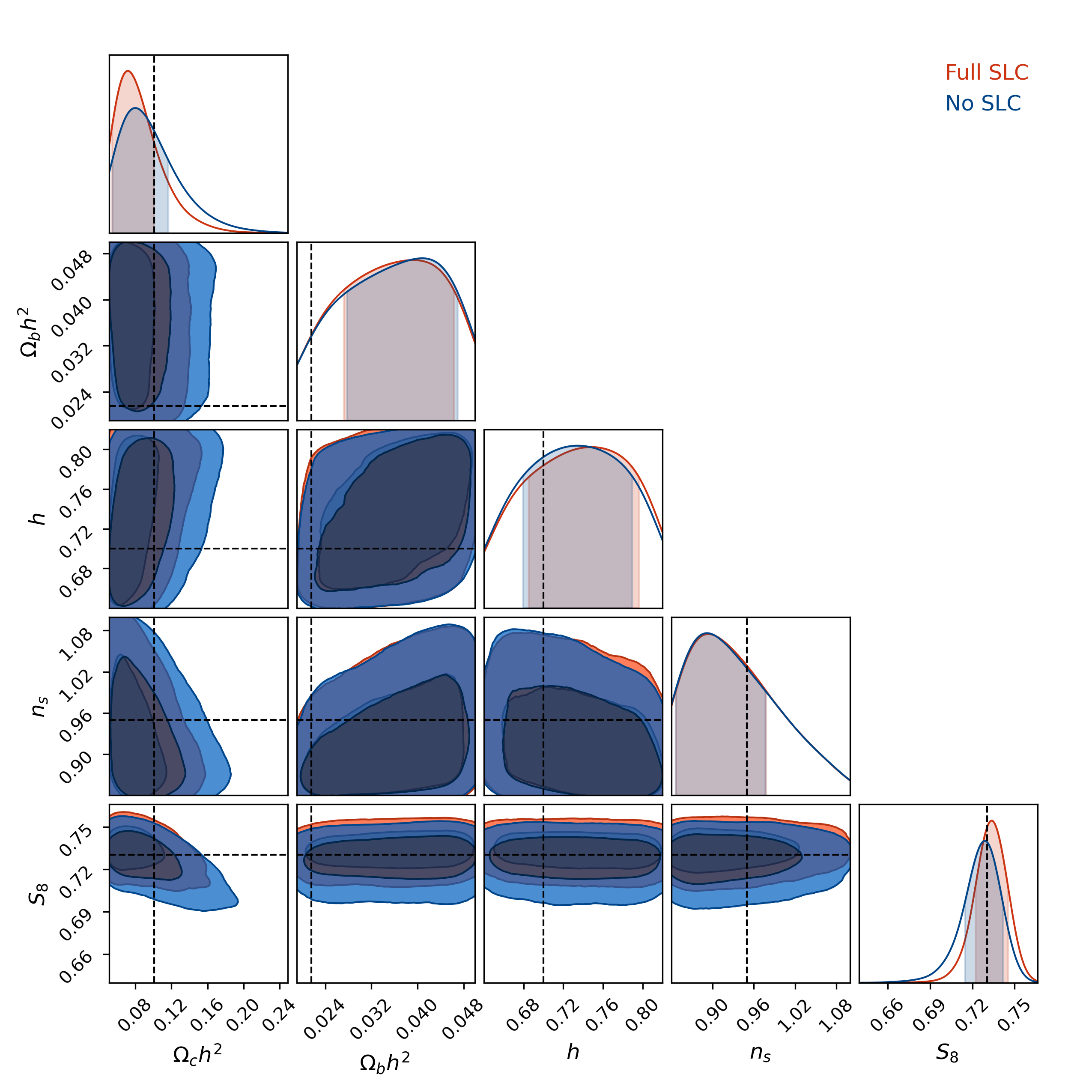}
    \caption{Cosmological parameter constraints for the KiDS-like set-up without nuisance parameters.}
\end{figure}

\begin{figure}
    \centering
    \includegraphics[width=\linewidth]{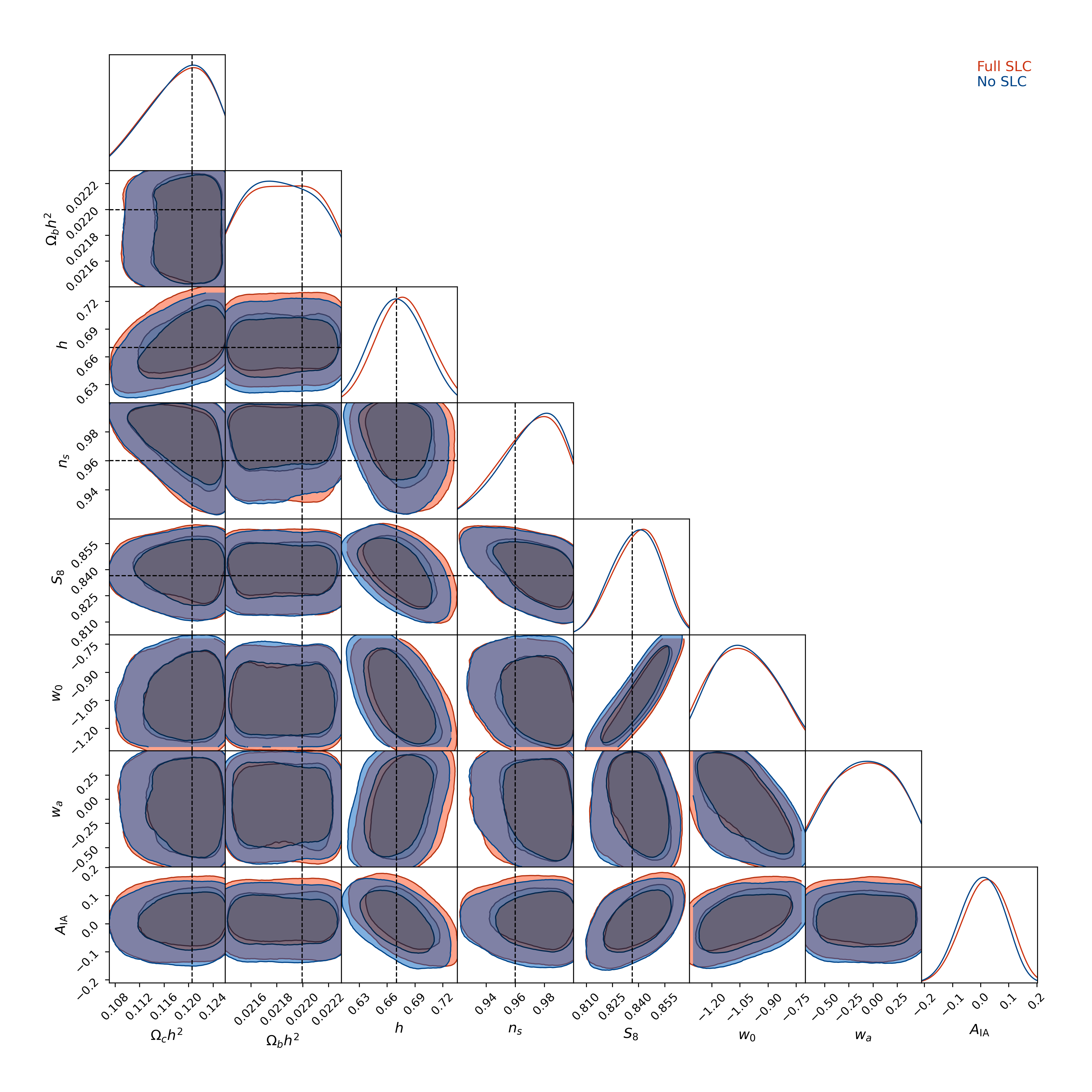}
    \caption{Cosmological parameter constraints for the \Euclid-like set-up with nuisance parameters.}
\end{figure}

\begin{figure}
    \centering
    \includegraphics[width=\linewidth]{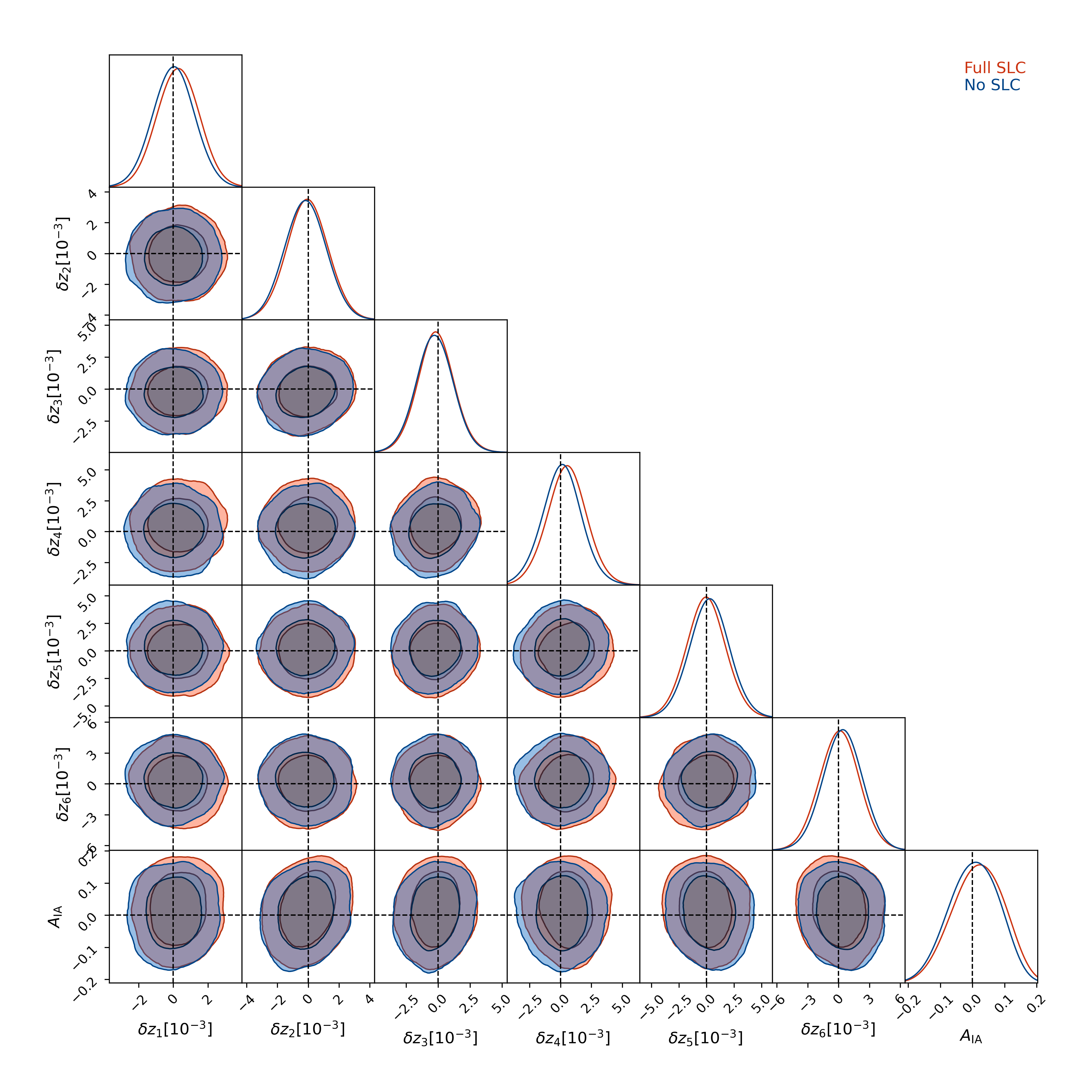}
    \caption{Nuisance parameter constraints for the \Euclid-like set-up.}
\end{figure}

\begin{figure}
    \centering
    \includegraphics[width=\linewidth]{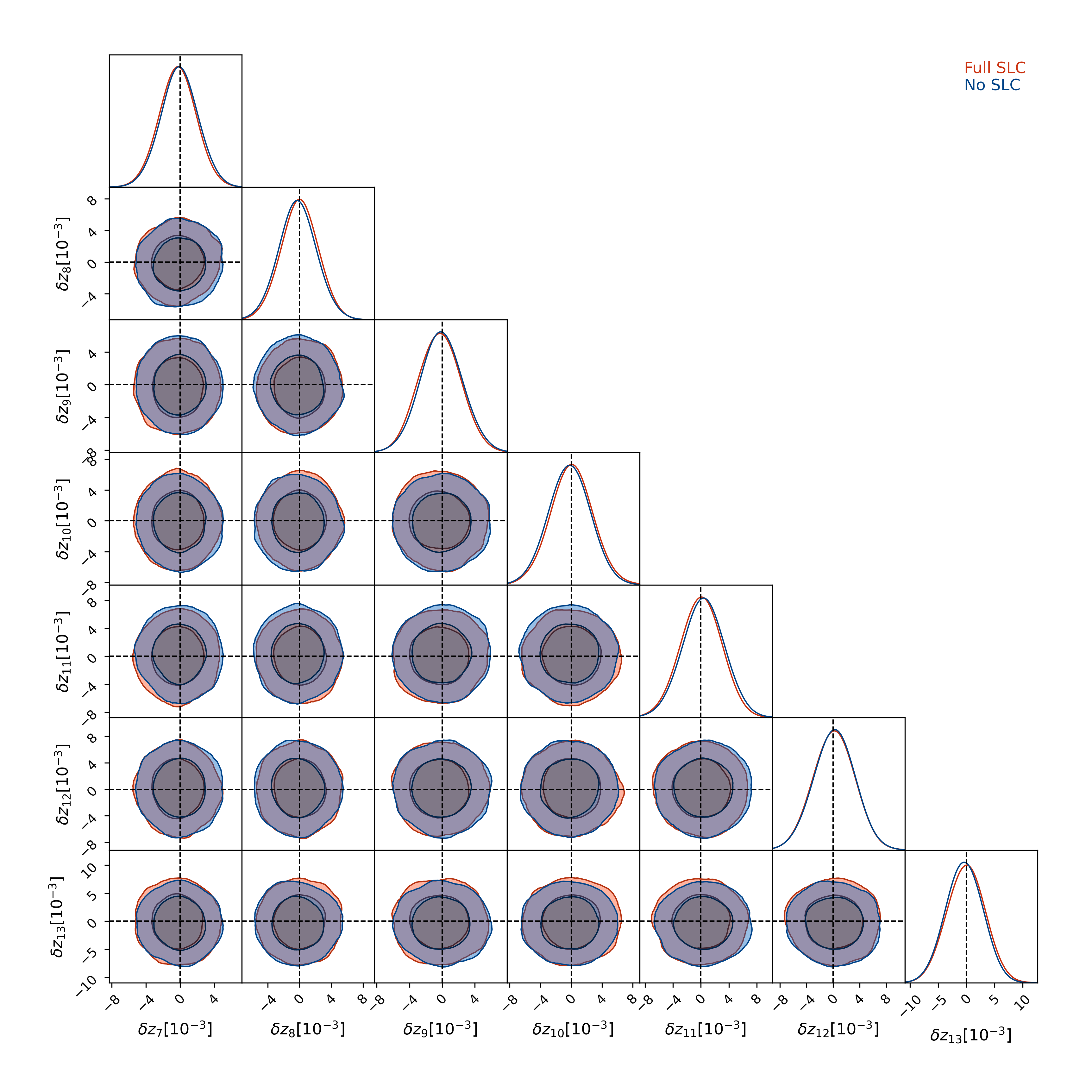}
    \caption{Further nuisance parameter constraints for the \Euclid-like set-up.}
\end{figure}

\begin{figure}
    \centering
    \includegraphics[width=\linewidth]{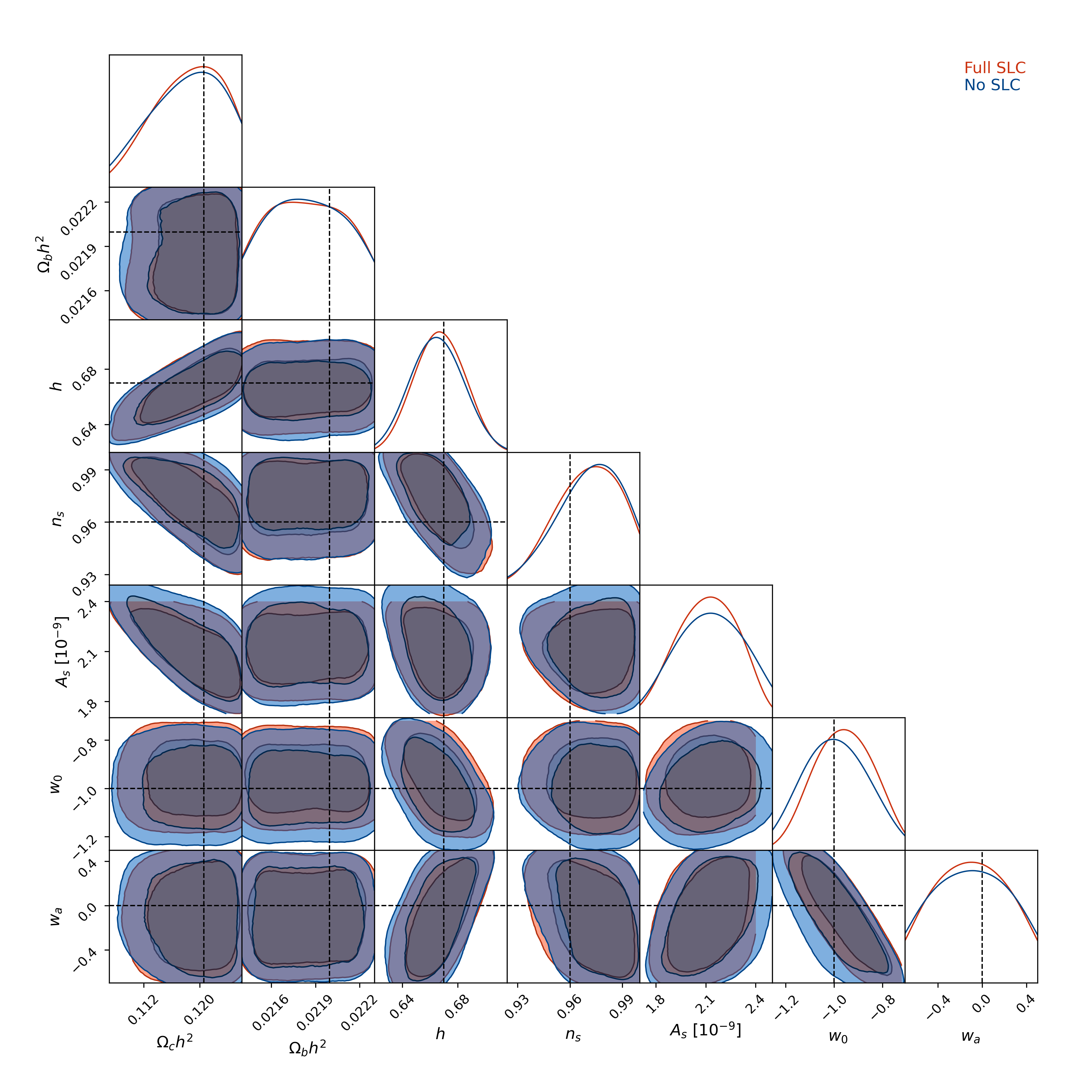}
    \caption{Cosmological parameter constraints for the \Euclid-like set-up without nuisance parameters.}
\end{figure}

\begin{figure*}
    \includegraphics[width=\textwidth, trim={4.5cm 4.5cm 2cm 4.5cm},clip]{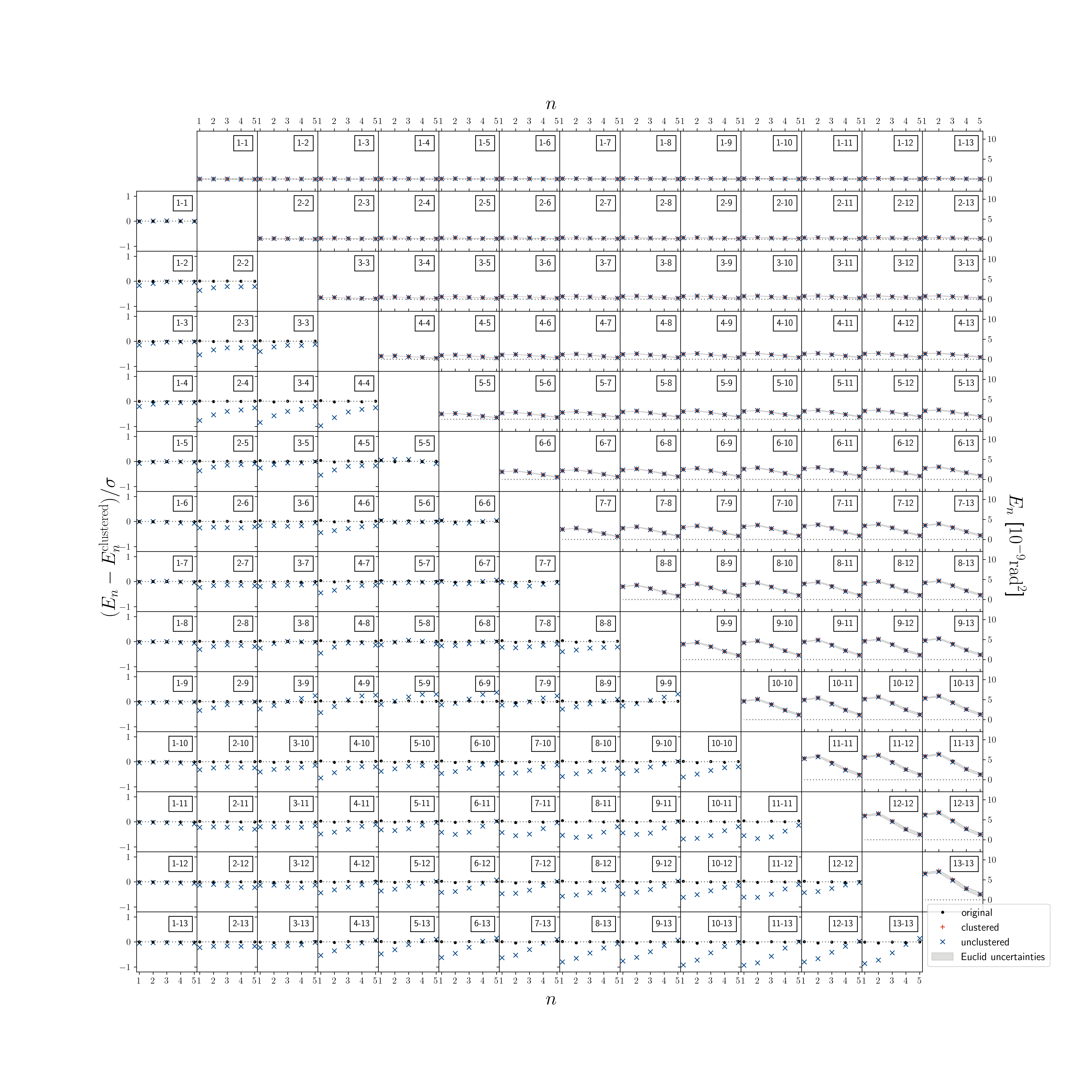}
    \caption{Measurement for the \Euclid-like set-up. \textit{Upper right}: COSEBI $E$-modes $E_n$ measured for the original FS2 galaxies (black points), the clustered catalogue (red plusses), and the unclustered catalogue (blue crosses) with the \Euclid uncertainty (grey band). \textit{Lower left}: Difference between $E_n$ for the original and unclustered sources (black points), and between the clustered and unclustered sources (blue crosses), divided by \Euclid uncertainty.}
    \label{fig: cosebisEuclid_all}
\end{figure*}

\end{appendix}

\end{document}